\documentclass[12pt]{article}

\usepackage[utf8]{inputenc}
\usepackage[T1]{fontenc}
\usepackage{amsmath, amssymb, amsfonts}
\usepackage{graphicx}
\usepackage{caption}
\usepackage{authblk}
\usepackage{geometry}
\usepackage{hyperref}
\usepackage{booktabs}
\usepackage{mhchem}
\usepackage{orcidlink}
\usepackage{lmodern}
\usepackage{setspace}
\usepackage{url}
\usepackage{subcaption}
\usepackage{float} 
\usepackage{multirow}
\usepackage{tikz}
\usetikzlibrary{arrows.meta,positioning,fit,backgrounds}

\usepackage{algorithm}
\usepackage{algpseudocode}
\algrenewcommand\algorithmicindent{1.0em}  

\title{ClusTEK: A grid clustering algorithm augmented with diffusion imputation and origin-constrained connected-component analysis: \\ Application to polymer crystallization}

\author[1]{Elyar Tourani\orcidlink{0009-0003-5181-6835}}
\author[1]{Brian J. Edwards\orcidlink{0000-0002-2378-5627} \thanks{Email: \texttt{bje@utk.edu}}}
\author[1]{Bamin Khomami\orcidlink{0000-0002-0091-2312} \thanks{Email: \texttt{bkhomami@utk.edu}}}
\affil[1]{Materials Research and Innovation Laboratory, Department of Chemical and Biomolecular Engineering, University of Tennessee, Knoxville, TN 37996, USA}

\date{}

\begin{document}
\maketitle

\newpage 

\section*{abstract}
Grid clustering algorithms are valued for their efficiency in large-scale data analysis but face persistent limitations: parameter sensitivity, loss of structural detail at coarse resolutions, and misclassifications of edge or bridge cells at fine resolutions. Previous studies have addressed these challenges through adaptive grids, parameter tuning, or hybrid integration with other clustering methods, each of which offers limited robustness. This paper introduces a grid clustering framework that integrates Laplacian-kernel diffusion imputation and origin-constrained connected-component analysis (OC-CCA) on a uniform grid to reconstruct the cluster topology with high accuracy and computational efficiency. During grid construction, an automated preprocessing stage provides data-driven estimates of cell size and density thresholds. The diffusion step then mitigates sparsity and reconstructs missing edge cells without over-smoothing physical gradients, while OC-CCA constrains component growth to physically consistent origins, reducing false merges across narrow gaps. Operating on a fixed-resolution grid with spatial indexing ensures the scaling of \( \mathcal{O}(n \log n) \). Experiments on synthetic benchmarks and polymer simulation datasets demonstrate that the method correctly manages edges, preserves cluster topology, and avoids spurious connections. Benchmarking on polymer systems across scales (9k, 180k, and 989k atoms) shows that optimal preprocessing, combined with diffusion-based clustering, reproduces atomic-level accuracy and captures physically meaningful morphologies while delivering accelerated computation.

\vspace{4mm}
Keywords: Molecular simulation data, Grid clustering, Diffusion imputation, Laplacian kernel, Connected component analysis, Polymer crystallization

\begin{figure}[b]
    \centering
    \includegraphics[width=0.65\linewidth]{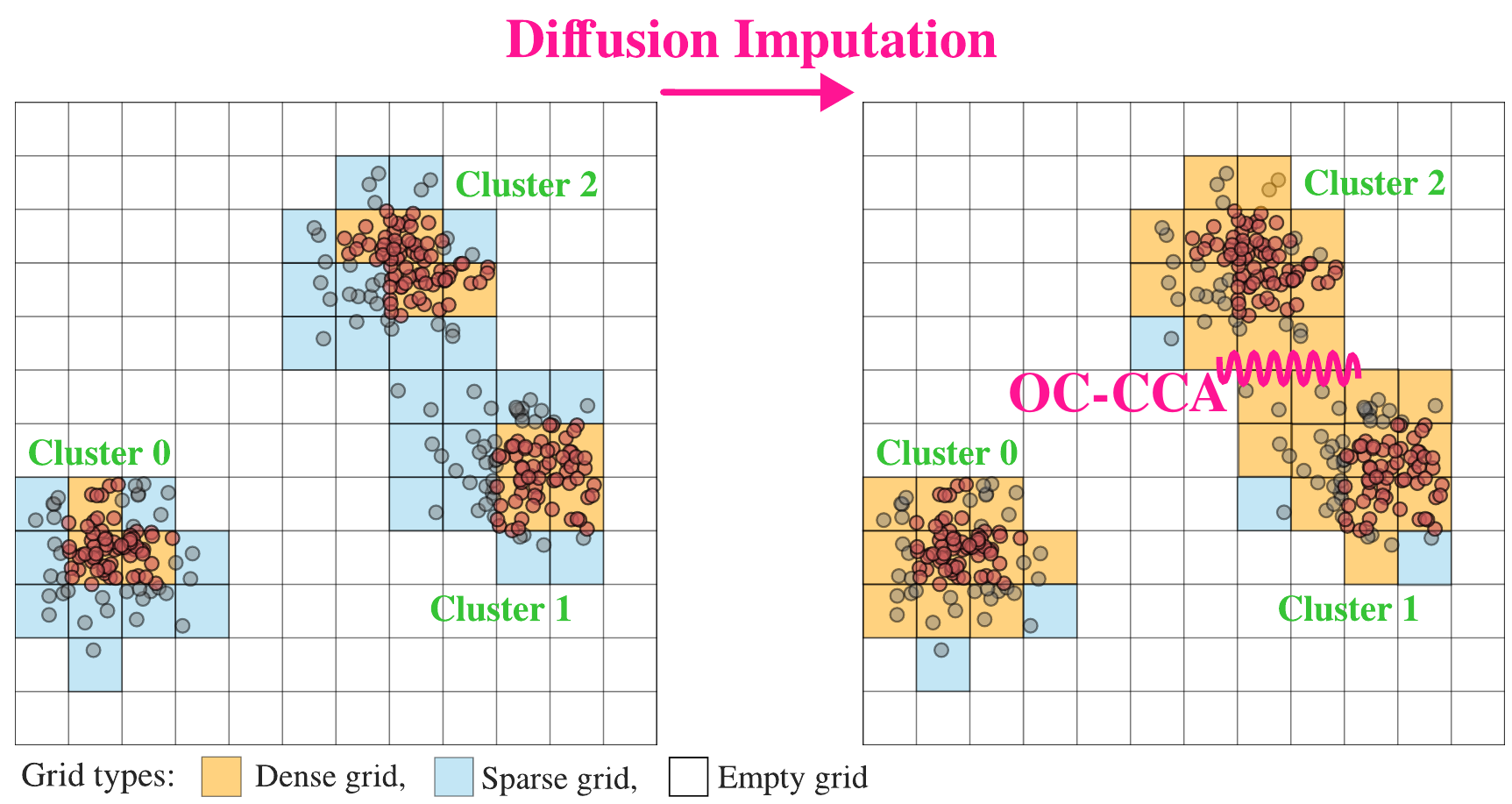}
    \caption*{}
    \label{fig:absfig}
\end{figure}

\newpage


\section{Introduction}
\label{sec:intro}

Grid-based clustering is a widely used approach in large-scale data analysis that partitions a bounded data domain into discrete cells and clusters rather than individual points \cite{cheng2018grid}. By aggregating statistics per cell, grid methods deliver favorable runtime and memory scaling, enable linear data passes, support cheap adjacency queries, and allow highly parallel updates, making them attractive for large datasets and high-throughput workflows. In practice, grid approaches are robust to point-wise noise and map cleanly to GPUs and streaming/distributed settings \cite{aggarwal2014data,al2020review}.

These advantages come with well-known limitations \cite{cheng2018grid,tareq2021systematic}. The results are sensitive to grid resolution and density thresholds: coarse grids suppress fine geometric features, while overly fine grids produce sparsity and fragment connectivity. Fixed grid spacing struggles to accommodate heterogeneous data distributions, often under-resolving dense regions while overemphasizing noise in sparse ones. As resolution increases, the majority of cells become empty or have low occupancy, yielding highly sparse maps where genuinely connected structures may appear artificially broken. The curse of dimensionality further intensifies this sparsity, making meaningful patterns harder to detect. Clusters with strong density variations or thin bridges are easily misclassified, and complex boundaries are only imperfectly represented when membership is based solely on cell occupancy. Together, these issues create a fundamental trade-off between computational efficiency and geometric fidelity.

Following the early canonical grid-based algorithm GRIDCLUS~\cite{schikuta1996grid}, a long series of research has addressed these issues.
Multi-resolution hierarchical grids (e.g., STING~\cite{wang1997sting}) and adaptive local refinements (e.g., MAFIA~\cite{goil1999mafia}, AMR~\cite{liao2004grid}) adjust bin widths or refine regions to improve flexibility, at the cost of increased algorithmic complexity~\cite{aggarwal2014data,wang1997sting,liao2004grid,nagesh2001adaptive}.
Axis-shifting methods (e.g., NSGC~\cite{ma2004new}, GDILC~\cite{yanchang2001gdilc}, ADCC~\cite{lin2007adaptable}) translate the grid and fuse results from displaced coordinate systems to reduce boundary artifacts.
Hybrid approaches combine grids with density or subspace searches (e.g., CLIQUE and its derivatives~\cite{agrawal1998automatic,cheng1999entropy,goil1999mafia}).

 WaveCluster~\cite{sheikholeslami1998wavecluster} applies wavelet filtering in feature space to expose high-density regions across multiple resolutions, while projection- and partition-based methods, such as OptiGrid~\cite{hinneburg1999optimal}, O-Cluster~\cite{milenova2002cluster}, and the Cell-Based Filtering (CBF) method~\cite{chang2002new}, recursively partition the data using axis-parallel hyperplanes or space-partitioning filters to locate dense subregions in high-dimensional spaces.
Although effective, these resolution- and threshold-dependent strategies often compromise the simplicity, interpretability, and algorithmic clarity of a single fixed uniform grid~\cite{chen2007density,he2011mr,DONG2018103,lee2024grid}.

Grid-based analysis has also recently become increasingly relevant in the physical sciences, where spatially resolved data often arise from simulations or imaging. In molecular dynamics (MD) simulation, gridding is used for coarse-grained atomic fields, for example, for crystallinity maps \cite{yamamoto2019molecular,sommer2010molecular,luo2011growth,knox2010probing}. Similar grid strategies are common in electron microscopy and tomography to identify domains, pores, or defects in materials \cite{da2019computations,rabbani2023deepangle,barakati2024physics,barakati2025reward}. Across these systems, challenges such as resolution sensitivity, sparse edges, and fragmented connectivity persist.

Large-scale MD simulations exemplify these challenges: modern trajectories generate terabytes of spatiotemporal data with atomic resolution, demanding scalable tools to identify and track structural motifs across space and time.
Conventional atom-based clustering, while precise, is computationally expensive and poorly suited for the repeated frame-by-frame analysis required for long trajectories. Grid-based clustering offers a scalable alternative, but excessive coarsening risks obscuring physically meaningful structures. 
The locality problem~\cite{cheng2018grid} is acute: a static grid can misalign with true boundaries when multiple structures coexist, leading to artificial fragmentation or spurious merges. 
Axis-shifting ensembles \cite{ma2004new,aggarwal2014data} partially alleviate this issue by averaging results over displaced grids (sliding windows), yet such post-hoc corrections risk introducing nonphysical attachments or detachments of clusters in molecular systems and become computationally prohibitive for long trajectories.

Polymer crystallization is a representative and demanding case study. Nucleation and growth involve complex morphologies, such as cylindrical domains, anisotropic fronts, and transient bridges, which are sensitive to thermodynamic and flow conditions. Despite extensive research \cite{rutledge2013computer,zhang2016review,saalwachter2023recent,yamamoto2009computer,yamamoto2019molecular,yi2013molecular,graham2014modelling,graham2019understanding,Baig2010a,Baig2010b,nafar2020thermodynamically,hall2019divining,DEB}, resolving nucleus shapes, critical sizes, and interfacial morphologies in MD simulations remains a challenge. 
Previous MD analyzes often rely on costly per-atom clustering pipelines \cite{yi2013molecular,nafar2020thermodynamically,hall2019divining,Baig2010a,Baig2010b} or on grid-based clustering approaches \cite{yamamoto2019molecular}, which remain sensitive to resolution and edge classification. 
For example, averaging orientational order within mesh cells, followed by thresholding and connected-component analysis (CCA) \cite{yamamoto2019molecular}, is efficient but can mishandle merges/splits and interfaces when grids are fine (sparse) or coarse (blurry).

To address these limitations, we retain the simplicity of a single fixed grid and introduce two physically motivated components:
(i) We introduce a diffusion-based imputation step. This physically motivated Laplacian convolution smooths scalar fields across neighboring cells, recovering contiguous domains and gradual transitions without over-smoothing sharp interfaces.
The diffusion-imputation step directly addresses grid sparsity by redistributing scalar information from dense cells to neighboring empty ones while preserving physical interfaces.
(ii) To address artifacts arising from post-diffusion bridging, we introduce the origin-constrained connected-component analysis (OC-CCA). This approach restricts merges by ensuring that any new connectivity must originate from cells that were dense prior to diffusion. Consequently, diffusion can repair boundaries, but the merged regions remain anchored to physically meaningful cores.
Together, diffusion improves the scalar field through the grids, and OC-CCA preserves the fine-scale topology, avoids spurious merges, and operates at fixed resolution with \( \mathcal{O}(n \log n) \) complexity. 

For the polymer crystallization case study, we use the crystallinity index ($C$-index)~\cite{Cindex}, a supervised scalar descriptor combining multiple structural features, as the grid field; however, the framework is agnostic to the choice of scalar property (i.e., density, order parameters, entropy or any other desired property). We further propose a lightweight, data-driven procedure to estimate cell size and density thresholds for unseen datasets using a composite of unsupervised criteria. Optionally, ground-truth atom-based labels can be used to tune these hyperparameters for known physical simulation datasets.

The main contributions of this work are:
(i) a physically motivated diffusion-imputation framework for grid-based clustering,
(ii) a novel origin-constrained connected-component analysis (OC-CCA) to prevent artificial merges,
(iii) a practical preprocessing stage for consistent parameter initialization, and
(iv) demonstration of the method on large-scale polymer crystallization datasets, achieving atomic-level accuracy and significant computational speedup.

This study presents a physically interpretable and computationally efficient alternative to traditional grid-based clustering methods. In Section \ref{sec:methods}, we detail the clustering framework, the simulation setup, and the parameter calculations. Section \ref{sec:results} evaluates the precision, efficiency, and morphological sensitivity of clustering between systems with varying complexities. Finally, Section \ref{sec:conclusion} summarizes the key findings and their broader implications.

\section{Methods}
\label{sec:methods}

\subsection{Grid Definition and Parameter Selection (preprocessing)}
\label{subsec:gridding}
We discretize the simulation domain into a uniform rectilinear grid and assign to each cell a scalar value \(C\) derived from the point-wise data. 
This scalar \(C\) may represent any physically or statistically significant quantity, such as a point density in synthetic data or a crystallinity index of atoms when crystallinity is examined from molecular dynamics trajectories.

Let the simulation domain be 
\(
\Omega = [x_{\min},x_{\max}] \times [y_{\min},y_{\max}] \times [z_{\min},z_{\max}],
\)
subdivided into \((n_x, n_y, n_z)\) bins along each axis. 
The corresponding cell widths are 
\(\Delta x = (x_{\max}-x_{\min})/n_x\), 
\(\Delta y = (y_{\max}-y_{\min})/n_y\), and 
\(\Delta z = (z_{\max}-z_{\min})/n_z\).
Each point \(p = (x_p, y_p, z_p)\) is assigned to its grid index,
\[
(i_p, j_p, k_p) = \Bigl(
\bigl\lfloor \tfrac{x_p - x_{\min}}{\Delta x} \bigr\rfloor,
\bigl\lfloor \tfrac{y_p - y_{\min}}{\Delta y} \bigr\rfloor,
\bigl\lfloor \tfrac{z_p - z_{\min}}{\Delta z} \bigr\rfloor
\Bigr),
\]
and the value of cell \((i,j,k)\) is given by the mean over all points within it.
For synthetic datasets, we define the per-cell field from occupancy counts,
i.e., $\text{count}_{i,j,k}=|\mathcal{P}_{i,j,k}|$ and construct the normalized initial field $C^{(0)}\in[0,1]$ by min--max scaling over non-empty cells (superscript $(0)$ denotes the initial state prior to diffusion iterations):
\[
C^{(0)}_{i,j,k} =
\begin{cases}
\dfrac{\text{count}_{i,j,k}-\min(\text{count}_{>0})}{\max(\text{count}_{>0})-\min(\text{count}_{>0})}, & \text{count}_{i,j,k}>0,\\[6pt]
0, & \text{count}_{i,j,k}=0.
\end{cases}
\]
For molecular dynamics data, each point carries a scalar attribute $C_p$
(e.g., a per-atom crystallinity index), and we set $C_{i,j,k}$ to the mean of $C_p$
over points in the cell:
\(
C_{i,j,k} = 
\frac{1}{|\mathcal{P}_{i,j,k}|}
\sum_{p\in\mathcal{P}_{i,j,k}} C_p,
\)
where \(\mathcal{P}_{i,j,k}\) is the set of points assigned to that cell.
Cells are categorized according to their scalar values relative to the threshold $C_{\mathrm{thr}}$: dense ($C_{i,j,k} > C_{\mathrm{thr}}$), sparse ($0 < C_{i,j,k} \le C_{\mathrm{thr}}$), or empty ($C_{i,j,k} = 0$).

Connections between cells are established using axis-aligned adjacency (4-neighbor in 2D, 6-neighbor in 3D), with optional support for corner-aligned adjacency and periodic wrapping.
For molecular dynamics data, cells that lack samples due to discretization are labeled \texttt{NaN}, distinguishing them from physically empty cells ($C_{i,j,k} = 0$), which are sampled but have a zero mean field.

\paragraph{Parameterization strategies (Stage~I).}
Stage~I determines the grid resolution and prediffusion selection threshold through two interchangeable, fully unsupervised strategies. 
Alternatively, users with prior domain knowledge can bypass Stage~I by specifying fixed parameters (\texttt{FIXED\allowbreak\_GRID}, \texttt{FIXED\allowbreak\_DENSE\allowbreak\_THR}).
For example, in molecular dynamics data, one may choose a grid size comparable to a characteristic correlation length or tune parameters using atom-based clustering from a representative snapshot as a reference ground truth.

\paragraph{Grid suggestion.}
Let $N$ be the number of points and $L_x=x_{\max}-x_{\min}$ (similarly \(L_y,L_z\)).
We estimate an isotropic cell edge length $h_0$ from three independent estimates implemented in \texttt{suggest\_grid\_size}: 
(i) \emph{k-nearest-neighbor (k-NN) spacing:} build a cKDTree, take the median $k$-NN distance $s_k$ (default $k{=}5$), and set $h_{\mathrm{k-NN}}=\alpha s_k$ with default $\alpha = 0.8$;
~(ii) \emph{Target occupancy:} given a target mean occupancy \(m\) (default \(\texttt{TARGET\_OCC}=2.5\)), choose the total number of cells \(\hat{G}=\max(1,\lfloor N/m\rfloor)\) and set \((n_x,n_y,n_z)\) proportionally to \((L_x,L_y,L_z)\) with \(n_x n_y n_z \approx \hat{G}\), implying voxel edges \((h_x,h_y,h_z)=(L_x/n_x,L_y/n_y,L_z/n_z)\) and \(h_{\mathrm{occ}}=(h_x h_y h_z)^{1/3}\) ~ (preserving aspect ratio);
~ (iii) \emph{Freedman–Diaconis backup (\texttt{FD\_BACKUP=True}):} widths per-axis \(b_\ell=2\,\mathrm{IQR}(\ell)/N^{1/3}\) for \(\ell\in\{x,y\}\) and \(h_{\mathrm{fd}}=\sqrt{b_x b_y}\) in 2D (geometric mean in 3D).

We take \(h_0 = \mathrm{median}\{h_{\mathrm{k-NN}},h_{\mathrm{occ}},h_{\mathrm{fd}}\}\) (ignoring the unreliable terms, if any), form
\((n_x^0,n_y^0,n_z^0) = (\lceil L_x/h_0\rceil,\lceil L_y/h_0\rceil,\lceil L_z/h_0\rceil)\),
and sweep \(h\) in a band around \(h_0\): \(h \in [(1{-}\rho)h_0,(1{+}\rho)h_0]\) with \(\rho\in[0.2,0.4]\) (\(\rho=\)\texttt{SWEEP\_PCT}=0.2) generates several candidate grids. 
Each $n$ is capped by \(n_{\max}\) (\texttt{MAX\_BINS}=200) to avoid pathologically fine partitions and form a small sweep around \(h_0\) with relative half-width \(\rho=\texttt{SWEEP\_PCT}=0.2\) to ensure distinct candidates \((n_x,n_y)\) during warm-start evaluation.

\paragraph{Unsupervised parameter tuning.}
Two interchangeable unsupervised tuning modes are implemented in Stage~I, differing in the way the candidate grid parameters are proposed and evaluated.
\begin{itemize}
\item \textbf{Method A (\texttt{tuning=grid}).}  
A deterministic grid search is performed on candidate resolutions $(n_x,n_y)$ and dense quantiles $q\in\{0.10,0.15,\dots,0.50\}$ on normalized cell counts.  
For each grid, non-empty cell counts are independently rescaled to $[0,1]$, and cells exceeding the quantile threshold are marked as dense.  
Dense cells are labeled via CCA and scored using the composite metric $\mathcal{Q}$ described below.  
Points in dense cells inherit component labels, whereas points in sparse or empty cells remain unlabeled and are excluded from silhouette and DBI evaluation.

\item \textbf{Method B (\texttt{tuning=bo}, default).}  
Alternatively, a Gaussian-process Bayesian optimization (BO) with an expected-improvement acquisition function, implemented by \texttt{scikit-optimize}, is used to explore the grid scale $h$ and the integer count threshold $R$, warm-started near the heuristic $h_0$ with multiple seeds $R$.
The search space is defined by $\log h \in [\log(\underline{\eta}h_0),\,\log(\overline{\eta}h_0)]$ and $R\in\{R_{\min},\ldots,R_{\max}\}$, with default bounds $(\underline{\eta},\overline{\eta})=(0.5,1.25)$ and $R\in[2,20]$.  
Degenerate configurations (e.g., near-empty or fully filled grids, or a single percolated cluster spanning the domain) are rejected during a sanity check.  
The same unsupervised score $\mathcal{Q}$ guides the optimization: for each proposed $(h,R)$, we (i) compute the dimensions of the grid $n_\ell=\lceil L_\ell/h\rceil$, (ii) classify the cells as dense if their raw count is $\ge R$, (iii) perform CCA on the dense mask, (iv) assign cluster labels to points, and (v) evaluate $\mathcal{Q}$.
\end{itemize}

\paragraph{Composite score function.}
Both tuning strategies (i.e., grid search or BO) maximize the same composite quality metric,
\begin{equation}
\mathcal{Q} = w_{\mathrm{sil}}\cdot \mathrm{sil}
+ w_{\mathrm{dbi}}\cdot \frac{1}{1+\mathrm{DBI}}
+ w_{\mathrm{cov}}\cdot \mathrm{cov},
\end{equation}
where $\mathrm{sil}$ is the silhouette coefficient, $\mathrm{DBI}$ the Davies–Bouldin index, and $\mathrm{cov}$ the coverage fraction (labeled points divided by total points).
The number of detected clusters is limited to $K\in[K_{\min},K_{\max}]=[1,50]$.  
Alternative metrics can be integrated through the modular \texttt{score\_partition} interface.  
By default, the weight triplet $(w_{\mathrm{sil}},w_{\mathrm{dbi}},w_{\mathrm{cov}})=(0.33,0.33,0.33)$ is fixed (\texttt{BO\_OPT\_WEIGHTS\allowbreak=False}), although it may optionally be included as BO parameters, forming a five-dimensional search on $(h,R,w_{\mathrm{sil}},w_{\mathrm{dbi}},w_{\mathrm{cov}})$.  
The best configuration, $(n_x,n_y,q)$ for \texttt{grid} or $(h,R)$ for \texttt{bo}, is then passed to the diffusion and OC-CCA stage (Sec.~\ref{subsec:impute-cca}).

\paragraph{Transition to Stage~II.}
Stage~I concludes once the optimal grid resolution and prediffusion threshold have been identified using either Method~A (grid search) or Method~B (Bayesian optimization). Therefore, Stage~I returns $(n_x,n_y)$ and $C_{\mathrm{thr}}$. 
This selected configuration is passed to Stage~II, where diffusion imputation and OC-CCA are applied. In Stage~II, diffusion imputation is performed on the fixed grid, exploring the values $(\beta, C_{\mathrm{sel}})$ to maximize the composite score $\mathcal{Q}$, and OC-CCA is conducted.

\subsection{Weighted Diffusion Imputation and Origin-Constrained CCA}
\label{subsec:impute-cca}
Finite grid resolution induces sparsity and locality artifacts that can fragment physically connected domains. 
We mitigate these artifacts using a weighted diffusion-based imputation that propagates information from dense cells into adjacent sparse cells while preserving the original dense support and rejecting empty cells. 
In practice, we diffuse a normalized per-cell field $C^{(0)}\in[0,1]$ computed from point counts (for synthetic datasets) or a scalar atom field (for MD crystallinity), as described in Sec.~\ref{subsec:gridding}.

\paragraph{Thresholds carried from Stage~I.}
Stage~I supplies the \emph{dense threshold} $C_{\mathrm{thr}}$, which defines the
prediffusion dense set and sets the update scale for sparse cells.
In \texttt{tuning=grid}, $C_{\mathrm{thr}}$ is chosen as a quantile of the normalized
scalar field. In \texttt{tuning=bo}, a count cutoff $R$ is converted to an equivalent
normalized threshold by taking the minimum $C^{(0)}$ over cells with
$\mathrm{count} \ge R$, guaranteeing an identical dense mask in the normalized domain.

A distinct \emph{selection threshold} $C_{\mathrm{sel}}\in(0,1)$ (denoted \texttt{cthr} in Stage~II's code) is tuned in Stage~II alongside the diffusion coefficient $\beta$; it is applied after diffusion to admit imputed sparse cells.
$C_{\mathrm{sel}}$ functions as a \emph{post-diffusion gate}: lower values favor recall (admitting more imputed cells), higher values favor precision (suppressing halos and spurious bridges).
For example, if $C_{\mathrm{thr}}=0.40$ and a sparse cell has
$C^{(0)}=0.18$ but rises to $C^{(\mathrm{final})}=0.27$ after diffusion,
the cell is admitted for $C_{\mathrm{sel}}=0.20$ (improving coverage) but rejected
for $C_{\mathrm{sel}}=0.30$ (preventing weak halo connections).

\begin{figure}[h!]
\centering
\includegraphics[width=0.6\textwidth]{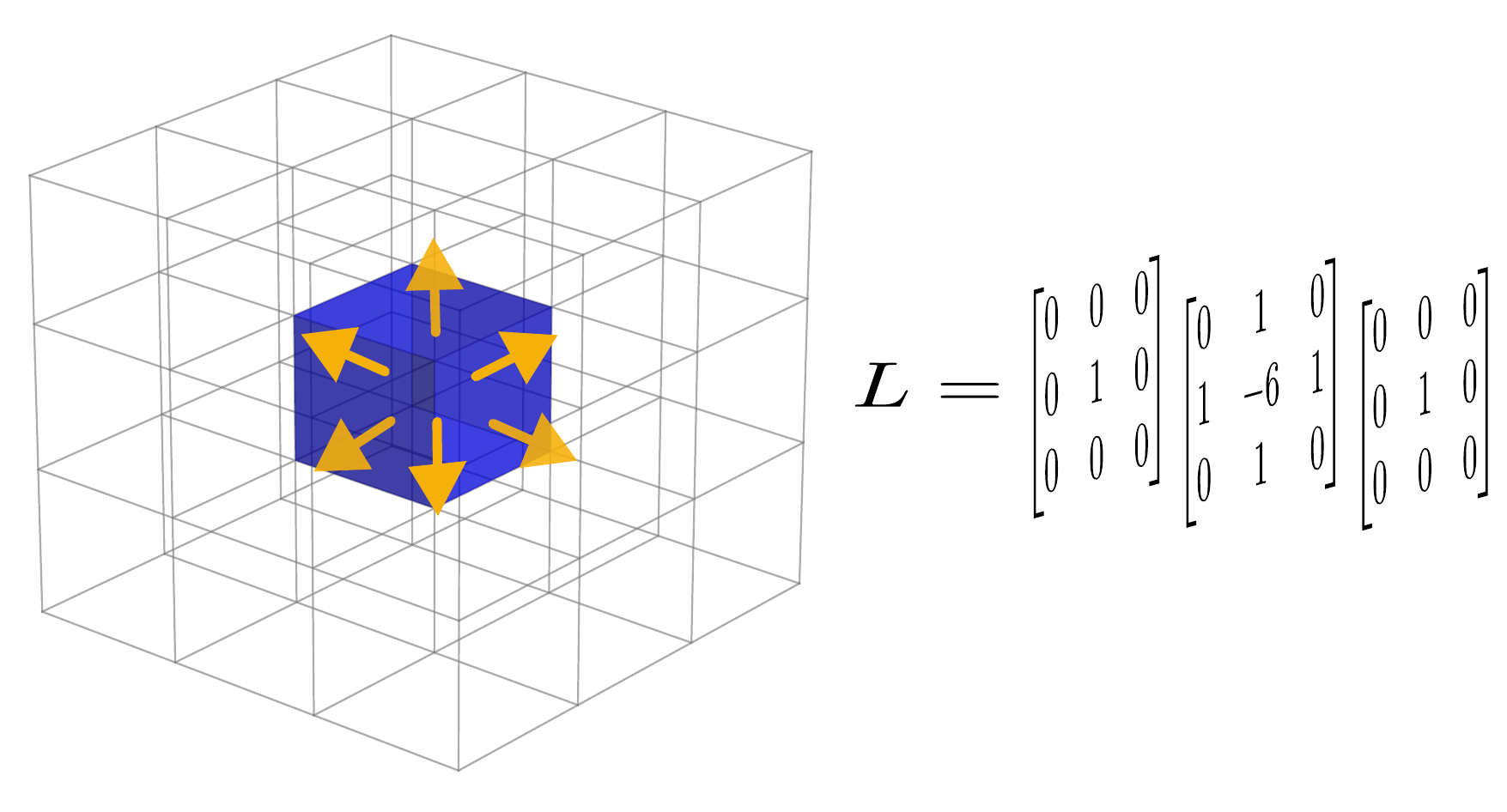}
\caption{
  Conceptual illustration of the weighted diffusion imputation. 
  Each dense cell (blue) propagates its normalized field value 
  to its neighboring sparse cells (orange arrows) through the discrete Laplacian operator, while empty cells remain clamped at zero. The diffusion step reconstructs continuity across sparse regions prior to the selection threshold $C_{\mathrm{sel}}$ being applied.
}
\label{fig:diffusion_schematic}
\end{figure}

\paragraph{Diffusion formulation (weighted).}
We evolve a discrete diffusion on the grid (periodic or nonperiodic boundary conditions (BCs) to match CCA):
\begin{equation}
\begin{split}
\partial_t C =& \, D\nabla^2 C~, \\
\quad
C^{(n+1)}_{i,j,k} =& \,
\begin{cases}
C^{(0)}_{i,j,k}~, & C^{(0)}_{i,j,k} > C_{\mathrm{thr}} \quad \text{(dense: clamped)}\\[4pt]
C^{(n)}_{i,j,k} + \beta\, w_{i,j,k}\,(L * C^{(n)})_{i,j,k}~, & 0 < C^{(0)}_{i,j,k} \le C_{\mathrm{thr}} \quad \text{(sparse)}\\[4pt]
0~, & C^{(0)}_{i,j,k}=0 \quad \text{(empty)}
\end{cases} ~,
\label{eq:diffusion}
\end{split}
\end{equation}
where $\beta>0$ is a tunable diffusion coefficient, $*$ denotes convolution, and $L$ is the standard discrete Laplacian stencil (5-point in 2D; 7-point in 3D; see Fig.~\ref{fig:diffusion_schematic}):
\[
L^{2D} = 
\begin{bmatrix}
0 & 1 & 0 \\
1 & -4 & 1 \\
0 & 1 & 0
\end{bmatrix},
\quad
L^{3D} = 
\begin{bmatrix}
0 & 0 & 0 \\
0 & 1 & 0 \\
0 & 0 & 0
\end{bmatrix}
\begin{bmatrix}
0 & 1 & 0 \\
1 & -6 & 1 \\
0 & 1 & 0
\end{bmatrix}
\begin{bmatrix}
0 & 0 & 0 \\
0 & 1 & 0 \\
0 & 0 & 0
\end{bmatrix} ~.
\]
The weighting factors modulate updates only on sparse cells,
\begin{equation}
w_{i,j,k} \;=\;
\begin{cases}
\min\!\big(1,\; C^{(0)}_{i,j,k}/C_{\mathrm{thr}}\big)~, & 0<C^{(0)}_{i,j,k}\le C_{\mathrm{thr}}\\[2pt]
0~, & \text{otherwise}
\end{cases} ~.
\label{eq:weights}
\end{equation}
Thus, dense cells are preserved, empty cells reject diffusion, and sparse cells accept diffusion proportionally to their initial strength.
For the explicit update in Eq.~\eqref{eq:diffusion}, a sufficient stability condition is
\(
\beta \le \frac{1}{2d}
\)
for a unit-spaced grid in $d$ dimensions.
In practice, we use $\beta \in [10^{-2},\,10^{-1}]$ and monitor convergence using a maximum-update tolerance criterion,
\(
\max_{\text{Sparse}}\bigl|C^{(n+1)}-C^{(n)}\bigr| < 10^{-4},
\)
after a minimum of $n_{\min}=50$ iterations or until a hard cap ($n=N_{\max}$) is reached.
This ensures stable and well-controlled diffusion convergence across datasets.

\paragraph{Selected set for connectivity.}
After imputation, the set used for connectivity is
\[
\mathcal{S} \;=\; \{(i,j,k): C^{(0)}_{i,j,k} > C_{\mathrm{thr}}\}
\;\cup\;
\{(i,j,k): 0< C^{(0)}_{i,j,k}\le C_{\mathrm{thr}} \ \wedge\ C^{(\mathrm{final})}_{i,j,k} > C_{\mathrm{sel}}\}.
\]
Thus, all pre-diffusion dense cells are retained; a sparse cell is admitted only if its imputed value exceeds $C_{\mathrm{sel}}$.

\paragraph{Origin-constrained CCA (OC-CCA).}\label{occa}
To prevent spurious cluster merges caused by imputation bridges, we introduce OC-CCA:
(i) Perform CCA on the pre-imputation dense set to obtain seed clusters.
(ii) Grow labels in the selected set \(\mathcal{S}\) under a no-merge rule: a sparse cell adopts a label only if its immediate neighborhood contains exactly one seed label. If multiple distinct seed labels are present, the cell remains unlabeled.
By growing clusters from initial seed points, the method maintains the original structure while allowing diffusion to recover the sparsity in edge connections.

The explicit algorithm implementing the entire workflow described in Sects.~\ref{subsec:gridding}–\ref{subsec:impute-cca} is presented below as Algorithm~\ref{alg:ClusTEK}.

\begin{algorithm}[H]
\caption{\small Diffusion-Enhanced Grid Clustering with Origin-Constrained CCA}
\label{alg:ClusTEK}
\begin{algorithmic}[1]
\Require Normalized per-cell field $C^{(0)}\!\in[0,1]$; dense threshold $C_{\mathrm{thr}}\in(0,1)$; selection threshold $C_{\mathrm{sel}}\in(0,1)$; diffusion coefficient $\beta>0$  
\Ensure Grid labels $L_{i,j,k}$ (optionally mapped to points)
\State Define masks from $C^{(0)}$: Dense $[C^{(0)} > C_{\mathrm{thr}}]$, Sparse $[0<C^{(0)}\le C_{\mathrm{thr}}]$, Empty $[C^{(0)}=0]$
\State Initialize $C^{(n)} \gets C^{(0)}$
\State Compute weights $w_{i,j,k}$ on Sparse cells as in Eq.~\eqref{eq:weights}; set $w_{i,j,k}\!=\!0$ on Dense and Empty
\For{$n=0,1,2,\dots$ \textbf{until convergence or } $n=N_{\max}$}
   \Comment{{\footnotesize explicit Laplacian update (5-pt/7-pt), BCs consistent with CCA (periodic or nonperiodic)}}
\State $\Lambda \gets L * C^{(n)}$ \Comment{{\footnotesize discrete Laplacian}}
\State $C^{(n+1)}\big|_{\text{Sparse}} \gets \mathrm{clip}\!\big(C^{(n)} + \beta\, w \odot \Lambda,\,0,\,1\big)\big|_{\text{Sparse}}$
  \State $C^{(n+1)}\big|_{\text{Dense}} \gets 1$, \quad $C^{(n+1)}\big|_{\text{Empty}} \gets 0$ \Comment{{\footnotesize clamp each step}}
  \State \textbf{if} $n\ge n_{\min}$ \textbf{and} $\max_{\text{Sparse}} |C^{(n+1)}-C^{(n)}|<\varepsilon$ \textbf{then break} \Comment{{\footnotesize or stop at $n=N_{\max}$}}
\EndFor
\State $\mathcal{S} \gets \{C^{(0)} > C_{\mathrm{thr}}\} \cup \{0<C^{(0)}\le C_{\mathrm{thr}} \ \wedge\ C^{(\mathrm{final})}>C_{\mathrm{sel}}\}$
\State Run CCA on dense cells to obtain seed labels $L_{\text{seed}}$ (respecting BCs/connectivity)
\State Initialize $L\gets -1$; set $L\big|_{\text{Dense}} \gets L_{\text{seed}}\big|_{\text{Dense}}$
\Repeat \Comment{{\footnotesize seeded, no-merge region growing into $\mathcal{S}$}}
  \ForAll{cells $p\in \mathcal{S}$ with $L(p)=-1$}
    \State $N \gets$ set of distinct seed labels in the face-neighborhood of $p$
    \If{$|N|=1$} \ \ $L(p)\gets$ the unique label in $N$
    \EndIf
  \EndFor
\Until{no assignments in a full pass}
\State \Return $L$ \Comment{{ \footnotesize (optional) map cell labels to points in $O(n)$}}
\end{algorithmic}
\end{algorithm}


All experiments were performed on a single-node CPU system (13th Gen Intel Core i9–13900K CPU, 64 GB RAM, NVMe SSD) using Python 3.12 with standard scientific libraries (NumPy, Pandas, SciPy, scikit-learn, Matplotlib, Seaborn). Grid generation, diffusion, and CCA labeling are implemented in vectorized NumPy kernels with explicit Laplacian stencils and optional periodic boundaries. Additional implementation details, version numbers, and reproducibility settings are provided in the Appendix ~\ref{app:config}.

\subsection{Complexity Analysis}
\label{subsec:complexity}

Let $n$ denote the number of points (or atoms), $g$ the total number of grid cells, 
$g_{\mathrm{d}}$ the number of dense cells, $g_{\mathrm{s}}$ the number of sparse or unsampled cells 
participating in diffusion, and $g_{\mathrm{sel}}$ the number of \emph{selected} cells 
retained for connectivity analysis after imputation 
(with $g_{\mathrm{sel}} \le g_{\mathrm{d}}+g_{\mathrm{s}}$).

\paragraph{Grid preprocessing and accumulation.}
If the grid resolution is fixed a~priori, assigning $n$ atoms to their corresponding cells requires 
only constant-time index arithmetic per atom, so the cost of computing the per-cell statistics 
$C_{i,j,k}$ scales as $O(n)$.  
When the grid resolution and density threshold are estimated automatically 
(e.g., by the $k$ nearest-neighbor spacing analysis described above), a one-time 
$O(n \log n)$ preprocessing step is required for KD-tree construction and neighbor queries, 
followed by $O(n)$ binning.  
In molecular simulation trajectories, grid parameters are typically determined once in a 
reference snapshot and reused for all subsequent frames; hence, the $O(n \log n)$ step 
does not contribute to the complexity of the clustering per-frame. However, 
for previously unseen datasets, this initialization cost may be included.

\paragraph{Diffusion-based imputation.}
Each explicit diffusion iteration updates only the sparse or unsampled cells.  
With $m$ iterations, the total cost is, therefore, $O(mg_{\mathrm{s}})$.  
Dense cells are clamped at~1 and empty cells at~0, contributing only minimal indexing overhead.

\paragraph{Adjacency structure on the selected grid.}
Two adjacency strategies are possible:
(i) Lattice indexing (array or hash).
In a dense network, a hash mapping integer indices $(i,j,k)$ to compact IDs can be constructed in 
$O(g_{\mathrm{sel}})$ time and memory.  
Face-sharing neighbors are obtained via constant-time modular index arithmetic.  
(ii) KD-tree (sparse centroids).
When the selected grid is sparse, storing the entire lattice is inefficient.  
Instead, a KD-tree is built on the centroids of selected cells, requiring 
$O(g_{\mathrm{sel}}\log g_{\mathrm{sel}})$ time and $O(g_{\mathrm{sel}})$ memory.  
Each cell performs a fixed-radius query equal to the face-to-face spacing, 
retrieving at most six neighbors, so the per-cell query cost is $O(\log g_{\mathrm{sel}})$ on average.  
In this work, the KD-tree strategy is employed, since simulation grids are typically sparse after 
thresholding, making it the more efficient and scalable option.

\paragraph{Connectivity labeling.}
Seeding the CCA constrained by origin in the dense subset requires $O(g_{\mathrm{d}})$ operations given the chosen adjacency structure.  
The subsequent region-growing phase visits each selected cell exactly once and inspects a neighborhood of constant size, 
for an overall cost of $O(g_{\mathrm{sel}})$.

\paragraph{Overall.}
The dominant costs are grid assignment, diffusion, adjacency construction, and connectivity labeling.  
If lattice indexing was used (i.e., on a dense regular grid), the total complexity would be
$O\!\big(n + m\,g_{\mathrm{s}} + g_{\mathrm{sel}}\big)$.  
However, in practical datasets with multiple clusters, or in molecular simulation datasets where the grid becomes sparse after thresholding, 
we employ the KD-tree strategy, resulting
in $O\!\big(n + m\,g_{\mathrm{s}} + g_{\mathrm{sel}}\log g_{\mathrm{sel}}\big)$.  
If grid resolution and threshold parameters are reestimated via the nearest-neighbor analysis $k$, an additional one-time $O(n\log n)$ initialization cost is incurred; 
otherwise, the clustering per-snapshot scales nearly linearly with $n$.  
Since $g_{\mathrm{sel}}, g_{\mathrm{s}} \ll n$ and $m$ are bounded (hundreds to thousands), the overall pipeline remains effectively nearly linear, 
with only a modest logarithmic factor from queries from KD-trees.  
Thus, the end-to-end framework maintains excellent scalability across a wide range of system sizes and resolutions typical of large-scale molecular simulation studies.

\subsection{Synthetic 2D Benchmarks} 
\label{subsec:synthetic}

To evaluate the parameterization stage (Stage~I) and then assess the entire pipeline with diffusion- and origin-constrained connectivity analysis (Stage~II), we used three standard 2D datasets with ground-truth labels: \texttt{Aggregation} and \texttt{R15}, \texttt{s\_set1}. 
These sets span compact, moderately anisotropic, and closely spaced clusters, providing controlled benchmarks with known cluster topology. 
For Stage~I, we used only coordinates $(x,y)$ and set \(C\) to the \emph{count} of points per cell. 
Ground-truth labels are kept out and later used to evaluate post-diffusion performance (ARI, NMI, and V-measure) in Sect.~\ref{sec:results}.

For each dataset, we run both Stage~I strategies from Sect.~\ref{subsec:gridding}:
(i) \texttt{tuning=grid} (heuristic grids around $h_0$ and quantile thresholds $q\in\{0.20,\ldots,0.50\}$), and
(ii) \texttt{tuning=bo} (Gaussian-process BO over $(\log h,R)$ with bounds $h\in[\underline{\eta}h_0,\overline{\eta}h_0]$, $R\in[R_{\min},R_{\max}]$).
Each proposal induces a dense mask on the grid (counts $\ge R$ in \texttt{bo}; normalized-count $\ge q$ in \texttt{grid}), followed by 4-neighbor CCA in dense cells. 
We score the resulting point partition using the unsupervised composite criterion $\mathcal{Q}$ defined in Sec.~\ref{subsec:gridding}.
The best configuration per strategy is then passed unchanged to the diffusion-imputation and OC-CCA stages.

Table~\ref{tab:stageI-final-used} summarizes the selected configurations \emph{before} diffusion:
for \texttt{grid}, $(n_x,n_y)$ and $q$; for \texttt{bo}, the optimized $(h,R)$ and the induced $(n_x,n_y)$.
We also show the associated aggregate $\mathcal{Q}$.
Figure~\ref{fig:stageI-overlays} provides visual overlays for the selected grids (dense CCA labels only). Each panel shows the grid configuration with the highest Q score identified for that
dataset (dense cells). 
Quantitative pre- and post-diffusion results (ARI, NMI, V-measure) on the same datasets, including pre-diffusion and post-diffusion, are reported in Sec.~\ref{sec:results}.

Across these datasets, both \texttt{grid} and \texttt{bo} strategies typically select comparable grid resolutions.
On sets with skewed local densities or highly uneven occupancy histograms,
\texttt{bo} may favor a slightly different $R$ and thus shift $(n_x,n_y)$, improving $\mathcal{Q}$ by balancing coverage with cluster separation.
The selected Stage~I configuration is carried forward intact to Stage~II, where diffusion-based imputation and topology-preserving OC-CCA are applied.

\begin{table}[t]
\centering
\caption{Stage~I selections actually used downstream (one per dataset). We report the winning strategy, parameters, induced grid, and composite score $\mathcal{Q}$. 
}
\label{tab:stageI-final-used}
\renewcommand{\arraystretch}{1.2}
\setlength{\tabcolsep}{6pt}
\small
\begin{tabular}{l c c l l c c}
\hline
\textbf{\footnotesize Dataset} & \textbf{\footnotesize \# Samples} & \textbf{\footnotesize \# Clusters} & \textbf{\footnotesize Strategy} & \textbf{\footnotesize Parameters} & \textbf{($n_x$, $n_y$)} & \textbf{$\mathcal{Q}$} \\
\hline
\multirow{2}{*}{Aggregation} & \multirow{2}{*}{700} & \multirow{2}{*}{7} 
& \texttt{grid} & $q=0.3$ & (15, 12) & 0.69 \\
& & & \texttt{bo} & $h=1.75,\ R=3$ & (19, 16) & 0.71 \\
\hline
\multirow{2}{*}{R15} & \multirow{2}{*}{600} & \multirow{2}{*}{15} 
& \texttt{grid} & $q=0.5$ & (26, 26) & 0.81 \\
& & & \texttt{bo} & $h=0.53,\ R=3$ & (26, 26) & 0.81 \\
\hline
\multirow{2}{*}{s\_set1} & \multirow{2}{*}{5000} & \multirow{2}{*}{15} 
& \texttt{grid} & $q=0.5$ & (36, 36) & 0.78 \\
& & & \texttt{bo} & $h=27709.3,\ R=5$ & (34, 34) & 0.80 \\
\hline
\end{tabular}
\end{table}

\begin{figure}[htbp]
  \centering
  \setlength{\tabcolsep}{4pt} 
  \renewcommand{\arraystretch}{1.0}
  \begin{tabular}{ccc}
    \includegraphics[height=4.4cm]{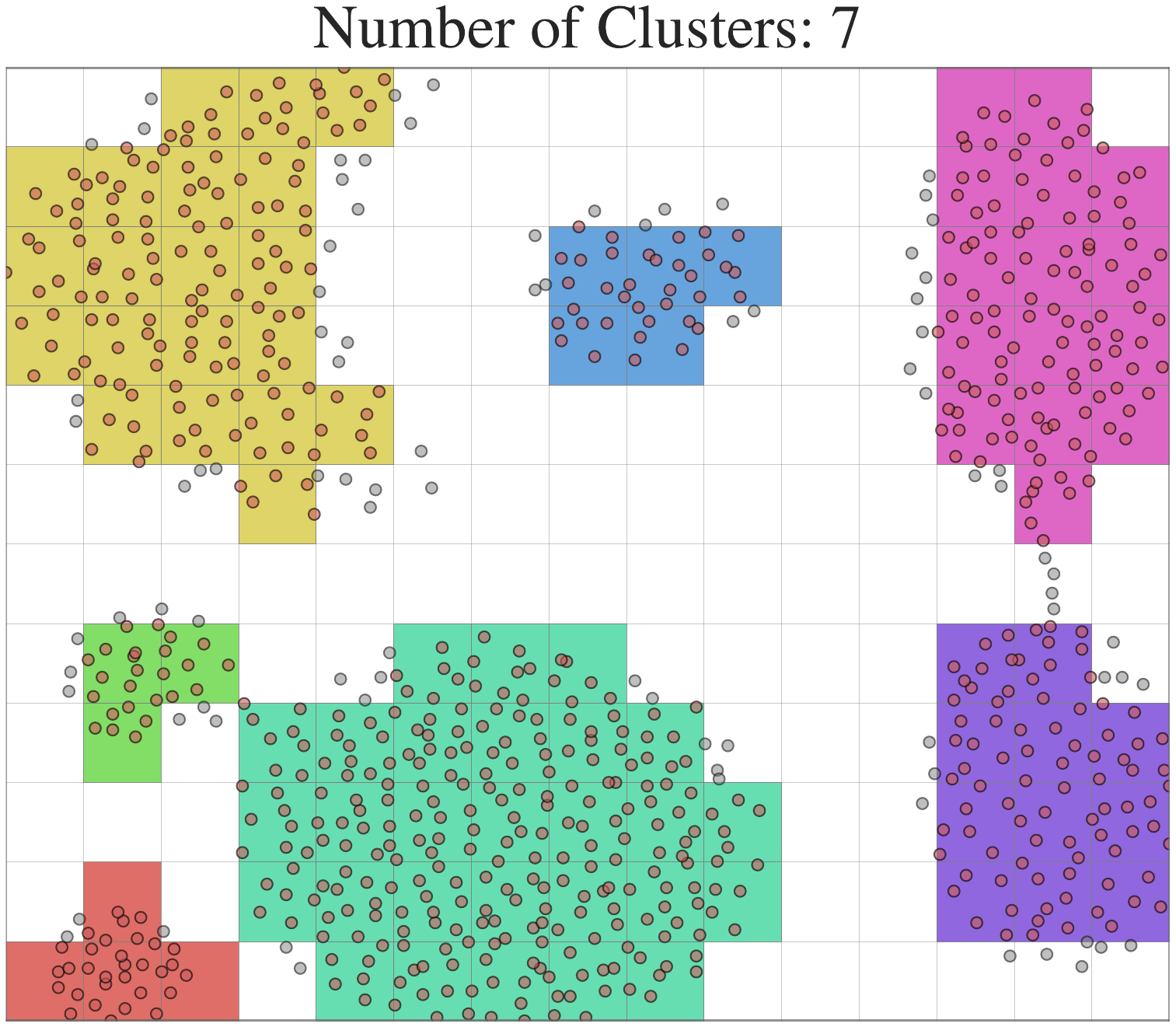} &
    \includegraphics[height=4.4cm]{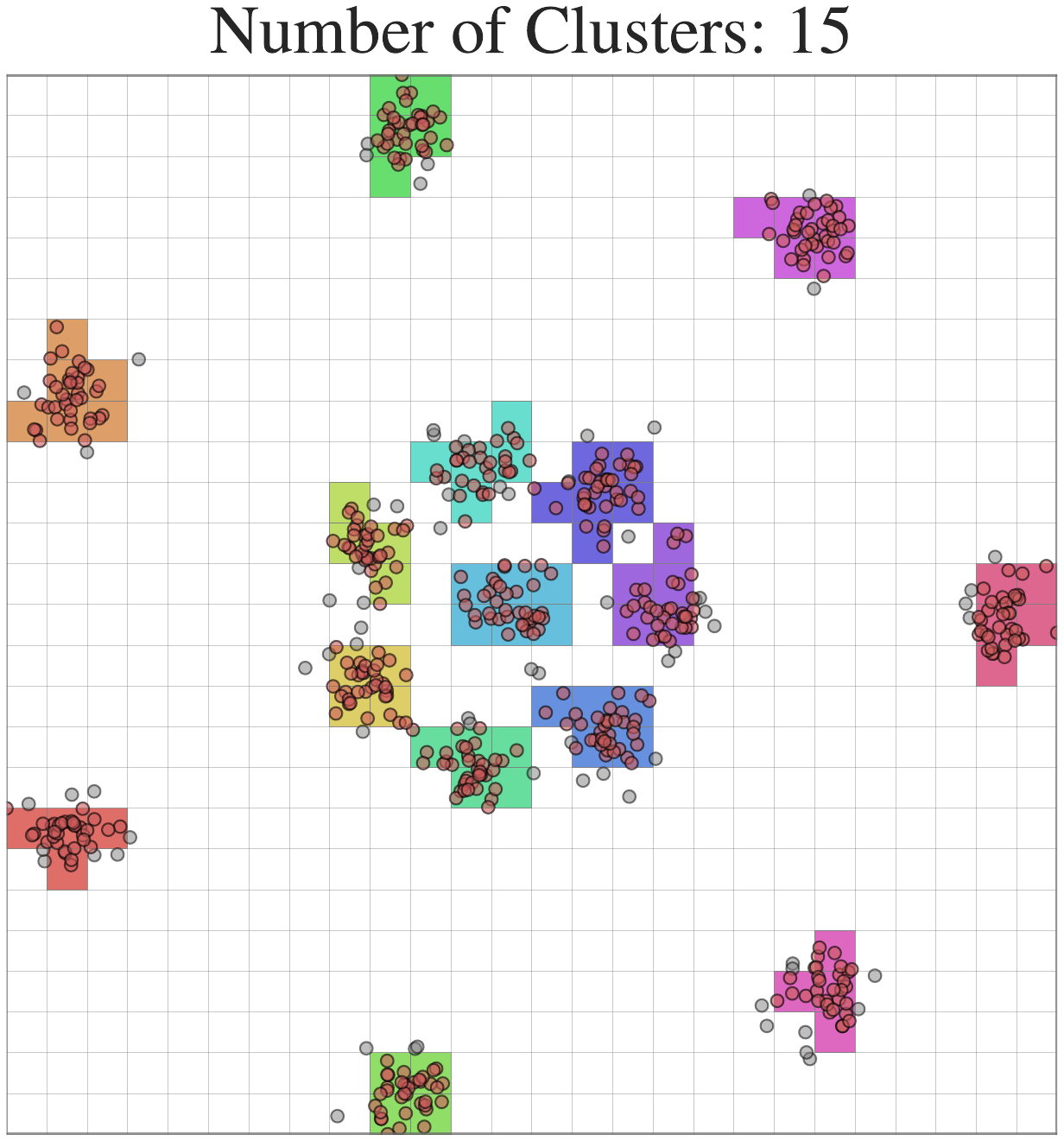} &
    \includegraphics[height=4.4cm]{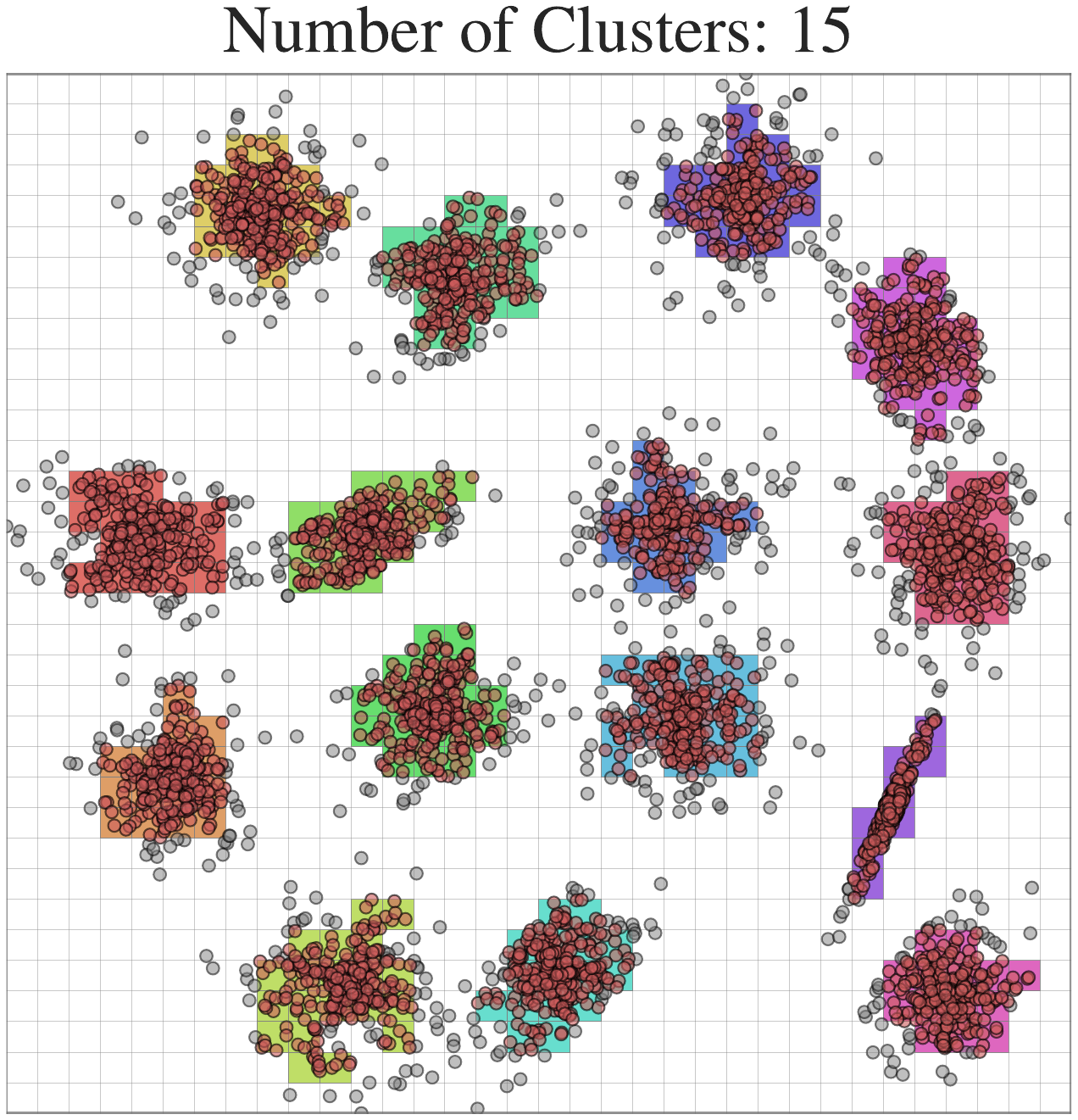} \\
    Aggregation & R15 & s\_set1
  \end{tabular}
  \caption{Stage~I overlays for the selected datasets. Grids represent the $(n_x, n_y)$ structure and each panel shows the grid configuration with the highest $Q$ score identified for that dataset (see Table~\ref{tab:stageI-final-used}). Axes and ticks are omitted for clarity. All panels share identical spatial extents.}
  \label{fig:stageI-overlays}
\end{figure}

\subsection{Molecular dynamics simulation details}
\label{sec:methods_md}

The Siepmann–Karaborni–Smit (SKS) unit atom potential (UA) \cite{siepmann1993simulating} was used to model polyethylene macromolecules, where the terminal \ce{CH3} methyl groups represent the chain ends and the internal \ce{CH2} methylene groups constitute the backbone units. To improve integration stability and avoid explicit bond constraints, the original rigid bonds were replaced with harmonic potentials \cite{moore2000molecular,baig2006rheological,cui1996multiple,baig2005rheological,ionescu2006structure}. 

Nonbonded intramolecular and intermolecular interactions were described using the 12-6 Lennard-Jones (LJ) potential: 
\begin{equation}
U_{LJ}(r_{ij}) = 4 \epsilon_{ij} \left[ \left( \frac{\sigma_{ij}}{r_{ij}} \right)^{12} - \left( \frac{\sigma_{ij}}{r_{ij}} \right)^6 \right],
\label{energy_lj}
\end{equation}
where $\epsilon_{ij}$ is the depth of the well and $\sigma_{ij}$ is the zero-potential separation between the particles $i$ and $j$. The LJ parameters were $\epsilon_i / k_B = 47$~K for \ce{CH2} and $114$~K for \ce{CH3}, with $\sigma_{i}=3.93$~\AA~for both species. Heterogeneous interactions follow the Lorentz–Berthelot mixing rules: $\epsilon_{ij} = (\epsilon_i \epsilon_j)^{1/2}$ and $\sigma_{ij} = (\sigma_i + \sigma_j)/2$. Nonbonded interactions were considered for pairs separated by at least three bonds, with a cutoff point of $2.5\,\sigma_{\rm CH_2}$. 

Bonded interactions were modeled using harmonic potentials. The stretching of the bonds was described as $U_{\rm str}(l) = \frac{k_l}{2} (l - l_0)^2$, with the equilibrium bond length $l_0 = 1.54$~\AA~and the stiffness $k_l/k_B = 452{,}900$~K/\AA$^2$. Bond bending used $U_{\rm bend}(\theta) = \frac{k_\theta}{2} (\theta - \theta_0)^2$, where $\theta_0 = 114^\circ$ and $k_\theta/k_B = 62{,}500$~K/rad$^2$. Torsional interactions were defined as $U_{\rm tor}(\phi) = \sum_{m=0}^3 a_m (\cos \phi)^m$, with coefficients $a_0/k_B = 1010$, $a_1/k_B = -2019$, $a_2/k_B = 136.4$ and $a_3/k_B = 3165$~K. Full details of the SKS force field are provided in Refs.~\cite{siepmann1993simulating,nafar2015individual,nafar2020thermodynamically}.

Simulations were performed with LAMMPS \cite{plimpton1995fast,thompson2022lammps} in the $NpT$ ensemble at 1~atm with periodic boundary conditions, using the Nosé–Hoover thermostat and barostat. 
For quiescent quenching simulations, we first studied a small system of 60 n-pentacontahectane chains (\ce{C150H302}) (hereafter referred to as 60 C150) in $T = 300$~K, corresponding to an undercooling $\sim25\%$, consistent with previous studies \cite{yi2013molecular,anwar2013crystallization,Cindex}. The larger quiescent systems contained 360 C500 chains. Both systems were equilibrated at $550$~K ($200$~ns for C150 and $10$~$\mu$s for C500) before quenching at $300$~K to induce nucleation. Single nucleation events were observed in the smaller chain C150 system, while multiple nuclei formed in the larger chain C500 system. Planar elongational flow (PEF) simulations were performed on a polydisperse melt with a polydispersity index $PDI = 1.8$, which includes chain lengths from C60 to C5000. These flow simulations were performed at $T = 450$~K (approximately 10\% above the melting temperature). For analysis, multiple configurations were selected at various Deborah number values ($\mathrm{De}$) to investigate nucleation and early cluster formation.

\subsection{Baselines and External Validation Metrics}
\label{subsec:baselines-metrics}

\paragraph{Baseline algorithms.}
We benchmark the proposed diffusion-enhanced grid clustering with OC-CCA (hereafter referred to as ClusTEK) against representative clustering paradigms spanning centroid-based, 
model-based, hierarchical, density-based, and grid-based approaches. 
These include \texttt{KMeans}, Gaussian Mixture Models (\texttt{GMM}), agglomerative hierarchical 
clustering, \texttt{DBSCAN}, \texttt{HDBSCAN}, and the canonical grid-based algorithm 
\texttt{CLIQUE}. 
For algorithms requiring a specified number of clusters (\texttt{KMeans}, \texttt{GMM}, 
\texttt{Agglomerative}), we provide the oracle cluster count \(k\) to provide a deliberately 
favorable comparison. 
For \texttt{CLIQUE}, we match the grid resolution to the selected \((n_x,n_y)\) used in the ClusTEK pipeline. 
For density-based methods, we sweep \texttt{min\_samples}, \texttt{min\_cluster\_size} 
and \(\varepsilon\) over standard recommended ranges, acknowledging their known sensitivity 
to parameterization in heterogeneous or time-varying data~\cite{Cindex}.


Other classical grid-based clustering algorithms (e.g., STING, WaveCluster, MAFIA) are not included in the present benchmark.
While these methods are historically important, they are not currently supported by
widely used, actively maintained Python libraries that integrate cleanly with modern
scientific computing workflows.
Including custom reimplementations would introduce additional sources of variability
related to software engineering choices, optimization strategies, and data handling,
thus confounding algorithmic comparisons.
To ensure methodological fairness, reproducibility, and ease of verification, we
therefore restrict our baselines to well-established methods with standardized,
publicly available reference implementations.

\paragraph{External validation metrics.}
In labeled synthetic datasets, we report standard external clustering metrics including 
Adjusted Rand Index (ARI), Normalized Mutual Information (NMI), V-measure, 
Fowlkes--Mallows score, and purity, together with unsupervised quality measures such as 
the silhouette coefficient, Davies--Bouldin index, and coverage whenever applicable. 
For molecular dynamics trajectories, where ground truth labels are unavailable, we employ manually tuned, high-precision atom-level clustering procedures to obtain a reliable reference and complement unsupervised scores with this pseudo-ground truth. 
Agreement with this reference is quantified using cluster-size distributions and
distributional discrepancies (Earth Mover’s Distance and Kolmogorov--Smirnov statistics), and is further supported by qualitative spatial overlays.
Additional physically motivated diagnostics (e.g., surface-based measures) are a natural extension of the present framework but are outside the scope of this study.

\paragraph{Protocol and reproducibility.}
All methods are tuned using constrained hyperparameter searches with fixed candidate
budgets per dataset to avoid unfair overfitting.
For grid-based approaches, the discretization $(n_x,n_y,n_z)$ is selected once on a
representative frame and reused throughout the trajectory.
Diffusion parameters, including the diffusion coefficient $\beta$ and the post-diffusion
selection threshold $C_{\mathrm{sel}}$, are tuned per dataset and then kept fixed across
all frames within that dataset, while iteration counts are determined by fixed
convergence criteria.
All code, parameter-sweep scripts, configuration files, random seeds, and library
versions are provided to ensure full reproducibility; see the Code Availability statement (or supplementary material) for access details.

\section{Results and Discussion}
\label{sec:results}

\subsection{Synthetic 2D Benchmarks}
\label{subsec:2d_results}

We begin by validating the diffusion-enhanced grid clustering framework on the labeled 2D datasets introduced in Sect.~\ref{subsec:synthetic}. Each dataset was parameterized using the Stage~I strategies (\texttt{tuning=grid} and \texttt{tuning=bo}), and the configuration that produces the highest aggregate score~$\mathcal{Q}$ was chosen for downstream diffusion imputation and connectivity analysis. Performance was evaluated using external metrics in Sect.~\ref{subsec:baselines-metrics}, including ARI, NMI, V-measure, Fowlkes--Mallows (FM), purity and coverage. 
Table~\ref{tab:2d_diffusion_metrics} reports the quality of clustering before diffusion, after diffusion with standard CCA and after diffusion combined with OC-CCA.

\begin{table}
\centering
\caption{
Clustering performance on synthetic 2D benchmarks before and after diffusion.
The \texttt{after\_std} columns correspond to diffusion followed by standard CCA,
while \texttt{after\_occa} denotes diffusion combined with OC-CCA.
Boldface indicates the best score within each dataset.
}
\label{tab:2d_diffusion_metrics}
\renewcommand{\arraystretch}{1.15}
\setlength{\tabcolsep}{6pt}
\small
\begin{tabular}{lccccccc}
\hline
\textbf{Dataset} & \textbf{$k$} & \textbf{Coverage} & \textbf{ARI} & \textbf{NMI} & \textbf{V-measure} & \textbf{FM} & \textbf{Purity} \\
\hline
\multicolumn{8}{l}{\textbf{Aggregation} \quad ($\beta^*=0.1$, iterations $=100$)} \\[2pt]
\quad before      & 7  & 0.9150 & 0.8884 & 0.8739 & 0.8739 & 0.9128 & 0.9137 \\
\quad after\_std  & 6  & 0.9581 & 0.8641 & 0.8844 & 0.8844 & 0.8947 & 0.9150 \\
\quad after\_occa & 7  & 0.9556 & \textbf{0.9340} & \textbf{0.9193} & \textbf{0.9193} & \textbf{0.9486} & \textbf{0.9530} \\
\hline
\multicolumn{8}{l}{\textbf{R15} \quad ($\beta^*=0.1$, iterations $=100$)} \\[2pt]
\quad before      & 15 & 0.8800 & 0.7646 & 0.8671 & 0.8671 & 0.7799 & 0.8733 \\
\quad after\_std  & 13 & 0.9467 & 0.8034 & 0.8984 & 0.8984 & 0.8201 & 0.8200 \\
\quad after\_occa & 15 & 0.9400 & \textbf{0.8960} & \textbf{0.9225} & \textbf{0.9225} & \textbf{0.9030} & \textbf{0.9333} \\
\hline
\multicolumn{8}{l}{\textbf{s\_set1} \quad ($\beta^*=0.1$, iterations $=190$)} \\[2pt]
\quad before      & 15 & 0.8726 & 0.7621 & 0.8701 & 0.8701 & 0.7780 & 0.8726 \\
\quad after\_std  & 11 & 0.9568 & 0.7336 & 0.8749 & 0.8749 & 0.7666 & 0.7118 \\
\quad after\_occa & 15 & 0.9536 & \textbf{0.9340} & \textbf{0.9433} & \textbf{0.9433} & \textbf{0.9387} & \textbf{0.9530} \\
\hline
\end{tabular}
\end{table}

\begin{figure}
\centering
\setlength{\tabcolsep}{2pt}
\renewcommand{\arraystretch}{1.0}
\begin{tabular}{ccc}
\includegraphics[width=0.32\linewidth]{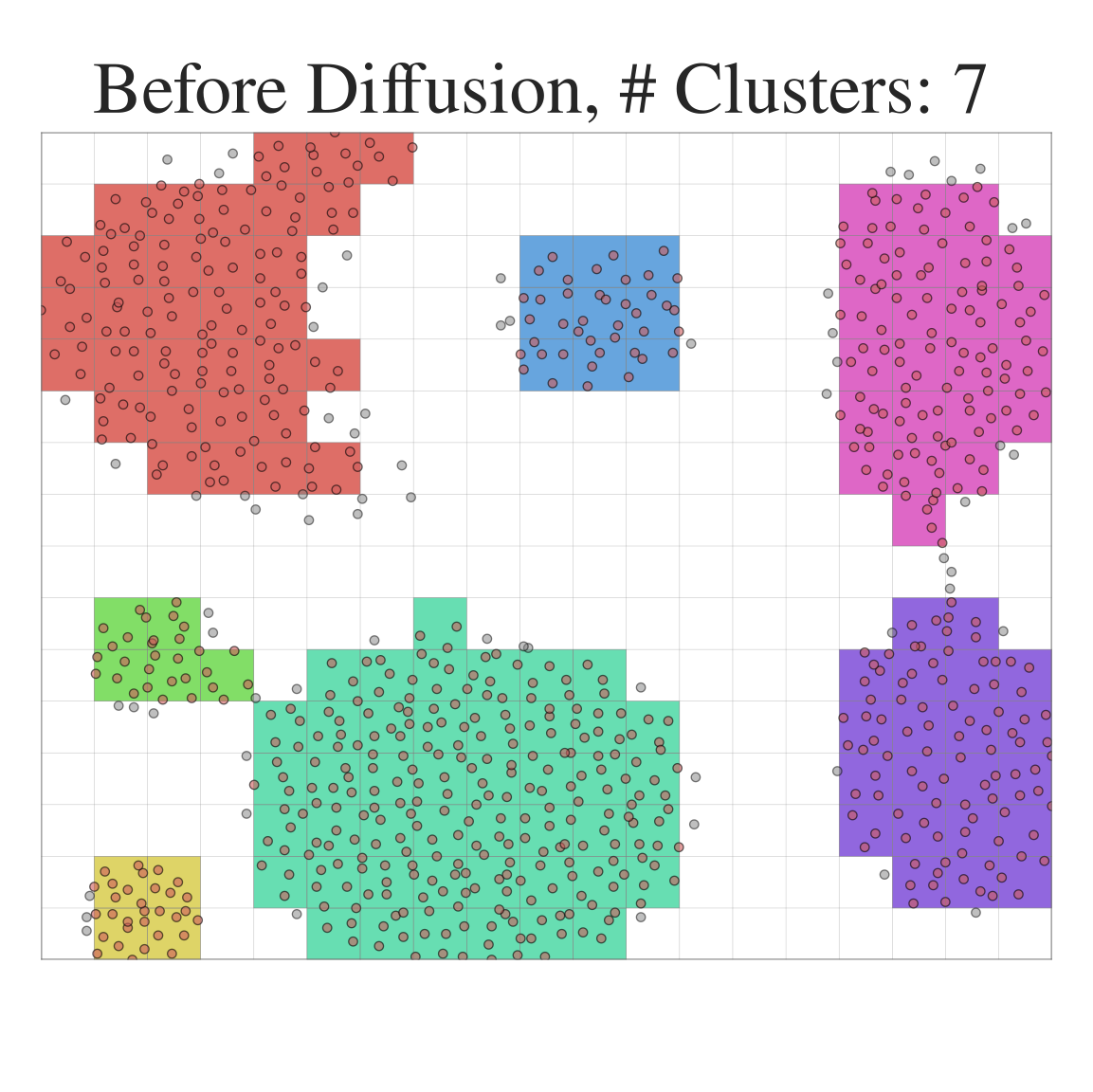} &
\includegraphics[width=0.32\linewidth]{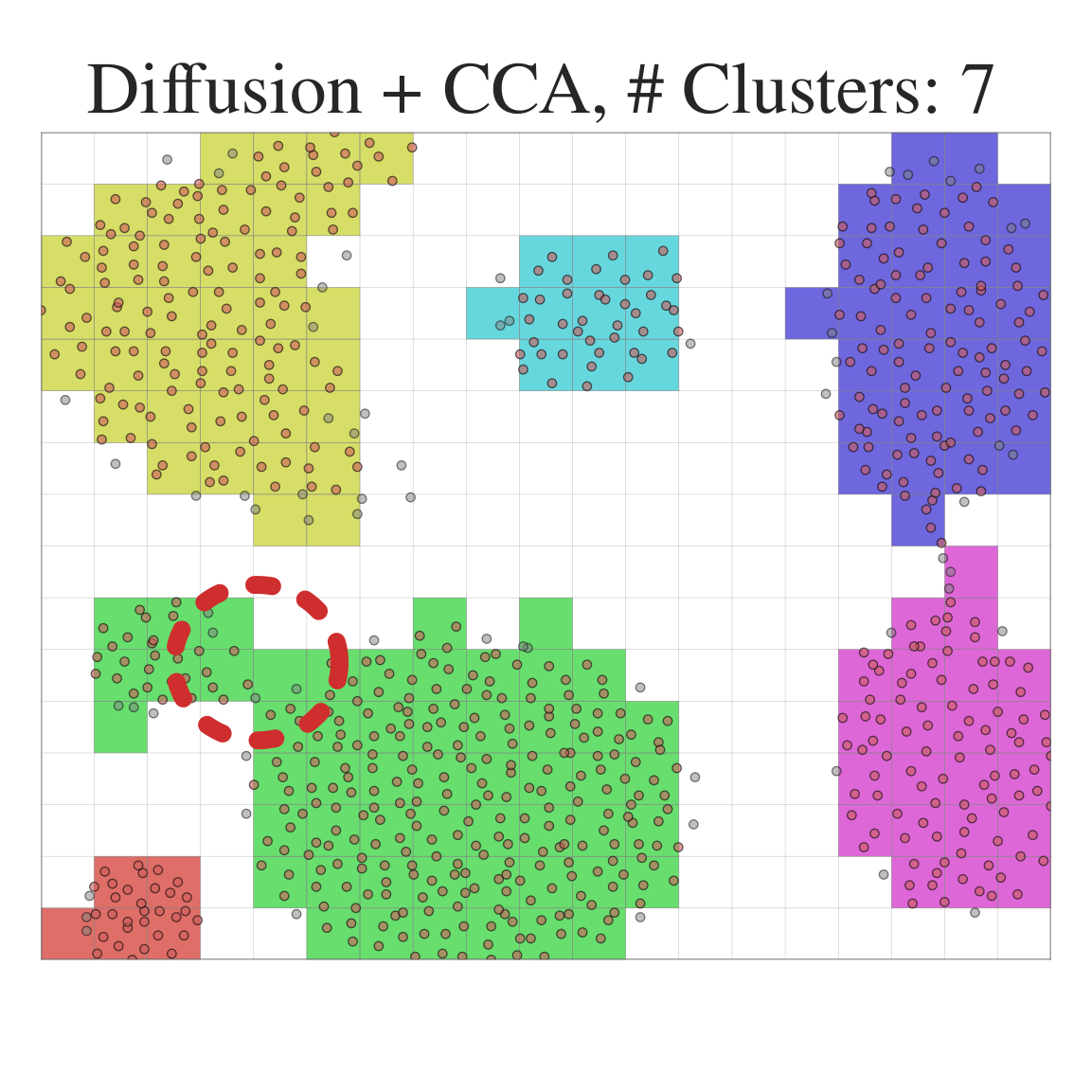} &
\includegraphics[width=0.32\linewidth]{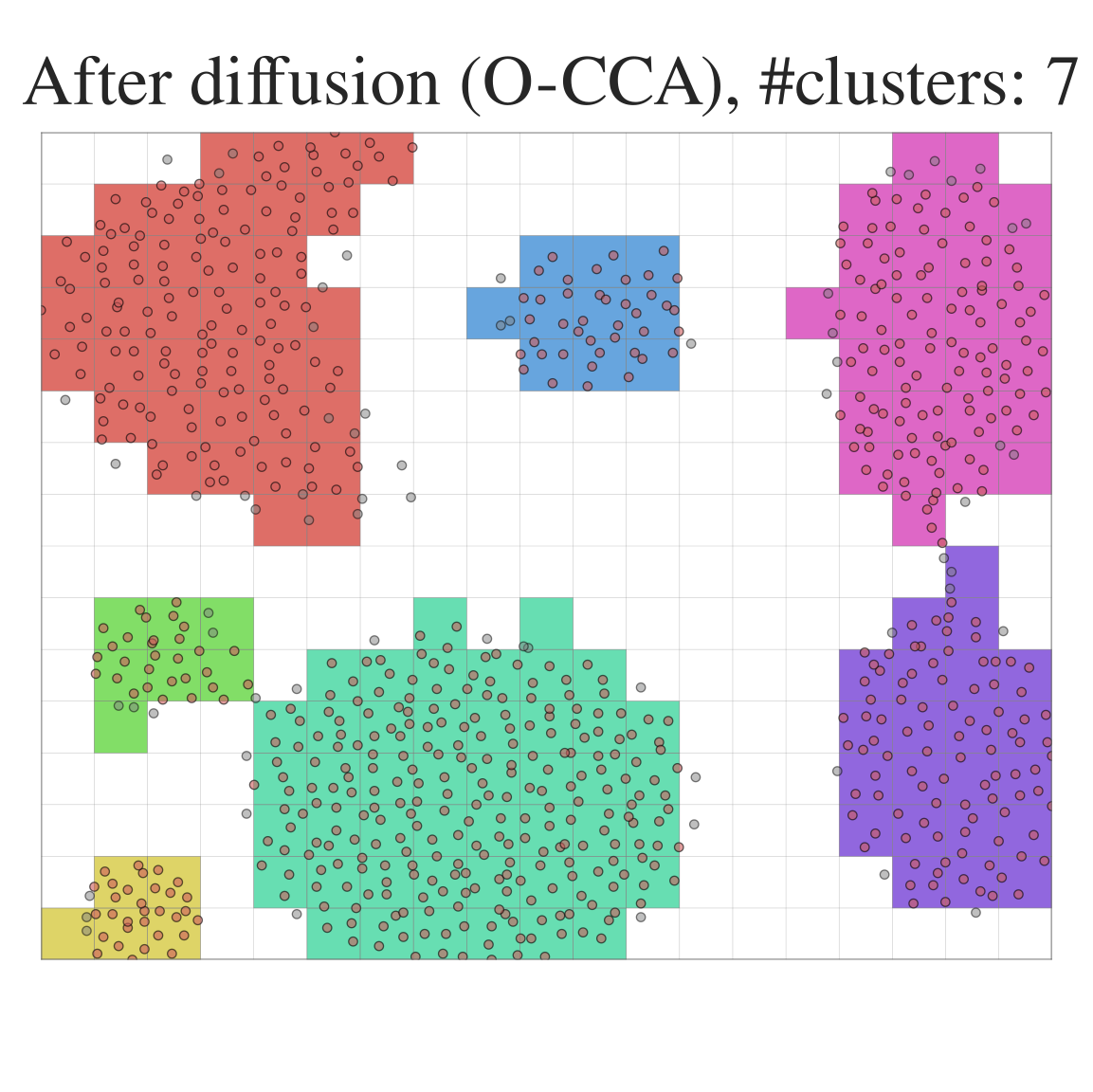} \\[-3pt]
(a) & (b) & (c) \\[4pt]

\includegraphics[width=0.32\linewidth]{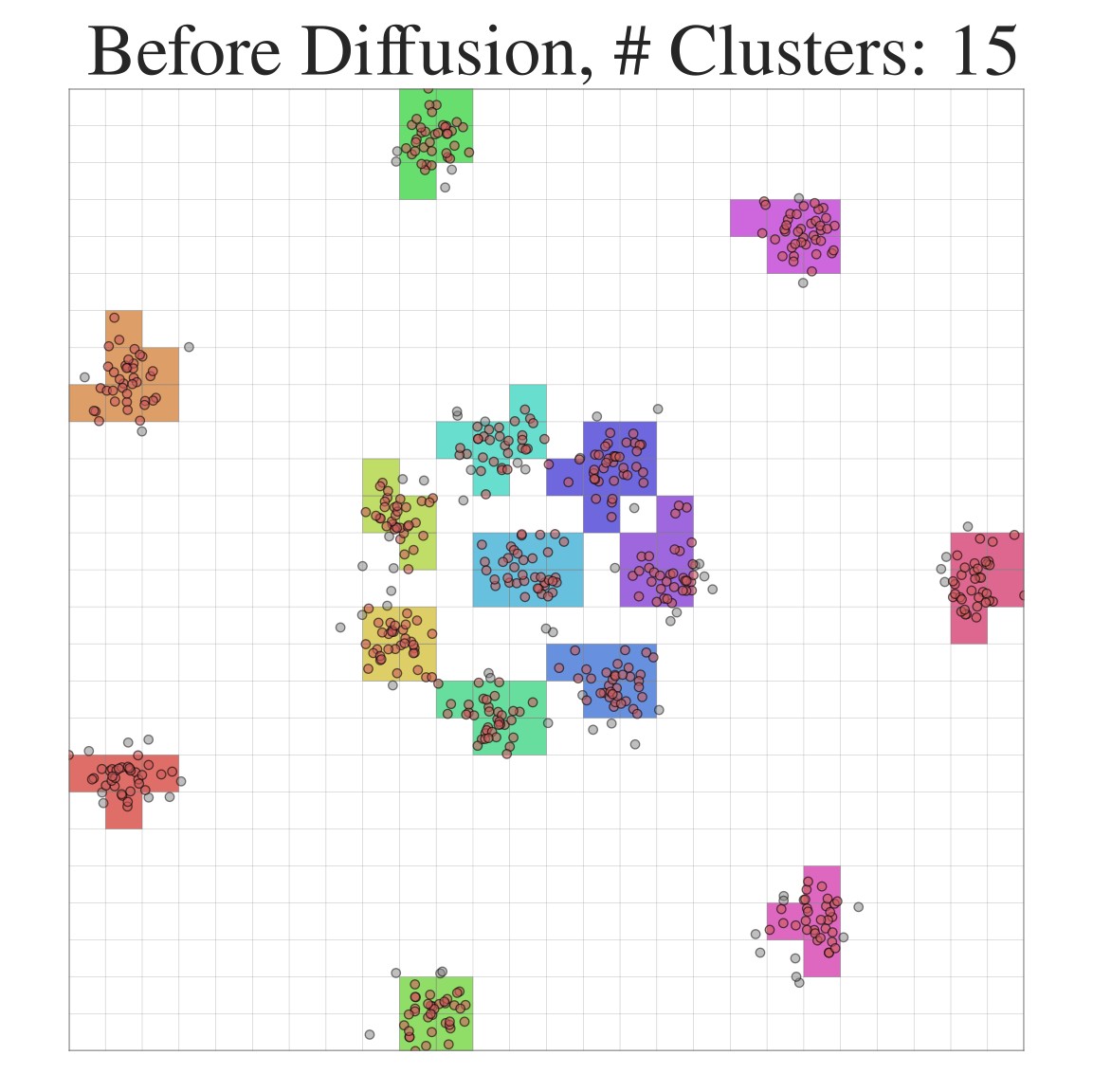} &
\includegraphics[width=0.32\linewidth]{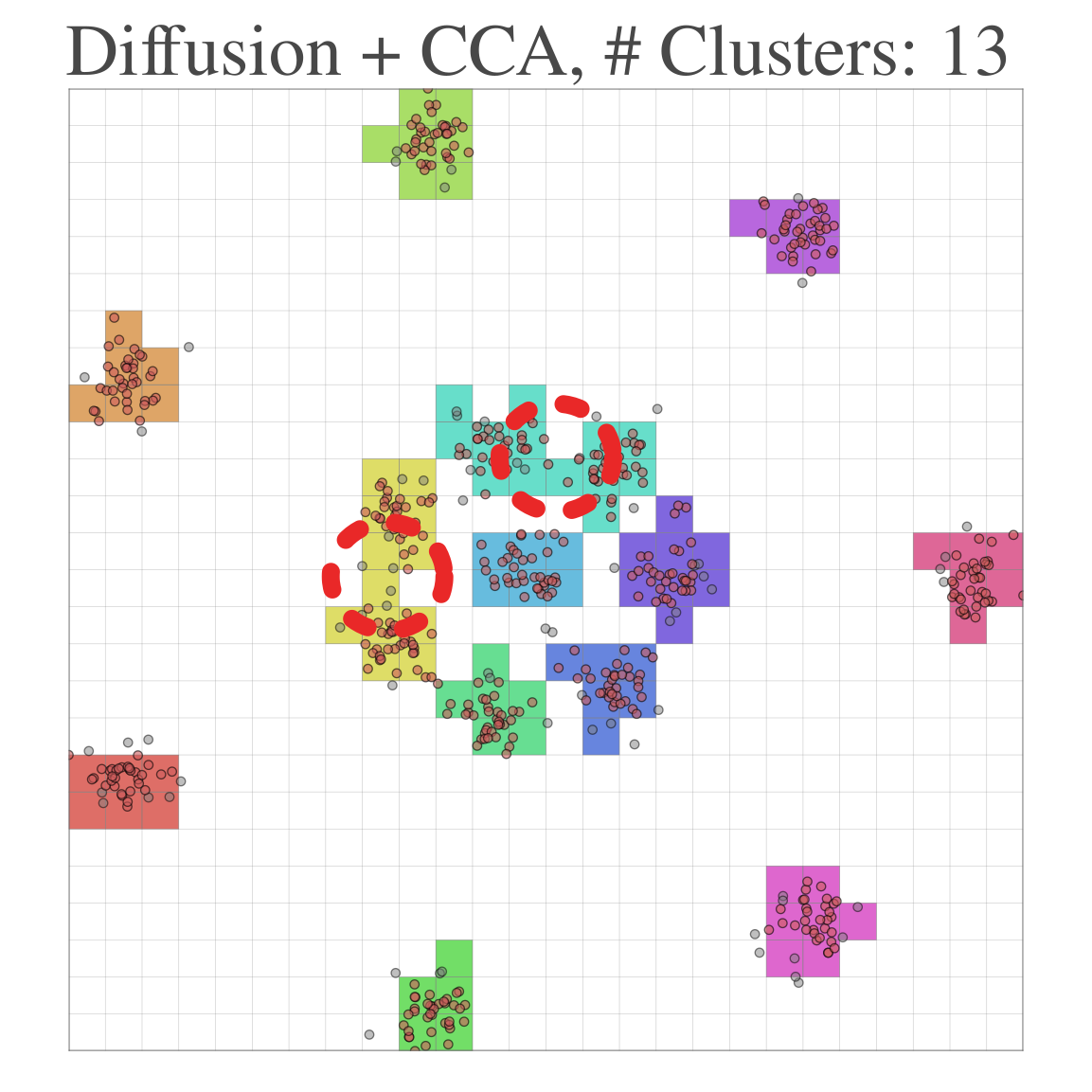} &
\includegraphics[width=0.32\linewidth]{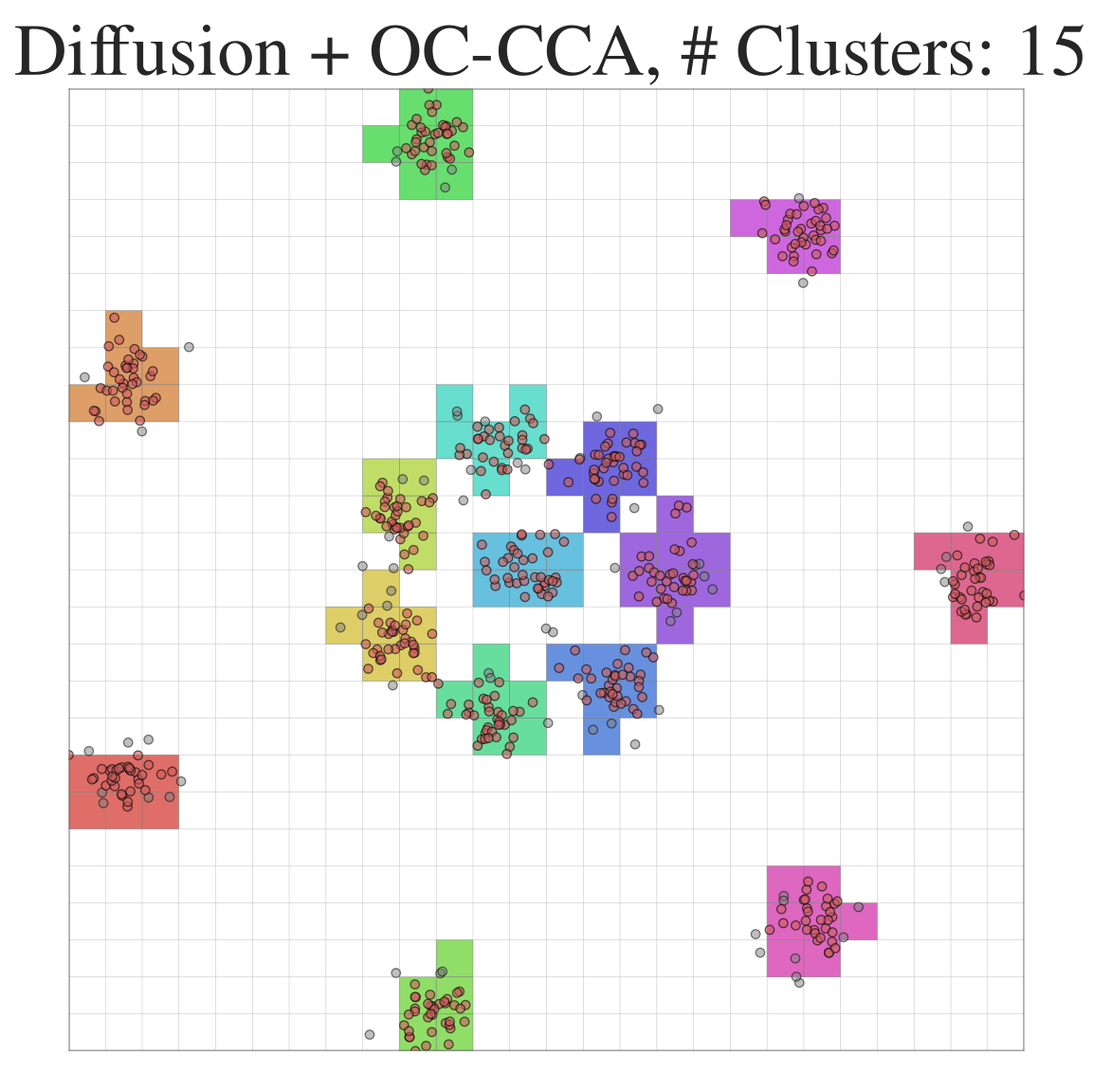} \\[-3pt]
(d) & (e) & (f) \\[4pt]

\includegraphics[width=0.32\linewidth]{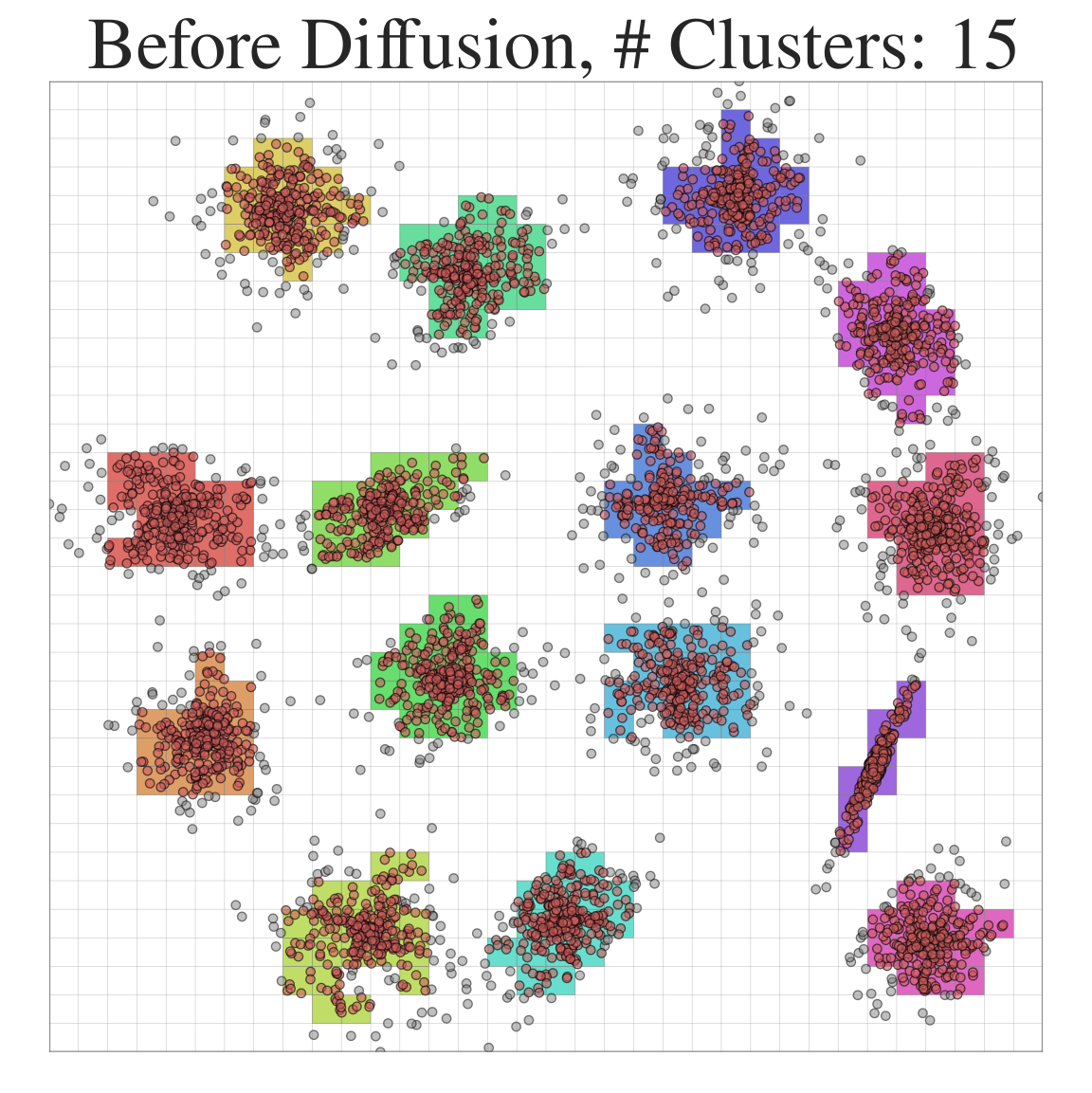} &
\includegraphics[width=0.32\linewidth]{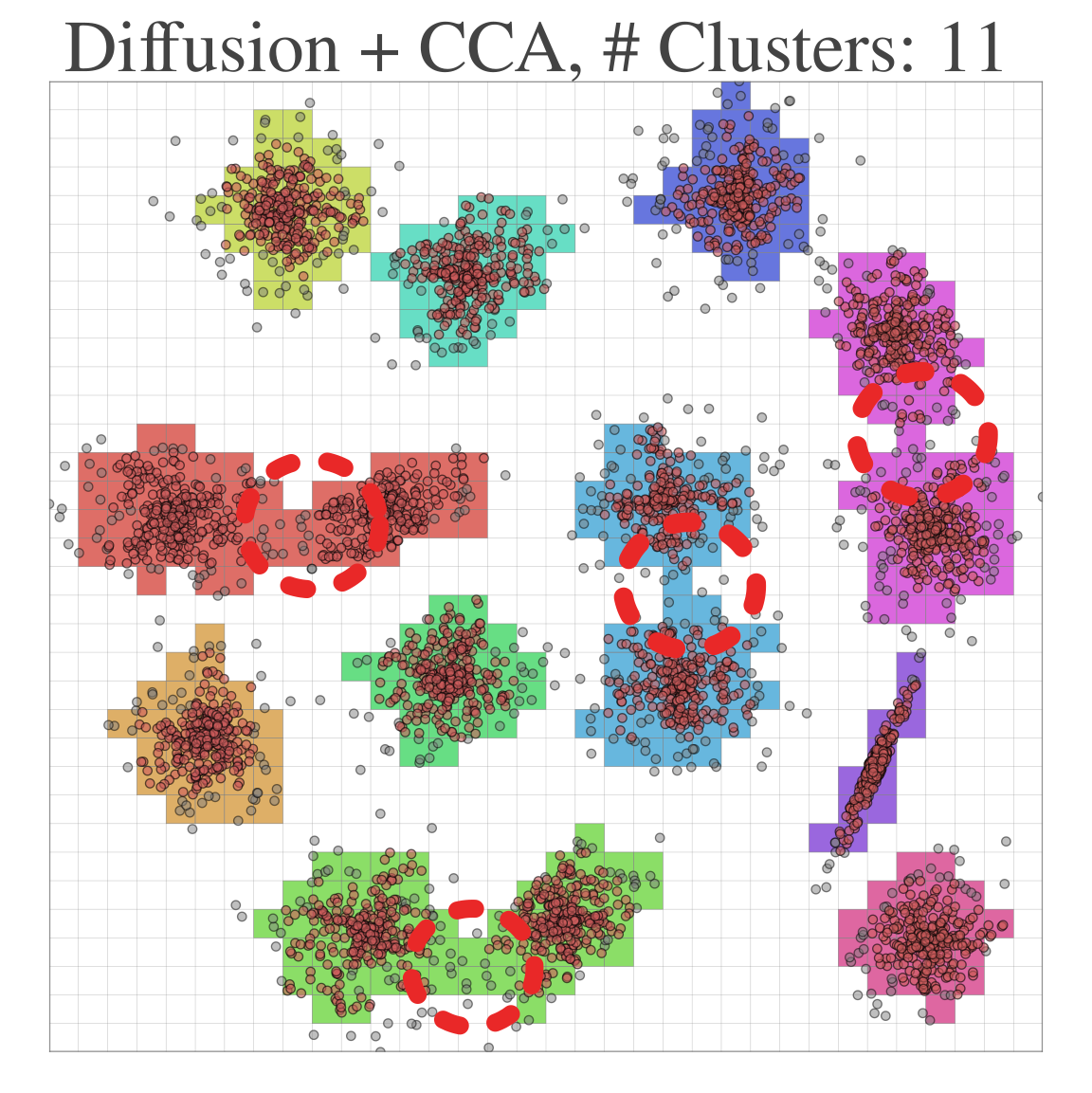} &
\includegraphics[width=0.32\linewidth]{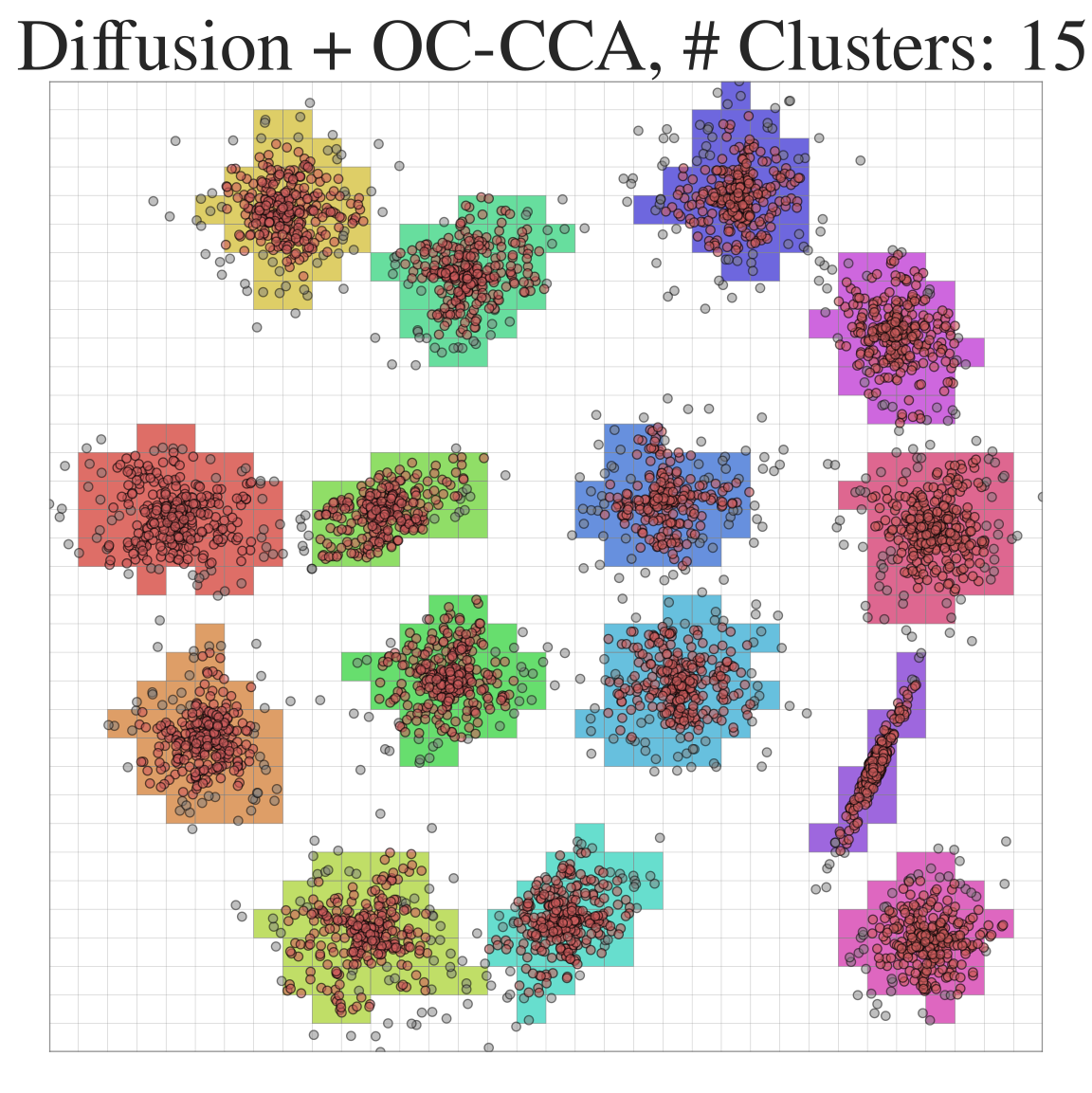} \\[-3pt]
(g) & (h) & (i) \\[4pt]
\end{tabular}
\caption{
Visual comparison of diffusion and connectivity stages on three synthetic benchmarks.
Each row corresponds to one dataset (Aggregation, R15, and s\_set1), whereas columns show the clustering
(a,d,g) before diffusion, (b,e,h) after diffusion with standard CCA, and (c,f,i) after diffusion with OC-CCA.
Diffusion improves continuity across sparse regions, but standard CCA may spuriously merge nearby clusters through diffusion halos (highlighted by dashed red circles in panels~b,e,h). OC-CCA removes these artificial bridges and restores correct cluster topology and count.
}
\label{fig:2d_diffusion_comparison}
\end{figure}

Figure~\ref{fig:2d_diffusion_comparison} visualizes the effect of diffusion and connectivity.
Diffusion imputation increases cell continuity by filling narrow gaps and smoothing sparsely populated boundaries, whereas OC-CCA prevents the resulting diffusion halos from bridging distinct structures. 
Standard CCA (panels~b,e,h) frequently merges nearby clusters across thin diffusion bands, reducing the recovered cluster count; falsely merged regions are highlighted by the dashed red circles in these panels. 
In contrast, OC-CCA (panels~c,f,i) restores the correct number and delineation of clusters by enforcing origin-constrained growth and rejecting spurious bridges.

Across all three datasets, diffusion alone (\texttt{after\_std}) increases coverage by approximately 4--8\% and slightly improves the NMI and V-measure, reflecting smoother intercluster transitions but possible occasional over-merging.
When coupled with OC-CCA, both ARI and purity increase substantially (up to~+0.17), suggesting that origin-constrained growth successfully prevents false merges while retaining the benefits of diffusion-based continuity.
In all cases, the recovered cluster number~$k$ matches the ground truth, demonstrating that diffusion and OC-CCA together preserve both the topology and the cluster count.

\subsubsection{Comparison with Other Clustering Algorithms}
\label{subsec:2d-comparison}

We benchmarked \texttt{ClusTEK} against representative clustering paradigms, including centroid-based (\texttt{KMeans}), model-based (\texttt{GMM}), bottom-up hierarchical (\texttt{Agglomerative}), density-based (\texttt{DBSCAN}, \texttt{HDBSCAN}) and grid-based (\texttt{CLIQUE}) methods. All baselines were tuned over standard hyperparameters using the same spatial extent and evaluation protocol described in Sect.~\ref{subsec:baselines-metrics}. For methods requiring a user-specified number of clusters, an oracle value equal to the true $k$ was supplied to provide a favorable comparison. For \texttt{KMeans} and \texttt{GMM}, the metrics were averaged over 10 random initializations. For \texttt{CLIQUE}, the grid resolution was matched with the selected \((n_x, n_y)\) used in ClusTEK. Density-based baselines were tuned using the known ground-truth cluster count to ensure a strong, advantageous reference.

The cost of ClusTEK corresponds to a clustering pass with fixed hyperparameters (grid, dense threshold or occupancy, diffusion coefficient, and post-diffusion threshold). The overhead for initial hyperparameter selection (e.g., Bayesian optimization over \(h\), \(R\) and scoring weights) is not included in the per-run timings in Tables~\ref{tab:aggregation-bench}--\ref{tab:s-set1-bench}.

\paragraph{Quantitative metrics.}
Tables~\ref{tab:aggregation-bench}--\ref{tab:s-set1-bench} summarize accuracy, coverage, and efficiency.
 In aggregation, ClusTEK achieves the highest ARI (0.9754) and purity (0.9734), outperforming even oracle-$k$ \texttt{GMM} and \texttt{KMeans}. ClusTEK also maintains excellent coverage (0.9734), surpassed only by algorithms that enforce full assignment, such as \texttt{KMeans}, 
\texttt{GMM}, and \texttt{Agglomerative}. Runtime remains competitive ($\sim$0.029~s) while using only 0.1~MB of additional memory.

In \texttt{R15}, oracle-$k$ \texttt{KMeans} and \texttt{GMM} unsurprisingly obtain near-perfect ARI/NMI, but these depend critically on prior knowledge of $k$. The density-based methods were also tuned using ground-truth information. ClusTEK, which does not require $k$, achieves ARI~0.8960 with much higher coverage (0.9400) than \texttt{CLIQUE} (0.4450) and with lower memory usage than all baselines.

On the more challenging \texttt{s\_set1} dataset, which contains narrow gaps and anisotropic cluster boundaries, ClusTEK maintains strong performance (ARI~0.9457, NMI~0.9487, purity~0.9618) with coverage~0.9670. True-$k$ \texttt{KMeans} and \texttt{GMM} achieve marginally higher ARIs ($\sim$0.995--0.997), again due to oracle knowledge of the correct number of clusters. \texttt{Agglomerative} clustering also performs well (ARI~0.9880), but incurs extremely high memory usage (127.6~MB) because it must store and manipulate the complete pairwise distance matrix. 
By contrast, ClusTEK requires only 0.21~MB, as all computations are carried out locally on a compact set of selected grid cells rather than on the complete point cloud.

\texttt{CLIQUE} performs weakest on all metrics (e.g., ARI 0.7873 on \texttt{Aggregation}, 0.4518 on \texttt{R15}, 0.6279 on \texttt{s\_set1}) and shows strong sensitivity to density thresholds. On fine grids it fragments, while on coarse grids it percolates. Matching its grid resolution to ClusTEK does not resolve these issues. Its Python-level cell bookkeeping (lists and dictionaries) leads to nontrivial overhead: runtime of 0.116 to 0.181~s and memory footprint of 1.7 to 6.6~MB, despite the small size of the dataset.
In contrast, ClusTEK is explicitly designed to minimize Python loops, relying instead on 
fixed-size NumPy arrays, 
vectorized diffusion (\texttt{ndimage} convolution), 
local masked operations per occupied cell, 
and a KD-Tree only on selected grid cells rather than on raw points.
These choices keep runtimes in the 0.01–0.05~s range and heap usage below 0.3~MB.

\begin{table}[ht!]
\centering
\caption{Benchmark on Aggregation: accuracy vs.\ efficiency. CPU time is wall-clock (s) and memory is peak Python heap (MB).}
\label{tab:aggregation-bench}
\renewcommand{\arraystretch}{1.15}
\resizebox{\textwidth}{!}{%
\begin{tabular}{l|c|c|c|c|c|c|c|c}
\hline
\textbf{Method} & \textbf{Coverage} & \textbf{ARI} & \textbf{NMI} & \textbf{V-measure} & \textbf{FM} & \textbf{Purity} & \textbf{Time (s)} & \textbf{Peak (MB)} \\
\hline
ClusTEK  & 0.9734 & 0.9754 & 0.9525 & 0.9525 & 0.9807 & 0.9734 & 0.029 & 0.1 \\
KMeans & 1.0000 & $0.7520\pm 0.01$ & $0.8535 \pm 0.007$ & $0.8535 \pm 0.01$ & $0.8045 \pm 0.01$ & $0.8949 \pm 0.005$ & 0.083 & 0.2 \\
GMM  & 1.0000 & 0.8142 & 0.8767 & 0.8767 & 0.8570 & 0.9075 & 0.094 & 0.7 \\
Agglomerative & 1.0000 & 0.8202 & 0.9074 & 0.9074 & 0.8452 & 0.9122 & 0.025 & 2.6 \\
DBSCAN & 0.9339 & 0.9231 & 0.9199 & 0.9199 & 0.9268 & 0.9196 & 0.009 & 0.2 \\
HDBSCAN & 0.8669 & 0.8883 & 0.8643 & 0.8643 & 0.8308 & 0.8390 & 0.096 & 0.9 \\
CLIQUE & 0.8533 & 0.7873 & 0.7928 & 0.7928 & 0.8202 & 0.8710 & 0.116 & 1.7 \\
\hline
\end{tabular}%
}
\end{table}

\begin{table}[ht!]
\centering
\caption{Benchmark on R15: accuracy vs.\ efficiency. CPU time is wall-clock (s) and memory is peak Python heap (MB).}
\label{tab:r15-bench}
\renewcommand{\arraystretch}{1.15}
\resizebox{\textwidth}{!}{%
\begin{tabular}{l|c|c|c|c|c|c|c|c}
\hline
\textbf{Method} & \textbf{Coverage} & \textbf{ARI} & \textbf{NMI} & \textbf{V-measure} & \textbf{FM} & \textbf{Purity} & \textbf{Time (s)} & \textbf{Peak (MB)} \\
\hline
ClusTEK  & 0.9400 & 0.8960 & 0.9225 & 0.9225 & 0.9030 & 0.9333 & 0.011 & 0.1 \\
KMeans & 1.0000 & $0.9928 \pm 0.001$ & $0.7630 \pm 0.007$ & $0.7630 \pm 0.01$ & $0.7531 \pm 0.01$ & $0.8949 \pm 0.005$ & 0.086 & 0.2 \\
GMM           & 1.0000 & 0.9928 & 0.9942 & 0.9942 & 0.9932 & 0.9967 & 0.017  & 0.5 \\
Agglomerative & 1.0000 & 0.9820 & 0.9864 & 0.9864 & 0.9832 & 0.9917 & 0.021  & 1.6 \\
DBSCAN        & 0.9733 & 0.9562 & 0.9631 & 0.9631 & 0.9592 & 0.9683 & 0.012  & 0.2 \\
HDBSCAN       & 0.9651 & 0.9617 & 0.9399 & 0.9399 & 0.9552 & 0.9733 & 0.096  & 0.7 \\
CLIQUE        & 0.4450 & 0.4518 & 0.4808 & 0.4808 & 0.4246 & 0.4450 & 0.101  & 1.5 \\
\hline
\end{tabular}%
}
\end{table}

\begin{table}[ht!]
\centering
\caption{Benchmark on s\_set1: accuracy vs.\ efficiency. CPU time is wall-clock (s) and memory is peak Python heap (MB).}
\label{tab:s-set1-bench}
\renewcommand{\arraystretch}{1.15}
\resizebox{\textwidth}{!}{%
\begin{tabular}{l|c|c|c|c|c|c|c|c}
\hline
\textbf{Method} & \textbf{Coverage} & \textbf{ARI} & \textbf{NMI} & \textbf{V-measure} & \textbf{FM} & \textbf{Purity} & \textbf{Time (s)} & \textbf{Peak (MB)} \\
\hline
ClusTEK  & 0.9670 & 0.9457 & 0.9487 & 0.9487 & 0.9496 & 0.9618 & 0.050 & 0.21 \\
KMeans & 1.0000 & $0.9950 \pm 0.004$ & $0.9930 \pm 0.007$ & $0.9930 \pm 0.01$ & $0.9931 \pm 0.01$ & $0.9969 \pm 0.005$ & 0.108 & 0.6 \\
GMM           & 1.0000 & 0.9970 & 0.9966 & 0.9966 & 0.9972 & 0.9986 & 0.035  & 4.5 \\
Agglomerative & 1.0000 & 0.9880 & 0.9894 & 0.9894 & 0.9881 & 0.9944 & 0.541  & 127.6 \\
DBSCAN        & 0.9776 & 0.9704 & 0.9695 & 0.9695 & 0.9725 & 0.9766 & 0.052  & 1.3 \\
HDBSCAN       & 0.9192 & 0.8686 & 0.9109 & 0.9109 & 0.8774 & 0.9182 & 0.396  & 5.4 \\
CLIQUE        & 0.8066 & 0.6279 & 0.8114 & 0.8114 & 0.6404 & 0.8066 & 0.181  & 6.6 \\
\hline
\end{tabular}%
}
\end{table}

\begin{figure}
\centering

\begin{minipage}[t]{0.3\linewidth}
    \includegraphics[width=\linewidth]{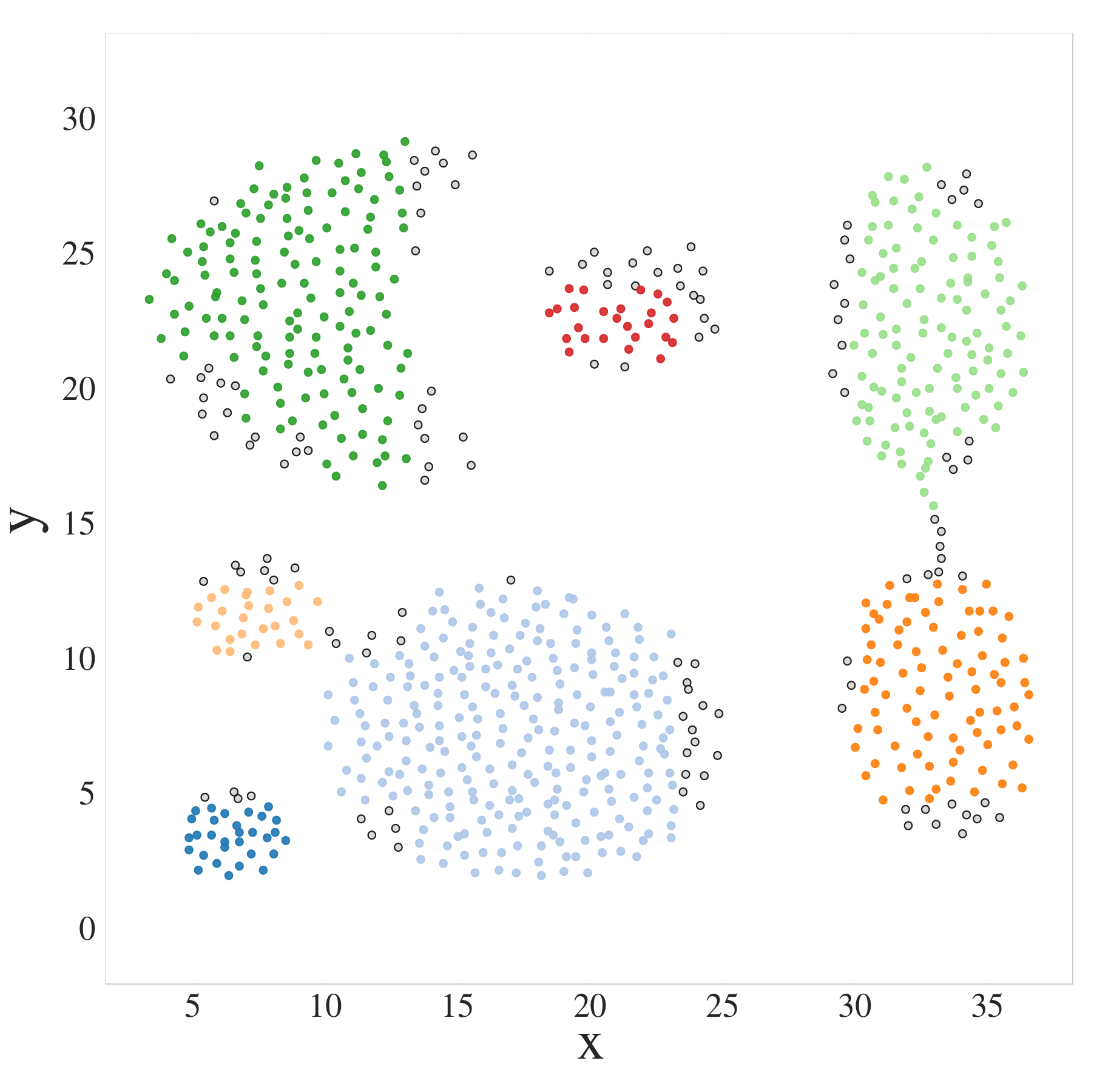}
    \centering
    {\small (a) Aggregation: CLIQUE}
\end{minipage}%
\hspace{0.02\linewidth}%
\begin{minipage}[t]{0.3\linewidth}
    \includegraphics[width=\linewidth]{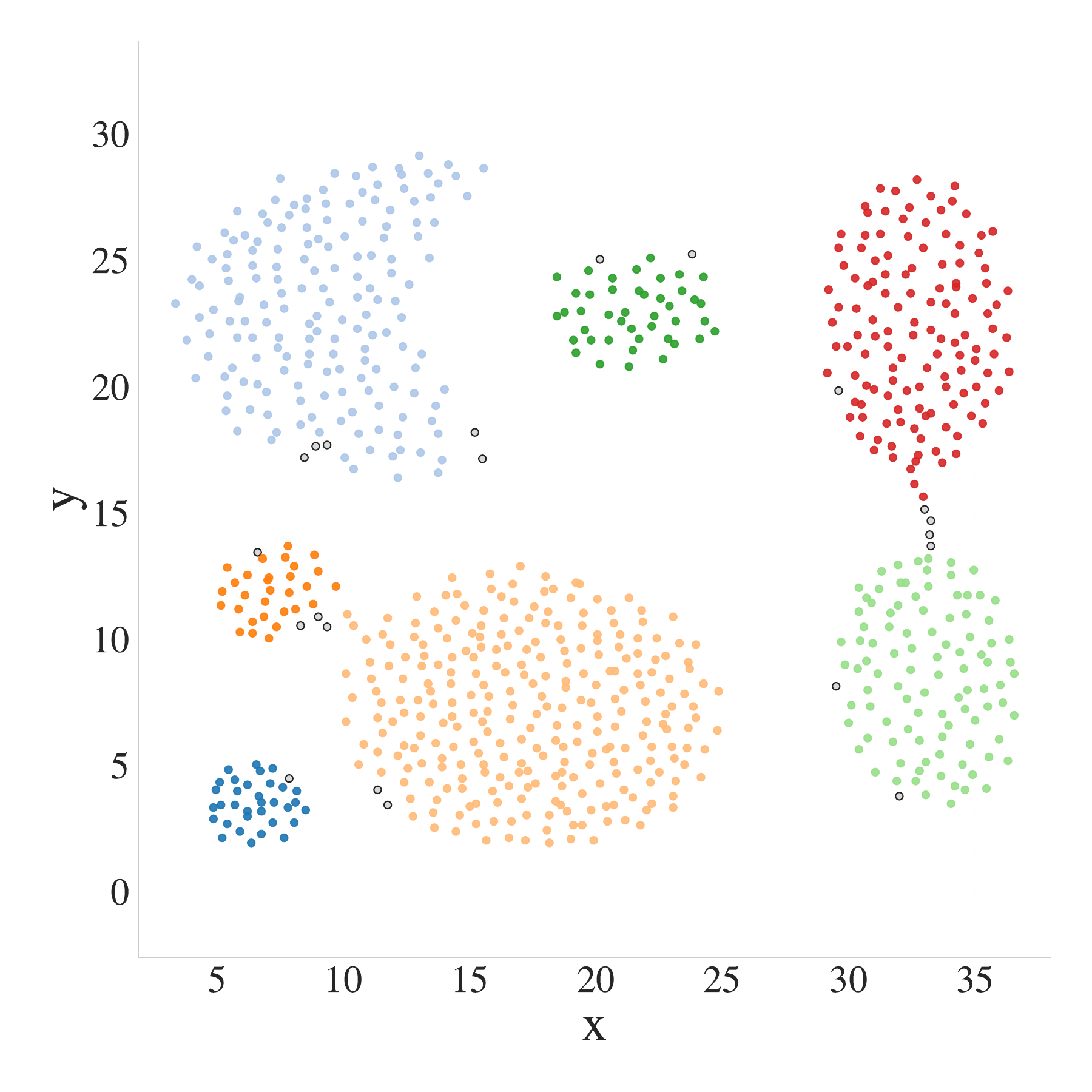}
    \centering
    {\small (b) Aggregation: ClusTEK}
\end{minipage}%
\hspace{0.02\linewidth}%
\begin{minipage}[t]{0.3\linewidth}
    \includegraphics[width=\linewidth]{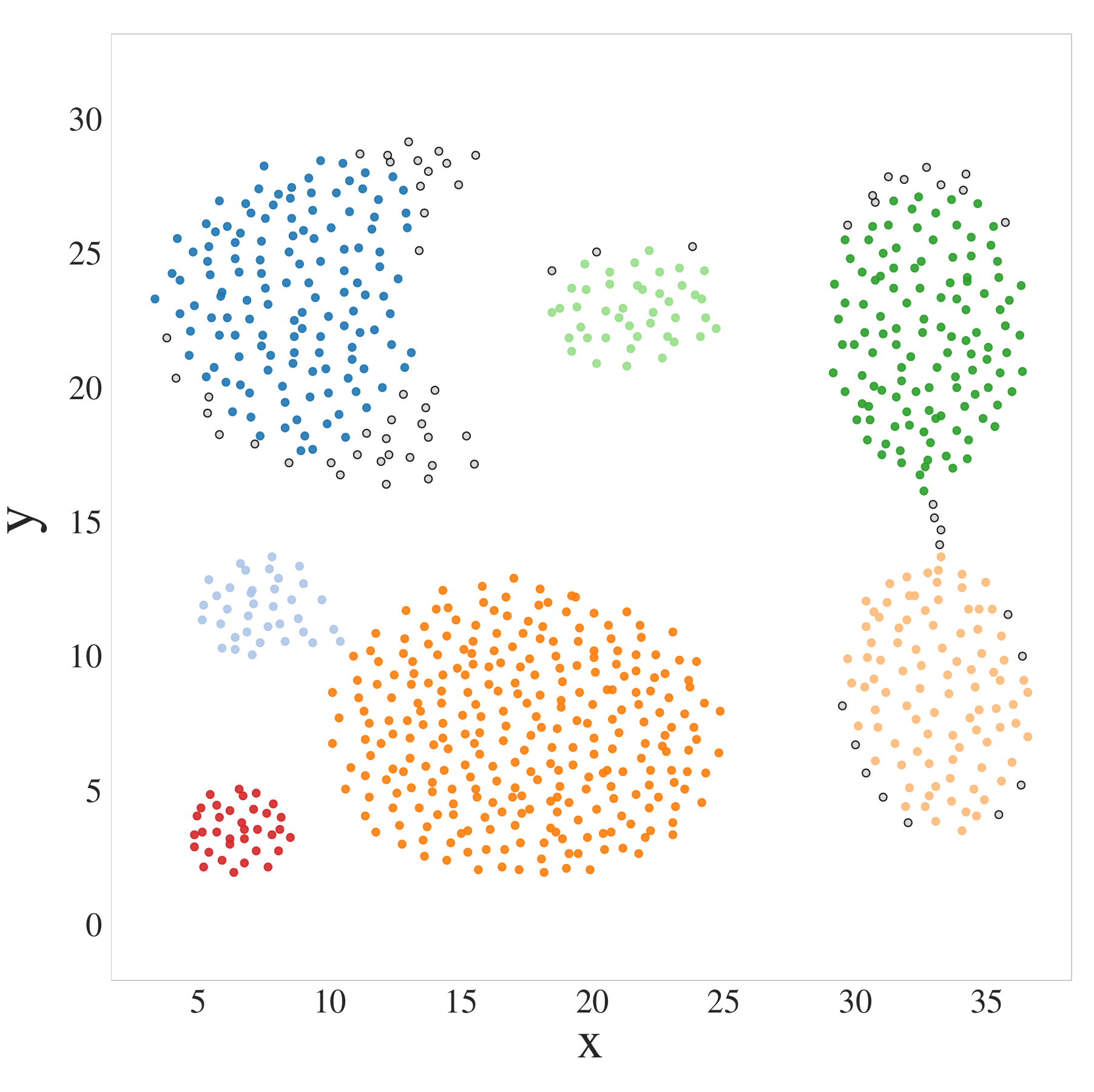}
    \centering
    {\small (c) Aggregation: DBSCAN}
\end{minipage}

\vspace{6pt}

\begin{minipage}[t]{0.3\linewidth}
    \includegraphics[width=\linewidth]{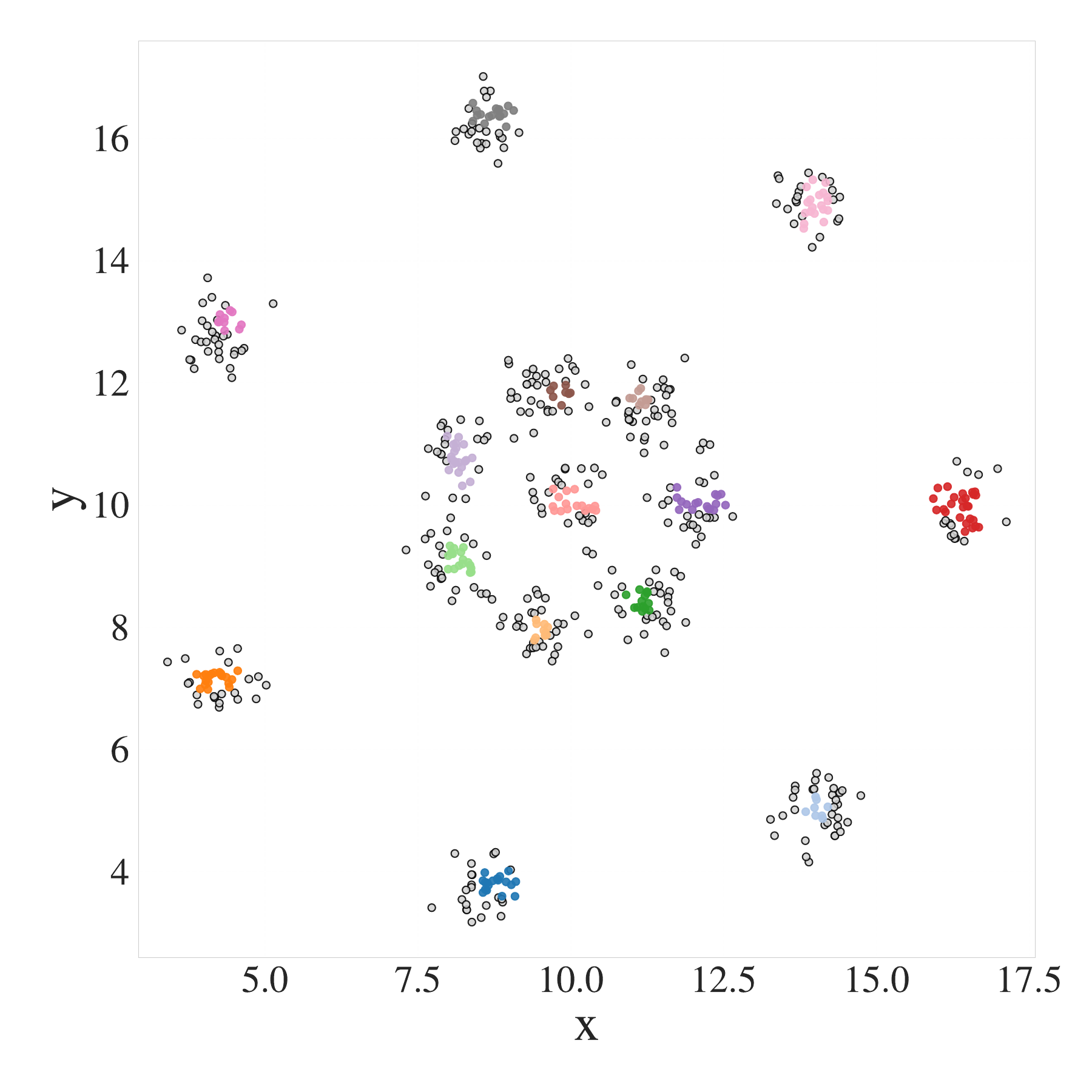}
    \centering
    {\small (d) R15: CLIQUE}
\end{minipage}%
\hspace{0.02\linewidth}%
\begin{minipage}[t]{0.3\linewidth}
    \includegraphics[width=\linewidth]{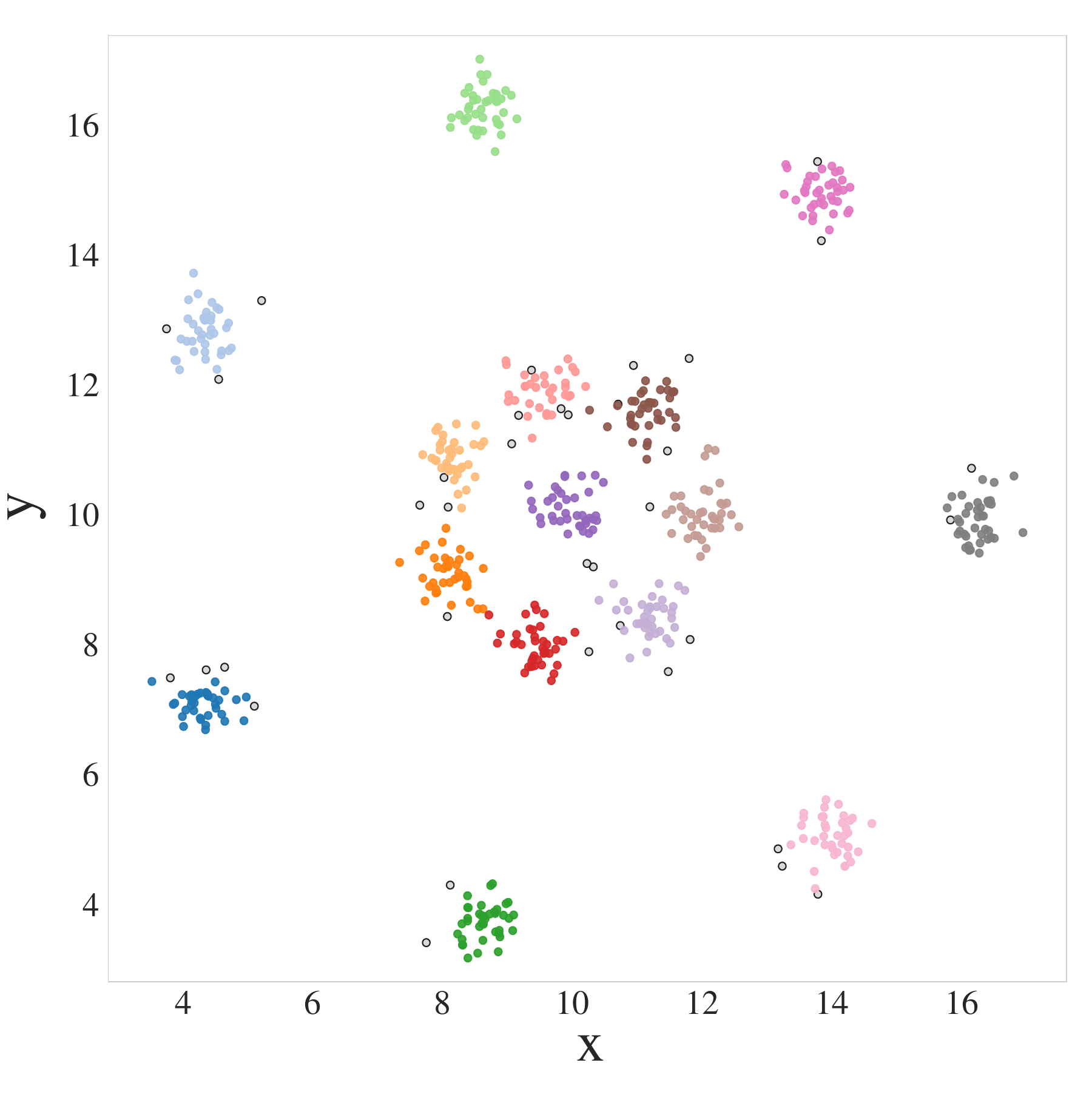}
    \centering
    {\small (e) R15: ClusTEK}
\end{minipage}%
\hspace{0.02\linewidth}%
\begin{minipage}[t]{0.3\linewidth}
    \includegraphics[width=\linewidth]{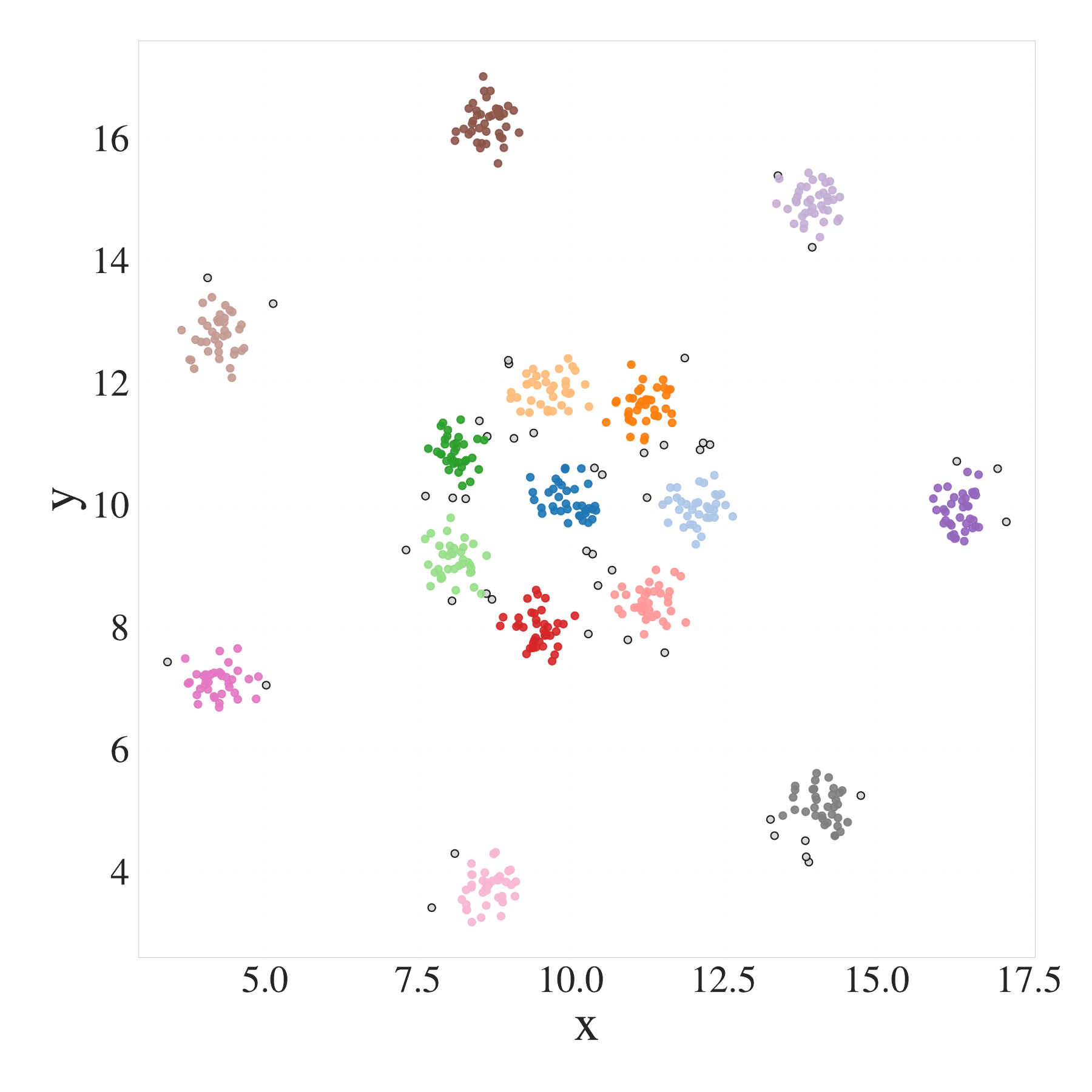}
    \centering
    {\small (f) R15: DBSCAN}
\end{minipage}

\vspace{6pt}

\begin{minipage}[t]{0.3\linewidth}
    \includegraphics[width=\linewidth]{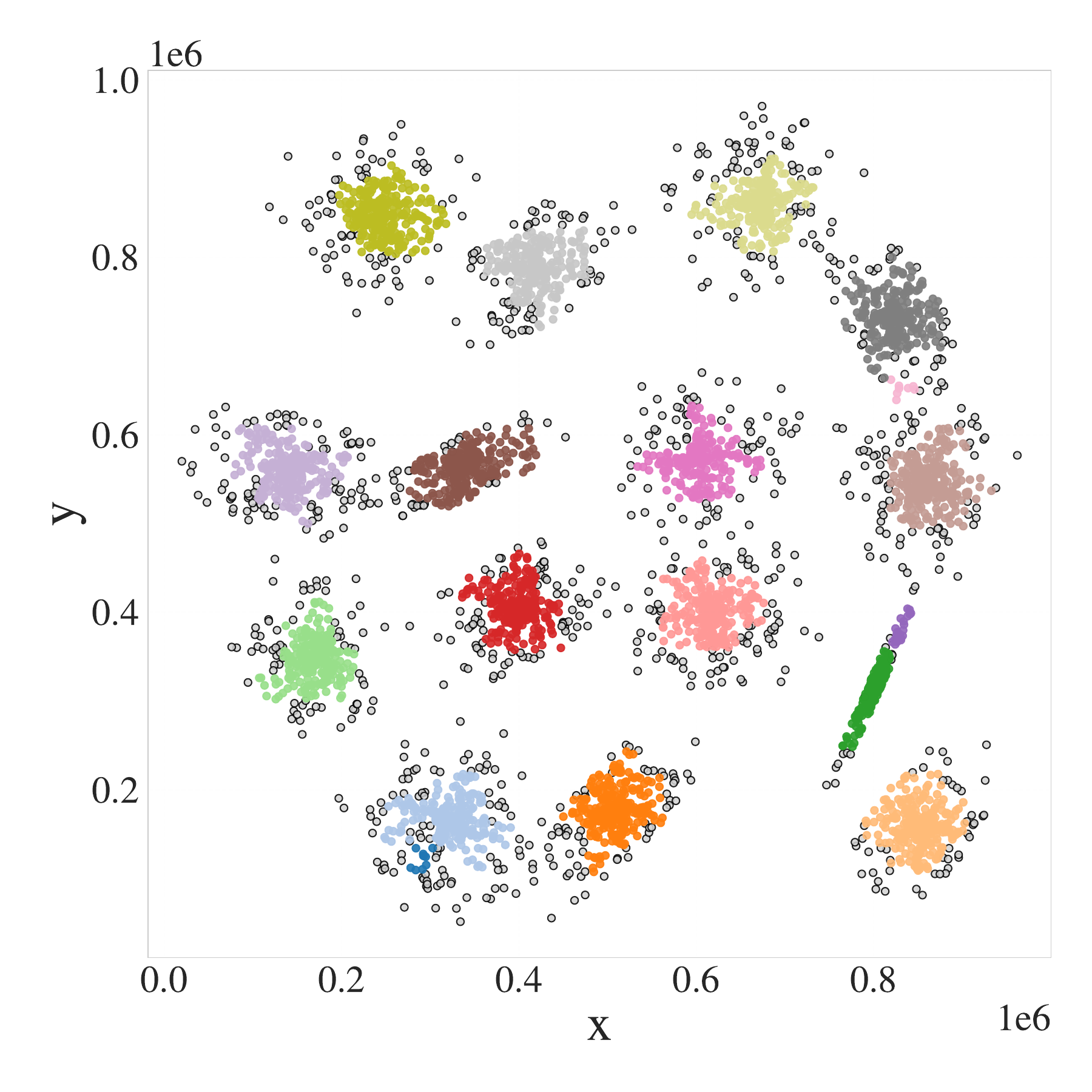}
    \centering
    {\small (g) s\_set1: CLIQUE}
\end{minipage}%
\hspace{0.02\linewidth}%
\begin{minipage}[t]{0.3\linewidth}
    \includegraphics[width=\linewidth]{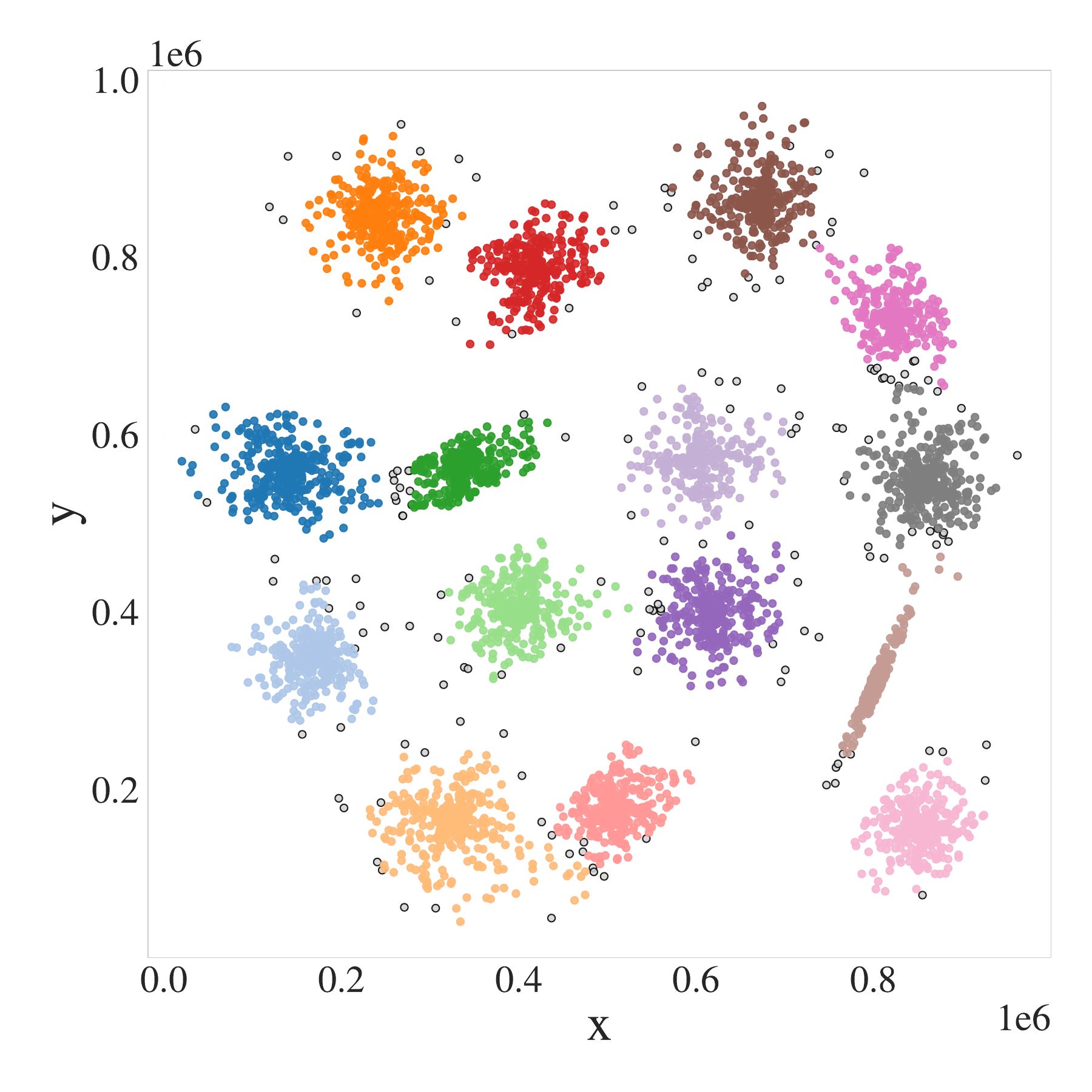}
    \centering
    {\small (h) s\_set1: ClusTEK}
\end{minipage}%
\hspace{0.02\linewidth}%
\begin{minipage}[t]{0.3\linewidth}
    \includegraphics[width=\linewidth]{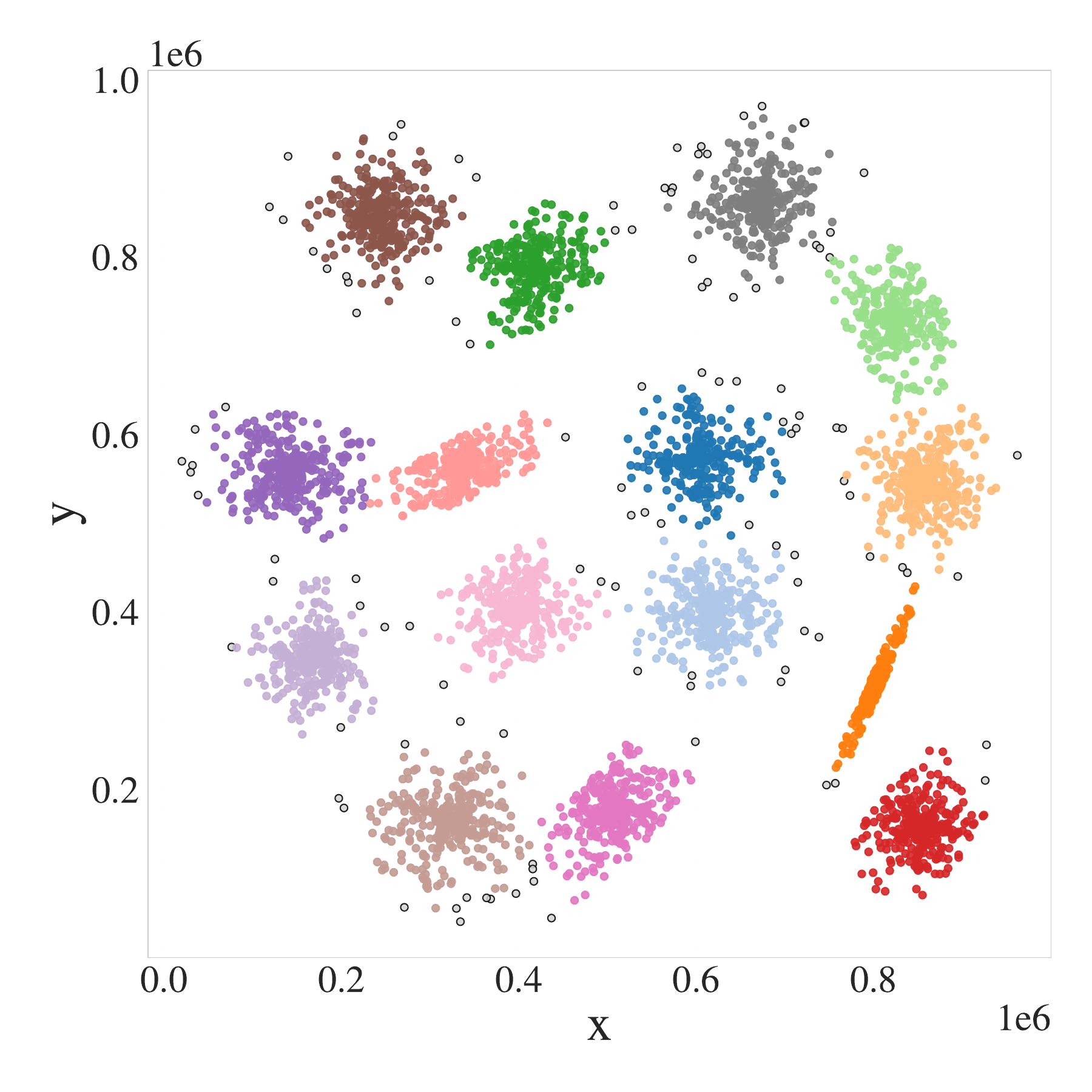}
    \centering
    {\small (i) s\_set1: DBSCAN}
\end{minipage}

\caption{
Qualitative comparison on synthetic 2D benchmarks using each method’s best-performing hyperparameters (Tables~\ref{tab:aggregation-bench}--\ref{tab:s-set1-bench}).
Columns: (left) \texttt{CLIQUE}, (middle) ClusTEK, (right) \texttt{DBSCAN}.
Rows: \texttt{Aggregation}, \texttt{R15}, and \texttt{s\_set1}.
ClusTEK preserves narrow intercluster gaps while maintaining continuity within clusters.
Density-based methods may over-connect crowded regions or absorb boundary points due to sensitivity to hyperparameter tuning.
\texttt{CLIQUE} displays strong resolution dependence.
Axes are omitted for clarity; all panels share identical spatial extents.
}
\label{fig:2d_clustering_comparisons}
\end{figure}

\paragraph{Qualitative comparison.}
Figure~\ref{fig:2d_clustering_comparisons} compares the outputs of \texttt{CLIQUE}, ClusTEK, and a high-quality \texttt{DBSCAN} configuration (selected as a strong baseline of accuracy–efficiency from Tables~\ref{tab:aggregation-bench}--\ref{tab:s-set1-bench}). Each method is shown using its best-performing hyperparameters. ClusTEK consistently preserves narrow gaps and fine-scale boundaries without over-connecting nearby structures, whereas density-based methods may erode thin separations or absorb boundary points due to their sensitivity to local density scales. \texttt{CLIQUE} exhibits characteristic fragmentation at finer resolutions and sparse halos around cluster edges, reflecting the limitations of its classical global grid discretization.

\paragraph{Remarks on fairness and robustness.}
The centroid- and model-based baselines perform extremely well in isotropic and well-separated clusters when supplied with the true $k$, as in \texttt{R15}. ClusTEK, by contrast, requires no prior knowledge of $k$ and is better aligned with datasets that exhibit anisotropy, density gradients, or thin bridges—conditions common in physical simulations. Density-based baselines remain competitive on uniform-density data, but are sensitive to local scale variations and require careful tuning, which is difficult to standardize across heterogeneous datasets or time-resolved trajectories. The grid-based \texttt{CLIQUE} method remains highly sensitive to grid resolution, and even under matched grids its ARI/NMI scores remain substantially lower.

\paragraph{Runtime and memory.}
Across all datasets, ClusTEK achieves runtimes of $3\times 10^{-2}$–$5\times 10^{-2}$~s 
for a full clustering pass (grid binning, dense-region selection, diffusion, and OC-CCA). 
Its memory footprint remains below 0.3~MB, substantially lower than grid-based \texttt{CLIQUE} (1.7–6.6~MB in datasets). 
These empirical trends corroborate the complexity analysis in Sect.~\ref{subsec:complexity} and highlight the suitability of ClusTEK for large-scale spatial datasets requiring memory locality and geometric fidelity.

\subsection{3D Molecular Dynamics Data: Grid Resolution and Diffusion-Imputation Analysis}
\label{sec:results_9k}
The 2D benchmarks in Sect.~\ref{subsec:2d_results} established the behavior of diffusion-enhanced grid clustering in controlled settings, including its robustness to narrow gaps and its favorable runtime–memory profile relative to classical clustering methods. We now transition to three-dimensional MD data, where the clustering task is substantially more demanding because of curved interfaces, heterogeneous local densities, and thermally-driven structural fluctuations.

To isolate the effect of grid resolution and diffusion on clustering fidelity, we begin with a small and interpretable system: a 9k-atom polyethylene configuration (60 C150 chains) quenched to 300~K. The chosen snapshot contains a single well-defined crystalline nucleus. The global density evolution for the 9k system is shown in Appendix
~\ref{app:density_evolution}, Fig.~\ref{fig:density_time_appendix}
This snapshot provides a clean reference for comparing atom-based clustering with grid-based clustering. By systematically varying cell size, crystallinity threshold $C_\mathrm{thr}$, and diffusion-imputation parameters, we identify the operating regime in which grid clustering accurately reproduces atom-level structure while maintaining its computational advantages.

Grid resolutions were selected to span coarse meshes, where each cell aggregates many atoms, to near-atomistic resolutions. A cell was labeled crystalline if its average $C$-index exceeded $C_\mathrm{thr}$, and CCA was used to extract contiguous clusters. The resulting grid-based clusters were compared with atom-based reference clusters to determine the optimal pair $(C_\mathrm{thr},\text{cell size})$. All cell sizes on the grid are reported in Lennard--Jones units; for polyethylene using the SKS model, \(\sigma \approx 3.93\,\text{\AA}\), which means that a cell size of \(1.0\sigma\) corresponds to a spatial resolution of approximately \(3.93\,\text{\AA}\).

\begin{figure}
\centering
\setlength{\tabcolsep}{2pt}
\renewcommand{\arraystretch}{1.0}
\begin{tabular}{ccc}
\includegraphics[width=0.30\linewidth]{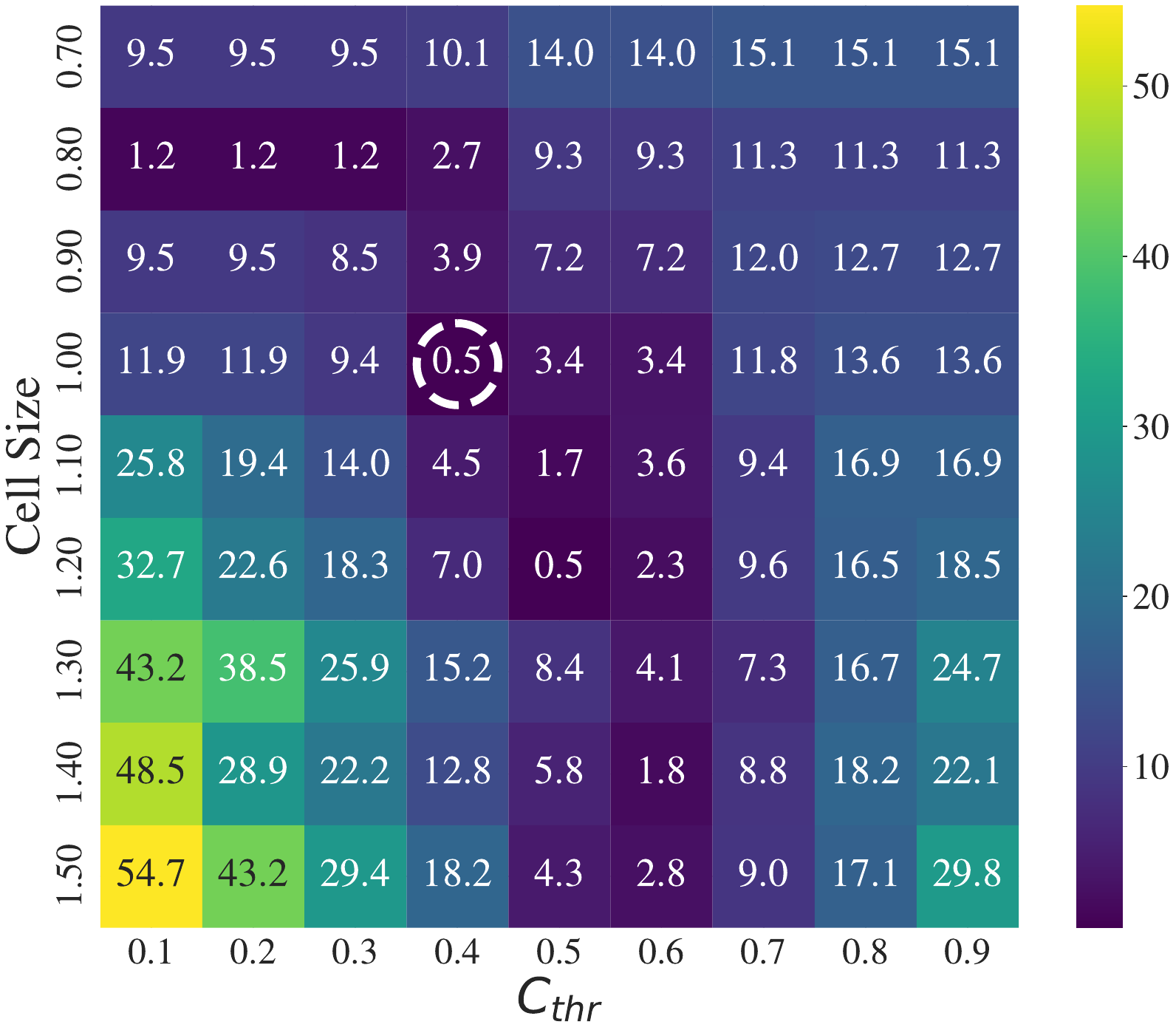} &
\includegraphics[width=0.30\linewidth]{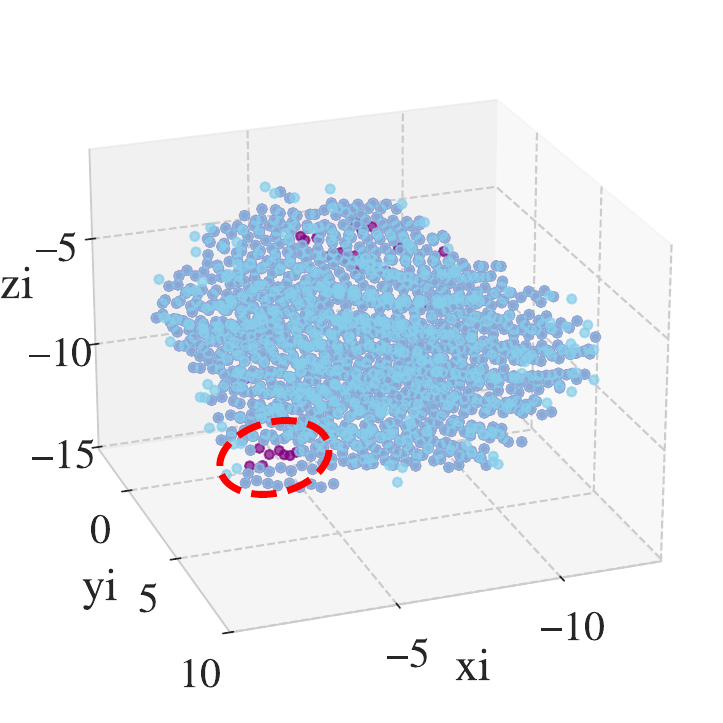} &
\includegraphics[width=0.30\linewidth]{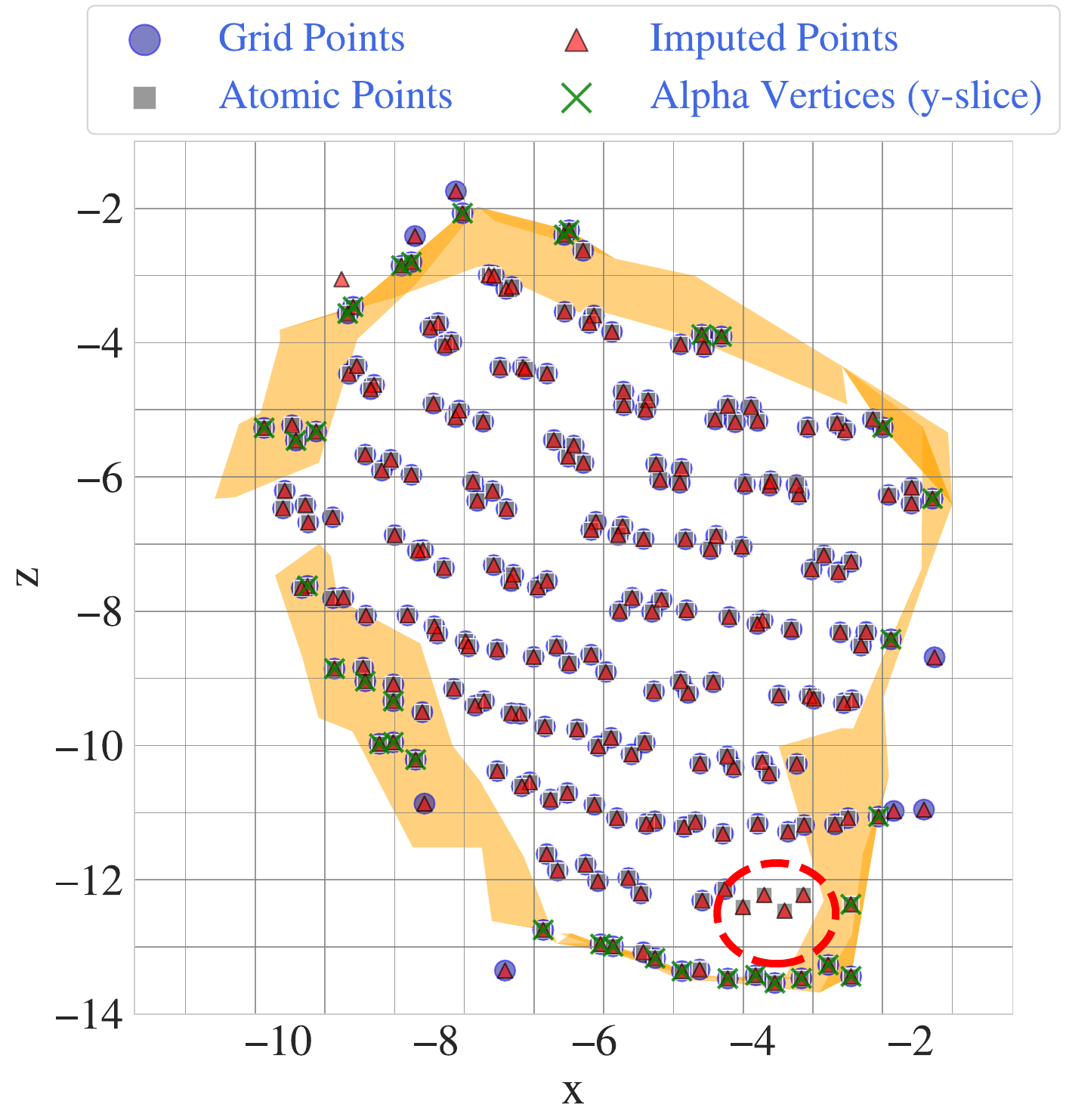} \\[-3pt]
(a)\small & (b)\small & (c)\small \\[4pt]

\includegraphics[width=0.30\linewidth]{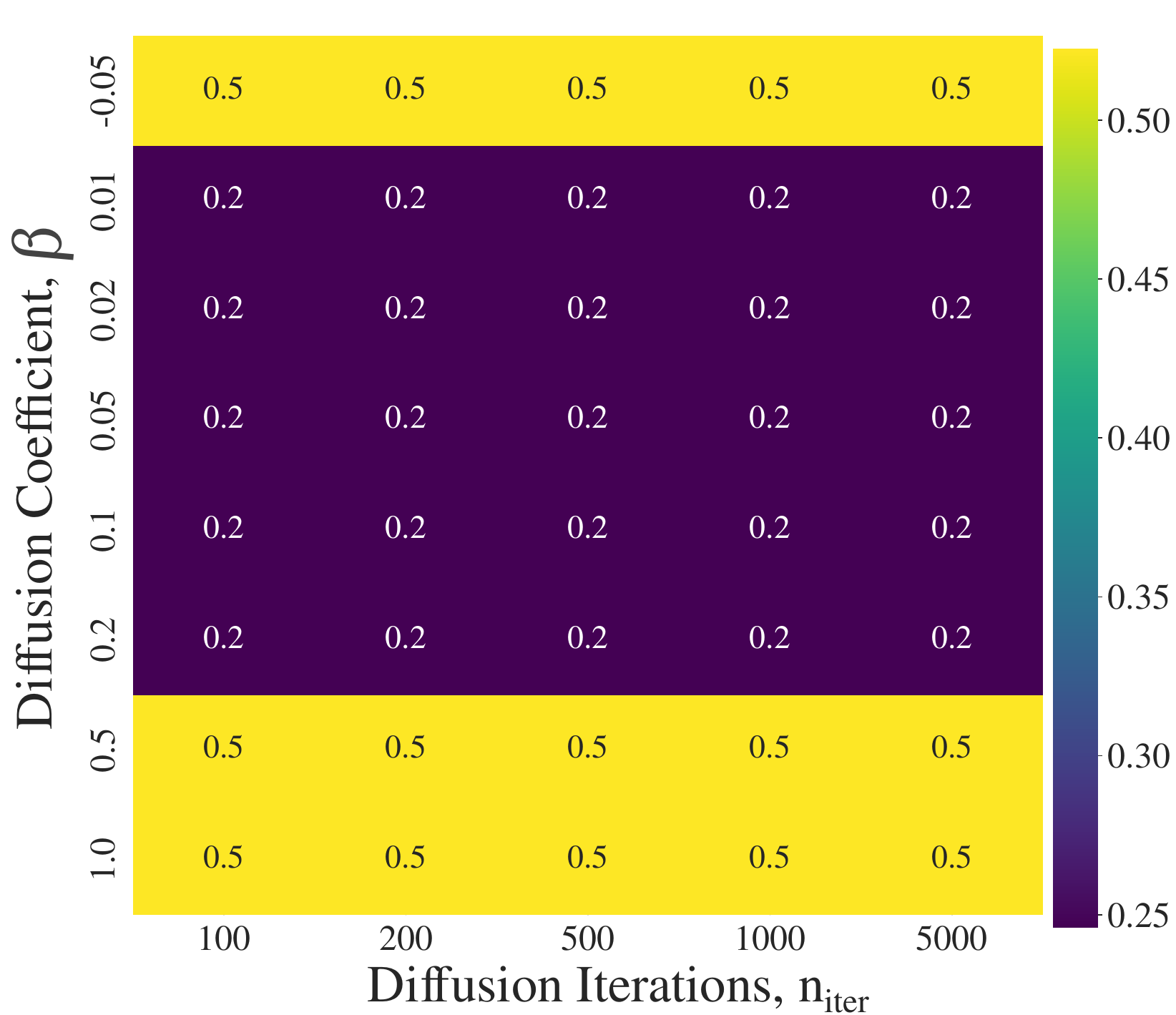} &
\includegraphics[width=0.30\linewidth]{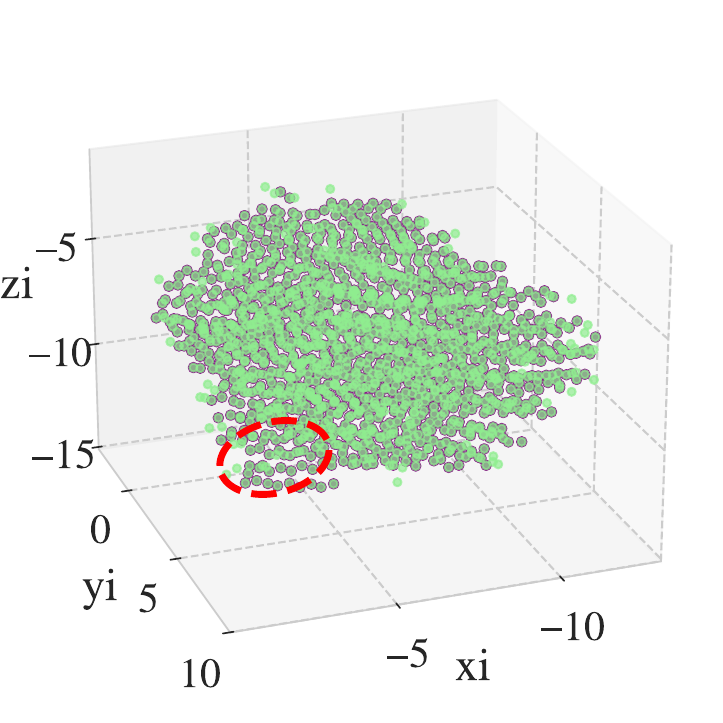} &
\includegraphics[width=0.30\linewidth]{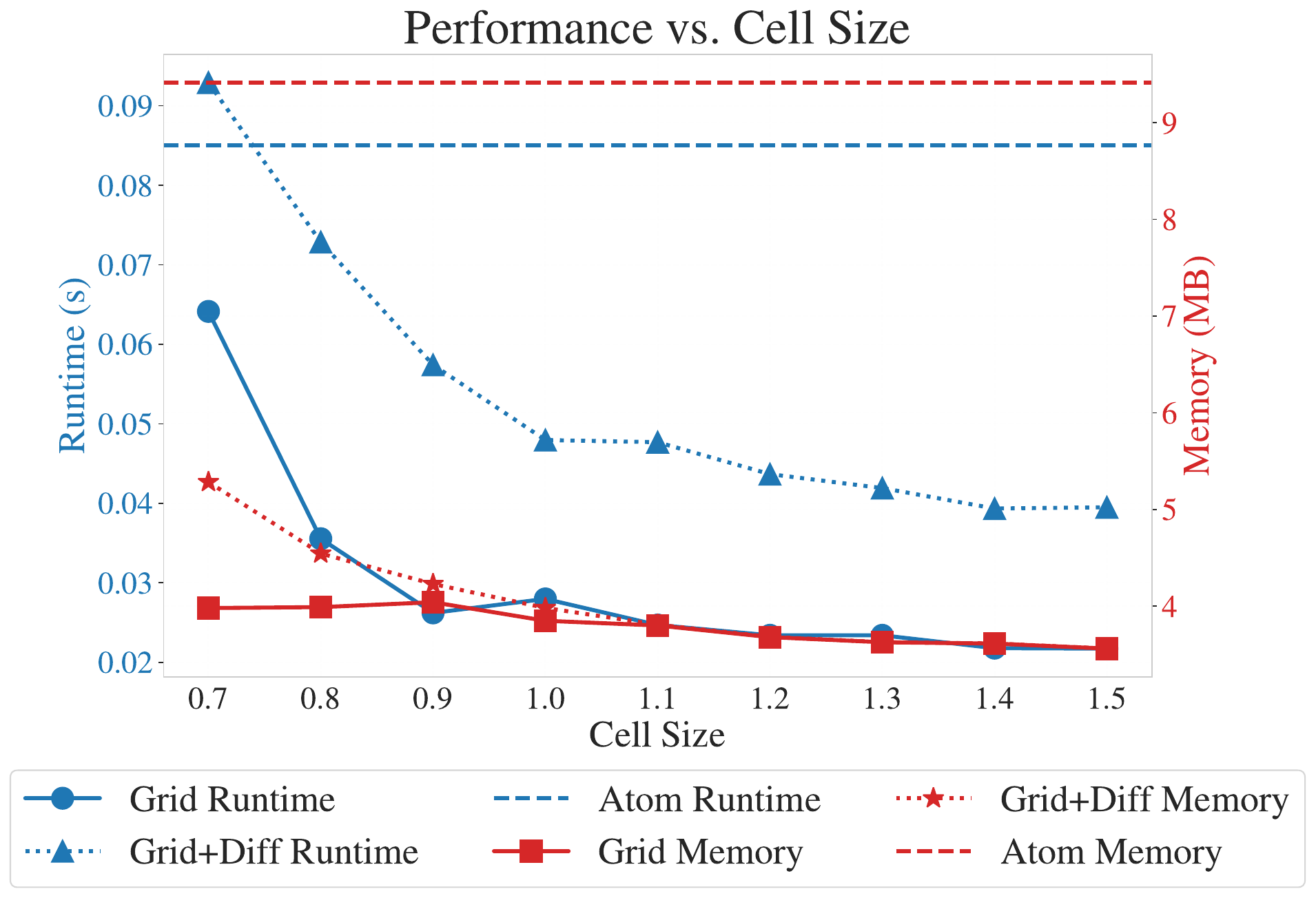} \\[-3pt]
(d)\small & (e)\small & (f)\small
\end{tabular}

\vspace{1ex}

\caption{Grid resolution benchmarking and effect of diffusion-based imputation for the 9k-atom quiescent system.
        (a) Percent volume difference between grid-based and atom-based cluster $\alpha$-shapes over a range of cell sizes and crystallinity thresholds $C_\mathrm{thr}$ for the non-imputed grid clustering. The red circle marks the near-optimal setting at $(C_\mathrm{thr}, \text{cell size}) = (0.4, 1.0)$.
        (b) 3D comparison of atom-based cluster atoms (purple) and grid-based cluster atoms (blue) at the optimal setting. The red-circled region highlights points consistently identified in the atom-based cluster but missed by the grid-based method.
        (c) Cross-sectional $x$--$z$ slice through the red-circled region. Atom-based points are shown as squares, grid-based points as circles, and diffusion-imputed points as triangles. Orange surfaces represent the local $\alpha$-shape polygon, and black crosses mark the slice vertices.
        (d) Volume-difference heatmap over the diffusion hyperparameter space $(\beta, n_{\mathrm{iter}})$ using the optimal grid resolution from panel~(a). The error remains low across a broad range of parameters, indicating robust imputation.
        (e) 3D comparison of atom-based (purple) and diffusion-enhanced grid-based (green) cluster atoms. The previously missed region is now recovered by imputation.
        (f) Wall-clock runtime (blue curves, left axis) and peak Python memory usage (red curves, right axis) for grid-based clustering at $C_\mathrm{thr} = 0.4$ across cell sizes. Solid lines with circular/square markers show the non-imputed grid runs, dotted lines with triangular/star markers show diffusion-enhanced grid runs, and dashed horizontal lines show the atom-based reference values.
}
\label{fig:gridsize_diffs}
\end{figure}

Figures~\ref{fig:gridsize_diffs}(a)--(f) summarize the grid resolution benchmark and the evaluation of diffusion-based imputation. The Panel~(a) presents the percentage volume discrepancy between the $\alpha$-shapes of grid-based and atom-based clusters in the parameter space $(C_\mathrm{thr},\text{cell size})$. The optimal configuration, highlighted in red, occurs near $(0.4, 1.0)$, where the grid-based cluster matches the atom-based reference more closely. The volume difference is computed as the percentage difference between the volume of the cluster enclosed by the grid-based $\alpha$-shape and that of the atom-based $\alpha$-shape. 
The choice of the parameter $\alpha$ is calibrated independently 
(Appendix~\ref{appendix:alpha_tuning}, Fig.~\ref{fig:appendix_alpha_tuning}). 
Similar heat maps based on surface-area discrepancies and diffusion-enhanced grid clustering are provided in Fig.~\ref{fig:appendix_heatmaps_9k} of the Appendix and exhibit the same optimal region in $(C_\mathrm{thr}, \text{cell size})$.

Panel~(b) compares the corresponding 3D point sets: cluster atoms based on atoms (purple) and 
cluster points based on the grid (blue) with optimal resolution. The red-circled region illustrates a characteristic failure mode of coarse grids: atoms located near cell boundaries may be missed due to spatial averaging within cells. These points are consistently included in the atom-based cluster but are excluded by the classical grid-based method.

Panel~(c) provides a detailed $x$--$z$ slice through this region, with grid cluster points shown as circles, imputed points as triangles, and atom-based points as squares. Cross symbols mark the vertices of the $\alpha$-shape in this slice, and orange surfaces represent the polygonal facets of the $\alpha$-shape within the plane. This view clearly shows that the imputed points fill in the region missed by the simple grid clustering. Diffusion-based imputation addresses precisely this scenario by propagating high-crystallinity information across neighboring cells (the full 3D effect is visible in panel~(e)).

The Panel~(d) evaluates the robustness of imputation by scanning the diffusion coefficient $\beta$ 
and the iteration count $n_{\mathrm{iter}}$ while fixing $(C_\mathrm{thr}, \text{cell size})$ to 
their optimal values from panel~(a). The volume discrepancy remains low across a broad parameter range, indicating stable and reliable imputation with a low sensitivity to the diffusion hyperparameters.
Panel~(e) repeats the 3D comparison of panel~(b), now replacing the simple grid-based clustering with the diffusion-enhanced grid clustering with hyperparameters $\beta$ and \texttt{num\_iter} chosen from panel (d), e.g. (0.1, 200). Here, purple points (masked by the green points) denote atom-based cluster atoms, and green points denote imputed grid-based cluster atoms. The same red-circled region from panel~(b) is now fully recovered by the imputation-enhanced method. This shows that imputation mitigates coarsening artifacts without overextending the cluster boundary.

The Panel~(f) reports the wall-clock runtime (blue) and the maximum usage of the Python heap (red) for grid-based clustering at $C_\mathrm{thr}=0.4$ in the cell sizes tested, with dashed horizontal lines indicating the atom-based clustering values. Atom-based clustering is benchmarked at a neighbor search cutoff of $1.5\sigma$, chosen to preserve physical connectivity (verified by visual inspection) while remaining as small as possible to maintain computational tractability. For the 9k system, grid-based clustering is already approximately three times faster than the atom-based method and uses roughly two times less memory. The diffusion-enhanced runs (dotted curves with triangular/star markers), shown here for a representative choice of 500 diffusion iterations, incur only a modest increase in runtime relative to the corresponding non-imputed grid runs, while exhibiting nearly identical memory usage. This confirms that diffusion-based imputation preserves the computational advantage of the grid-based approach at this scale. All memory values reflect \texttt{tracemalloc} measurements within Python, rather than total system memory consumption.

In general, the 9k-atom system identifies the operating regime in which grid-based clustering achieves atom-level fidelity: cell sizes of $0.8$ -- $1.0\sigma$ with $C_\mathrm{thr}\approx$ $0.4$, optionally enhanced with diffusion. Within this regime, the reconstructed cluster accurately matches the atom-based morphology while substantially reducing computational cost. These observations guide the selection of grid and diffusion parameters for the larger MD systems analyzed in Sec.~\ref{sec:results_large}.

\subsection{Validation on Large MD Systems: 180k and 989k Atoms}
\label{sec:results_large}
The 9k-atom baseline study in Sect.~\ref{sec:results_9k} identified an effective 
operating regime for diffusion–enhanced grid clustering: a cell size of 
$\approx 1.0\,\sigma$ with a crystallinity threshold $C_\mathrm{thr}\approx 0.4$. 
Within this range, grid-based clusters reproduced atom-resolved morphology with high 
fidelity, did not require additional parameter tuning, and introduced minor 
computational overhead. We now validate these settings on two substantially larger 
and more heterogeneous systems: (i) a 180k-atom quiescent configuration containing 
multiple simultaneously growing nuclei, with its density evolution shown in Appendix~\ref{app:density_evolution}, Fig.~\ref{fig:density_time_appendix} (b), and (ii) a 989k-atom polyethylene melt undergoing planar elongational flow (PEF), exhibiting elongated domains, thin bridges, and directional anisotropy. These datasets span the two regimes most relevant to polymer crystallization, quiescent nucleation and flow-induced crystallization, and simultaneously enable a direct evaluation of the scalability of diffusion-enhanced grid clustering in terms of runtime, memory usage, and robustness across heterogeneous structural environments.

Unless stated otherwise, the 3D diffusion-enhanced grid algorithm (ClusTEK3D) uses optimized parameters $(C_\mathrm{thr}, \text{cell size}) = (0.4,\,1.0\sigma)$ and the diffusion-imputation settings chosen from the broad low-error plateau identified in Fig.~\ref{fig:gridsize_diffs}(d), e.g.\ $\beta = 0.1$ with $n_{\mathrm{iter}} \approx 500$ to ensure convergence. Atom-based clustering employs a neighbor cutoff of $1.5\sigma$, which we verified by visual inspection to preserve crystalline connectivity while maintaining computational efficiency.

\begin{figure}
\centering
\setlength{\tabcolsep}{2pt}
\renewcommand{\arraystretch}{1.0}
\begin{tabular}{ccc}
\includegraphics[width=0.32\linewidth]{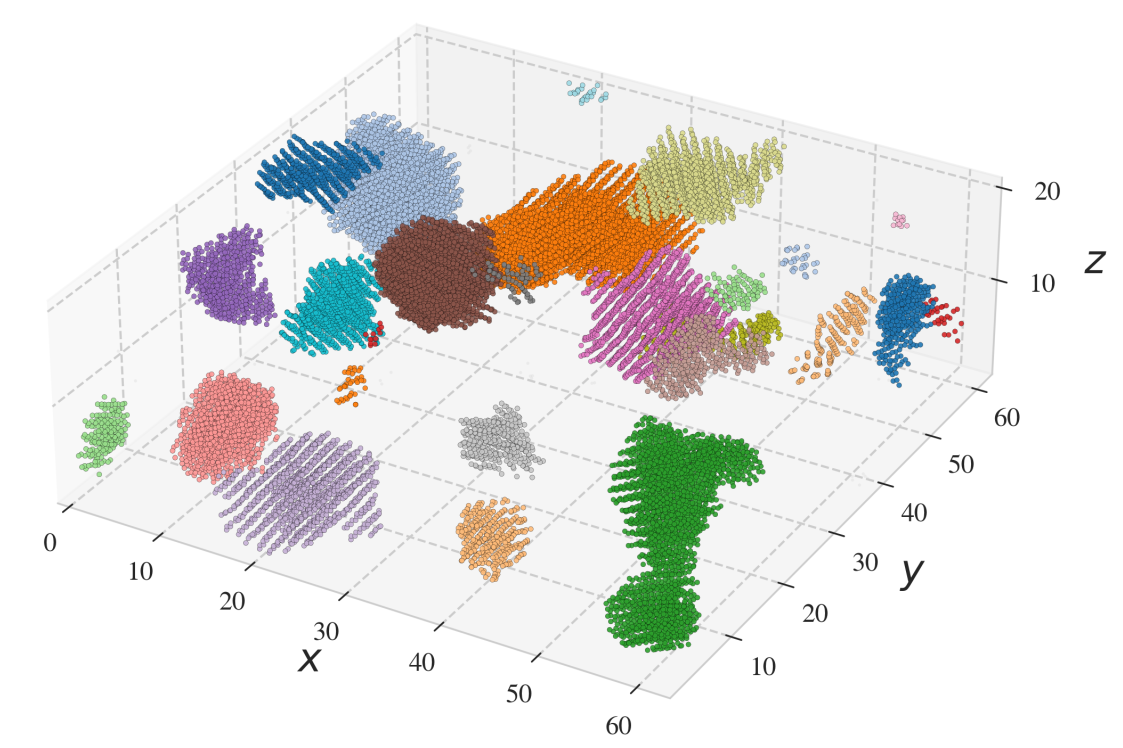} &
\includegraphics[width=0.32\linewidth]{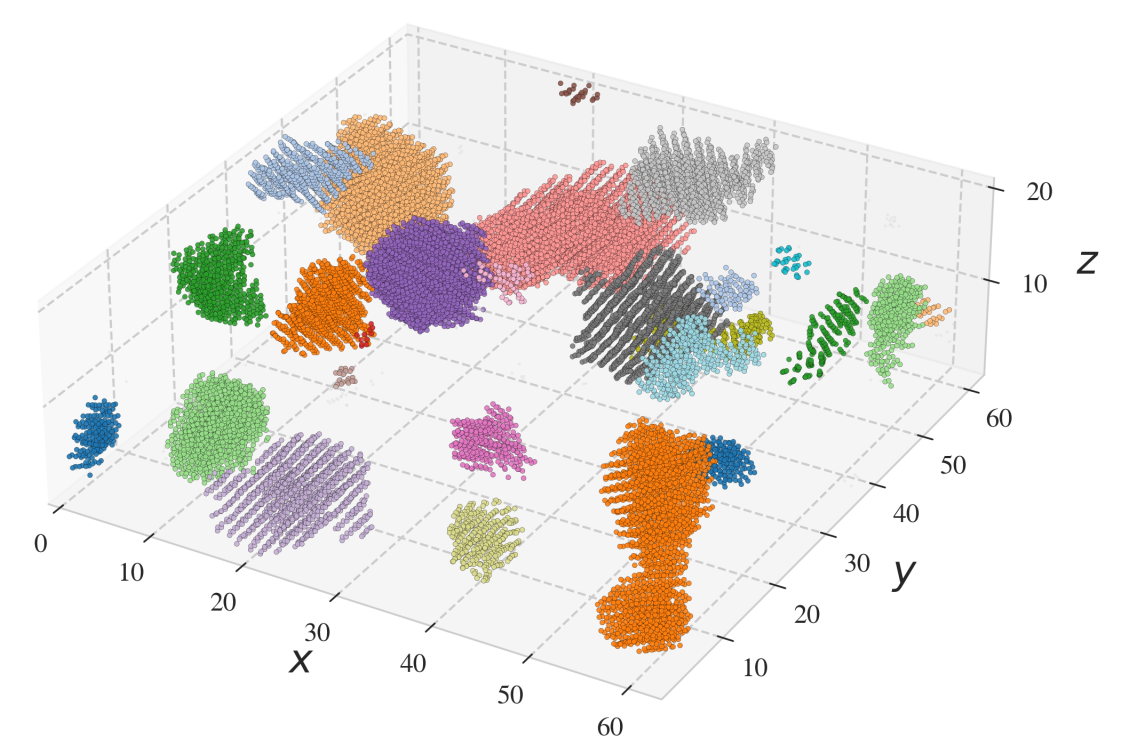} &
\includegraphics[width=0.32\linewidth]{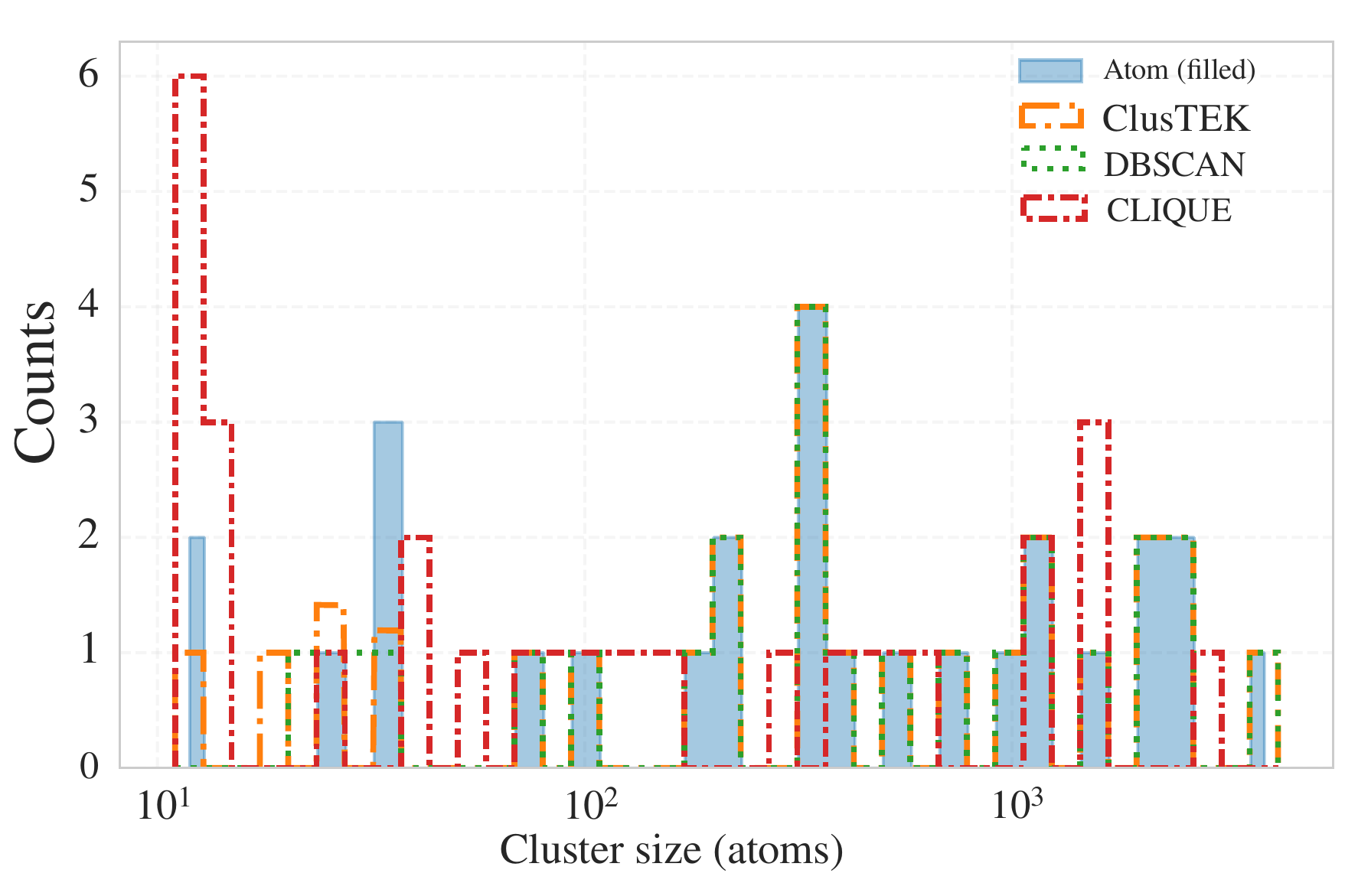} \\[-3pt]
(a)\small 180k atom-based clusters & (b)\small 180k ClusTEK3D clusters. & (c)\small 180k size distributions\\[4pt]

\includegraphics[width=0.32\linewidth]{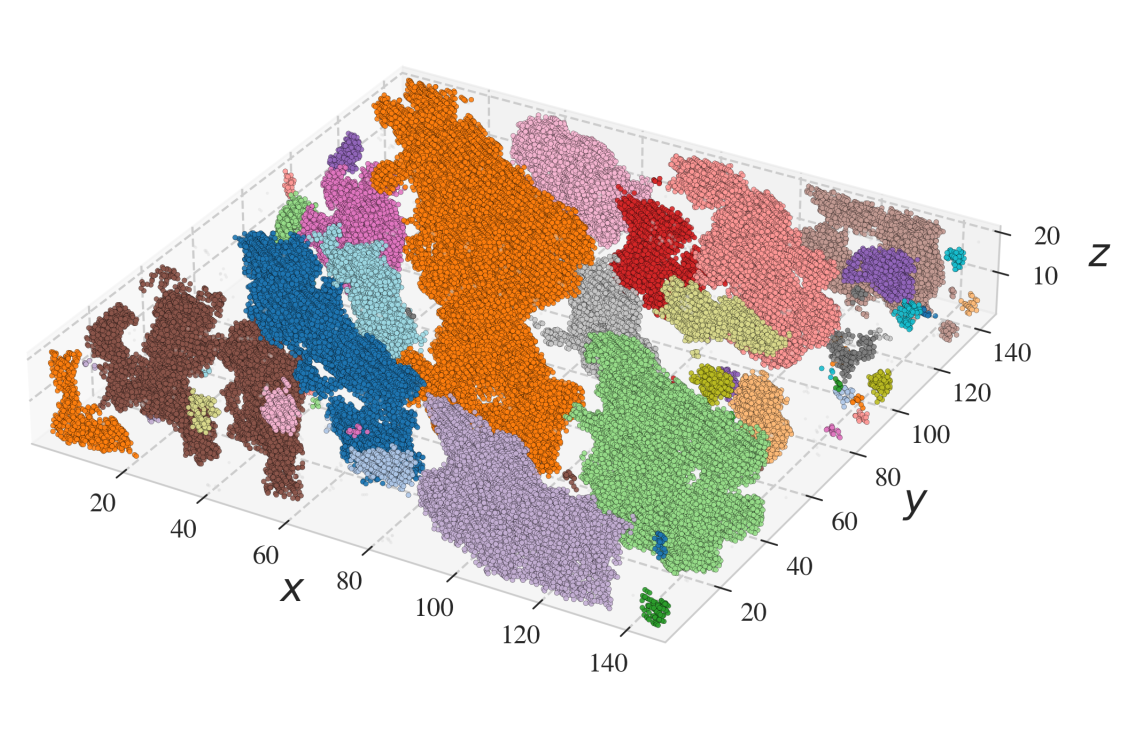} &
\includegraphics[width=0.32\linewidth]{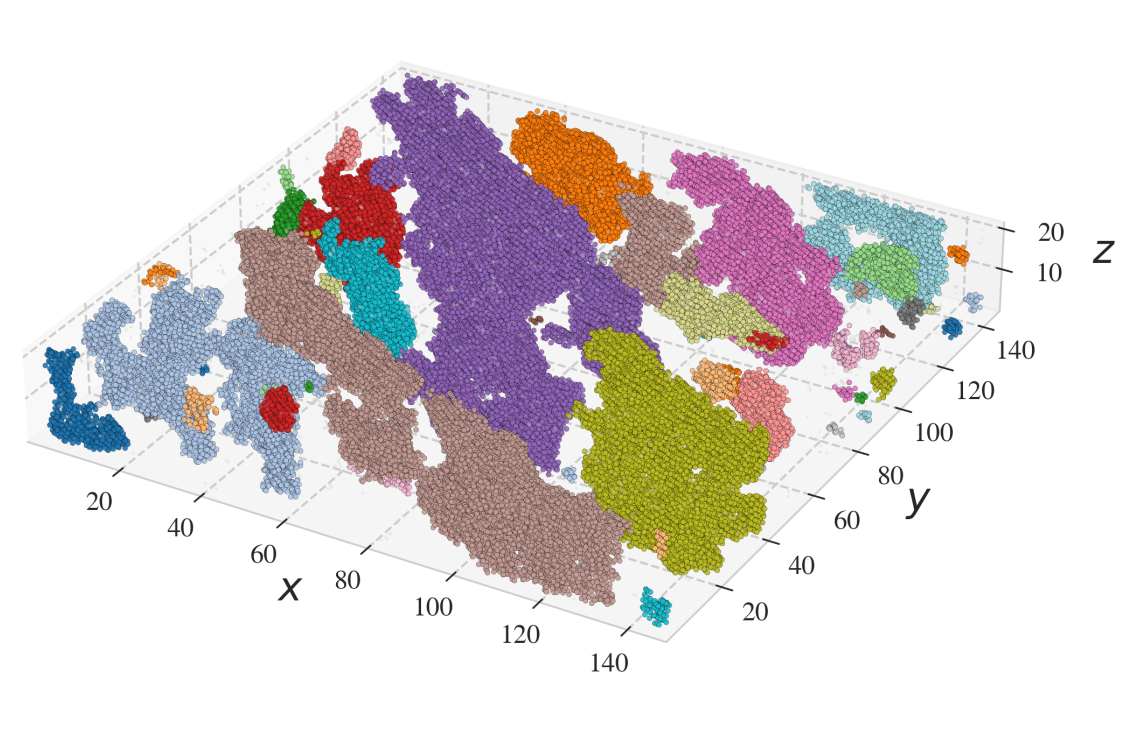} &
\includegraphics[width=0.32\linewidth]{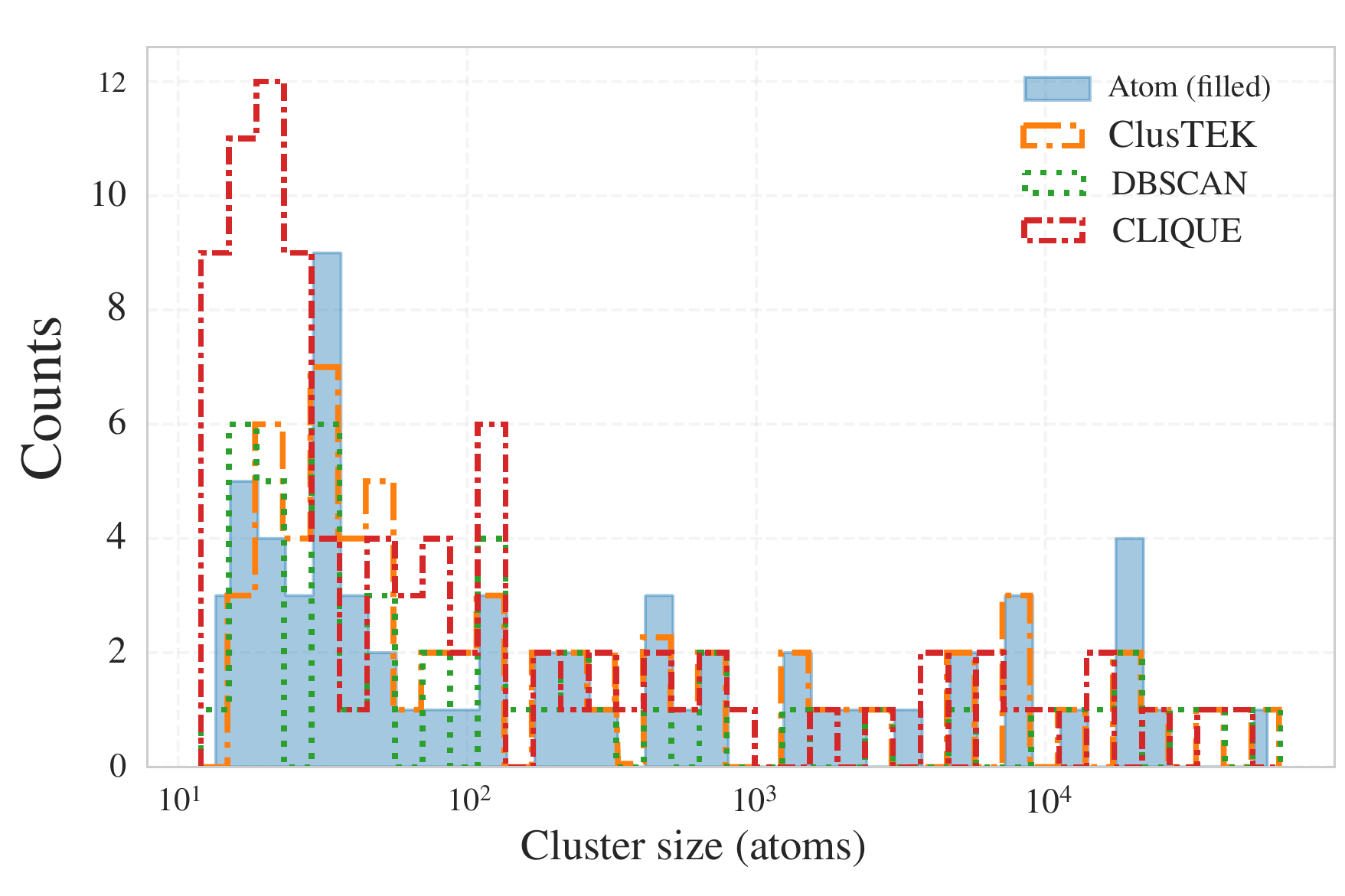} \\[-3pt]
(d)\small 989k atom-based clusters & (e)\small 989k ClusTEK3D clusters & (f)\small 989k size distributions
\end{tabular}

\vspace{1ex}

\caption{Overview of large-system validation. Top row: quiescent 180k-atom system. 
(a) Atom-based reference clusters obtained by atom CCA. 
(b) The ClusTEK pipeline output with $(C_\mathrm{thr},\text{cell size})=(0.4,1.0\sigma)$ and the diffusion parameters calibrated in Sec.~\ref{sec:results_9k}. 
(c) Cluster-size distributions for all methods: atom CCA, ClusTEK, \textsc{DBSCAN}, and 
\textsc{CLIQUE}. 
Bottom row: 989k-atom polyethylene melt under planar elongational flow. 
(d) Atom-based clusters. 
(e) Corresponding ClusTEK3D grid clusters. 
(f) Cluster-size distributions for the same set of methods. 
All panels correspond to a single representative snapshot for each system; 
distribution-based accuracy metrics are reported in Tables~\ref{tab:mean_emd_ks} and~\ref{tab:large_metrics}.}    
\label{fig:large_systems_overview}
\end{figure}

Figure~\ref{fig:large_systems_overview} compares atom-based reference clusters 
with ClusTEK3D grid clusters for representative snapshots of large-scale 
systems. The top row corresponds to the 180k-atom quiescent melt, while the 
bottom row reports analogous results for the 989k PEF-driven configuration.

In the 180k-atoms quiescent system, atom-based CCA identifies multiple well-separated 
crystalline nuclei spanning a broad range of cluster sizes. 
Panels~\ref{fig:large_systems_overview}(a)--(b) show that ClusTEK3D reproduces the atom-based morphology with high fidelity: 
each atom-level nucleus maps to a single grid component, including weakly percolating and branched structures. 
Notably, the grid parameters calibrated on the 9k system transfer directly to the 180k configuration without further adjustment.


The corresponding cluster-size distributions in Fig.~\ref{fig:large_systems_overview}(c) show close agreement between atom-based CCA and ClusTEK3D across the entire size range. 
\textsc{DBSCAN} also yields good agreement for this snapshot; however, its 
hyperparameters were explicitly tuned to optimize performance for this specific 
configuration. As discussed in our previous work~\cite{Cindex}, such a tuning does 
not guarantee robustness across different time regimes or heterogeneous 
snapshots within the same simulation. 
\textsc{CLIQUE}, whose grid resolution and density thresholds were also 
selected to provide a favorable comparison, exhibits larger discrepancies, especially for 
small and intermediate cluster sizes. 
Full three-dimensional renderings of the \textsc{DBSCAN} and \textsc{CLIQUE} 
cluster assignments are provided in Appendix ~\ref{app:dbscan_clique_3d}. 
Quantitative discrepancies between all methods are analyzed in subsequent 
paragraphs using distribution-based metrics.

The 989k configuration poses a more challenging test due to anisotropic 
crystalline domains.
Panels~\ref{fig:large_systems_overview}(d)--(e) show that ClusTEK3D preserves the 
topology of elongated domains, including thin necks and folded branches. 
Atom-based CCA identifies $k=68$ clusters in the representative snapshot, 
whereas ClusTEK3D detects $k=65$, indicating a nearly one-to-one correspondence; 
the small discrepancy arises from a single peripheral component near a major 
nucleus. 
The histograms in panel~(f) again show a close overlap between the atom-based CCA and 
ClusTEK3D. 
Both methods capture a small number of very large domains accompanied by a long 
tail of intermediate-sized clusters, while suppressing spurious tiny components. 
\textsc{DBSCAN} also performs well for this snapshot after hyperparameter tuning, 
detecting clusters $k=47$.
In contrast, \textsc{CLIQUE} fragments several elongated nuclei, producing a total 
of $k=92$ clusters, consistent with its sensitivity to local density variations 
and fixed spatial partitioning.
Three-dimensional renderings of the 989k-atom fragmentations are also provided in Appendix~\ref{app:dbscan_clique_3d}.

To quantify discrepancies between cluster-size distributions produced by 
different clustering algorithms, we evaluated distribution-based metrics.
Table~\ref{tab:mean_emd_ks} reports the mean Earth Mover’s Distance (EMD) and 
Kolmogorov--Smirnov (KS) discrepancies, averaged over three representative 
snapshots per system. 
For each snapshot, the EMD and KS statistics are computed relative to the 
atom-based reference cluster-size distributions. 

Across both large-scale systems, the 180k quiescent melt and the 989k 
flow-driven configuration, diffusion-enhanced grid clustering exhibits the 
closest agreement with the atom-based ground truth. 
ClusTEK3D consistently produces the lowest EMD and KS values, reducing the error 
of the non-imputed grid method by approximately $35$–$45\%$, and substantially 
outperforming the point-based baselines \textsc{DBSCAN} and \textsc{CLIQUE}. 
Detailed per-snapshot KS statistics (including $p$-values) and
EMD values for all methods and system sizes are reported
in Appendix~\ref{app:large_metrics}, Table~\ref{tab:large_emd_ks}.


\begin{table}[t]
    \centering
    \caption{Mean Earth Mover's Distance (EMD) and Kolmogorov--Smirnov (KS) 
    discrepancies with respect to atom-based clustering, averaged over three 
    representative snapshots for each large system. Lower values indicate 
    better agreement with the atom-based reference}
    \label{tab:mean_emd_ks}
    \begin{tabular}{l|c|c|c|c}
        \toprule
        \textbf{Method} & \textbf{180k EMD} & \textbf{180k KS} & \textbf{989k EMD} & \textbf{989k KS} \\
        \hline
        \midrule
        ClusTEK      & 43.620 & 0.076 & 314.264 & 0.149 \\
        \textsc{DBSCAN}      & 44.450 & 0.092 & 951.789 & 0.138 \\
        \textsc{CLIQUE}      & 228.161 & 0.254 & 907.478 & 0.221 \\
        \bottomrule
    \end{tabular}
\end{table}

To quantify further clustering performance and computational efficiency, 
Table~\ref{tab:large_metrics} reports a set of external accuracy metrics 
(coverage, ARI, NMI, V-measure, FM, and purity).
All accuracy metrics are computed with respect to the atom-based CCA reference 
clusters for each snapshot.
Across both systems, ClusTEK achieves the highest or near-highest accuracy scores 
across all metrics while maintaining near-full spatial coverage of the crystalline 
regions. 
In contrast, \textsc{CLIQUE} exhibits greater variability in 
accuracy, particularly for the larger and more heterogeneous 989k system, 
reflecting their sensitivity to hyperparameter selection and local density 
variations.

We do not report direct runtime and memory comparisons for the large-scale MD systems
in Table~\ref{tab:large_metrics}.
ClusTEK performs clustering on the full atomic configuration (180k and 989k atoms),
whereas the baseline methods (\textsc{DBSCAN} and \textsc{CLIQUE}) were applied only
to the subset of atoms pre-filtered as crystalline (approximately 10–20\% of the
system size).
This choice was made deliberately in favor of the baselines to ensure their feasibility
at this scale.
A fully fair performance comparison would require applying \textsc{DBSCAN} and
\textsc{CLIQUE} to the complete four-dimensional space $(x,y,z,C)$ for all atoms,
which would incur substantially higher computational cost and introduce additional
challenges in hyperparameter tuning across evolving time frames.
We therefore restrict large-system comparisons to accuracy, robustness, and
distributional agreement with atom-based references, while detailed runtime and
memory scaling are assessed separately in controlled settings
(Sects.~\ref{sec:results_9k} and~\ref{subsec:2d-comparison}).

\begin{table}[ht!]
\centering
\caption{Accuracy and efficiency metrics for the 180k and 989k systems. 
CPU time is wall‐clock (s); memory is peak Python heap usage (MB).}
\label{tab:large_metrics}
\renewcommand{\arraystretch}{1.15}
\resizebox{0.7\textwidth}{!}{%
\begin{tabular}{lcccccc}
\hline
\textbf{Method} & \textbf{Coverage} & \textbf{ARI} & \textbf{NMI} & \textbf{V-measure} & \textbf{FM} & \textbf{Purity}\\
\hline
\multicolumn{7}{l}{\textbf{180k}} \\
ClusTEK         & 0.9900 & 1.0000 & 1.0000 & 1.0000 & 1.0000 & 1.0000 \\
DBSCAN         & 0.9925 & 0.9999 & 0.9996 & 0.9996 & 0.9999 & 1.0000 \\
CLIQUE         & 0.5615 & 0.9952 & 0.9896 & 0.9896 & 0.9959 & 1.0000 \\
\hline
\multicolumn{7}{l}{\textbf{989k}} \\
ClusTEK         & 0.9894 & 0.9744 & 0.9850 & 0.9850 & 0.9784 & 0.9726 \\
DBSCAN         & 0.9895 & 0.8947 & 0.9399 & 0.9399 & 0.9186 & 0.9090 \\
CLIQUE         & 0.9088 & 0.9306 & 0.9579 & 0.9579 & 0.9454 & 0.9377 \\
\hline
\end{tabular}%
}
\end{table}

A central design principle of ClusTEK is to identify contiguous crystalline domains
directly from raw, possibly unthresholded physical fields, rather than relying on an explicit
prefiltering of atoms by a hard crystallinity cutoff.
In principle, one could trivially isolate crystalline atoms (e.g., by selecting those
with $C$-index $=1$) and subsequently apply a spatial clustering algorithm to their coordinates.
However, such a procedure bypasses the core challenges addressed by ClusTEK:
(i) selecting a physically meaningful threshold in a coarse-grained
representation, and (ii) recovering interfacial connectivity that is lost due to grid
discretization and local sparsity.
This formulation also generalizes naturally beyond crystallinity analysis, as the scalar
field $C$ may be replaced by any physically meaningful per-particle descriptor in
unseen datasets.

\subsection{Statistical Evaluation of Clustering Performance}
\label{sec:stats}

To assess whether the three clustering algorithms (ClusTEK, DBSCAN, and CLIQUE) exhibit equivalent performance across snapshots, we applied the nonparametric Friedman test to each accuracy metric. The Friedman test evaluates whether the median performance ranks of a set of algorithms are identical under repeated measurements, without assuming normality of the underlying distributions. The null hypothesis is that all algorithms achieve equivalent performance across snapshots.

Table~\ref{tab:friedman} reports the Friedman test statistics and corresponding $p$-values for both system sizes (180k and 989k atoms) across the principal accuracy metrics. For the 180k system, the coverage differences are statistically significant at the level $\alpha = 0.05$ ($p = 0.0498$). The remaining metrics (ARI, NMI, V-measure, FM)
display $p$-values in the range $0.059$--$0.061$, suggesting statistically significant differences at the less conservative $\alpha = 0.10$ threshold. Purity, as expected from its near-unity values across all algorithms, shows no significant differences.

For the 989k system, purity again shows significance at $\alpha=0.05$ ($p = 0.0498$), while coverage, NMI, and the V-measure exhibit moderate evidence of performance differences (with $p \approx 0.097$). ARI and FM yield higher $p$-values, reflecting the metric instability driven by the large morphological variability of the 989k snapshots.

Overall, the Friedman analysis provides consistent statistical evidence that the 
algorithms do not behave equivalently across snapshots, with ClusTEK generally 
attaining the top performance rank across all metrics.
Although post-hoc pairwise tests (e.g., the Nemenyi post-hoc test for Friedman rankings~\cite{nemenyi1963distribution}) can be applied when a larger number of datasets are available, their power is limited for the present sample size of three snapshots per system. We therefore refrain from pairwise
comparisons and instead rely on the stable and consistently superior ranking of
ClusTEK across metrics as evidence of its improved clustering fidelity relative to
\textsc{DBSCAN} and \textsc{CLIQUE}.

\begin{table}
\centering
\caption{Friedman test statistics and corresponding $p$-values for the
180k and 989k systems across all accuracy metrics. The null hypothesis
states that all clustering algorithms exhibit equivalent performance
across snapshots.}
\label{tab:friedman}
\renewcommand{\arraystretch}{1.25}
\resizebox{0.7\textwidth}{!}{%
\begin{tabular}{lcccccc}
\hline
\textbf{180k System} 
& \textbf{Coverage} & \textbf{ARI} & \textbf{NMI} & \textbf{V} 
& \textbf{FM} & \textbf{Purity} \\
\hline
Friedman $\chi^2$ 
& 6.0000 & 5.6000 & 5.6364 & 5.6364 & 5.6000 & 2.0000 \\
$p$-value           
& 0.0498 & 0.0608 & 0.0597 & 0.0597 & 0.0608 & 0.3679 \\
\hline
\hline
\textbf{989k System} 
& \textbf{Coverage} & \textbf{ARI} & \textbf{NMI} & \textbf{V}
& \textbf{FM} & \textbf{Purity} \\
\hline
Friedman $\chi^2$
& 4.6667 & 2.6667 & 4.6667 & 4.6667 & 2.6667 & 6.0000 \\
$p$-value
& 0.0970 & 0.2636 & 0.0970 & 0.0970 & 0.2636 & 0.0498 \\
\hline
\end{tabular}
}
\end{table}

\section{Conclusion}
\label{sec:conclusion}

This work presented a diffusion–enhanced grid clustering framework
(ClusTEK) for scalable analysis of large molecular
dynamics datasets.  
The method integrates three components:  
(i) grid-based coarse-graining of local structural properties  
(here, crystallinity via the $C$-index),  
(ii) diffusion-based imputation to stabilize sparse or partially sampled
cells, and  
(iii) origin-constrained connected-component analysis to ensure
physically consistent cluster connectivity.  
Together, these steps provide an efficient alternative to atom-based
clustering for systems containing hundreds of thousands to millions of
particles.

Synthetic 2D benchmarks showed that diffusion-enhanced imputation improves
cluster continuity without over-smoothing, enabling ClusTEK to recover thin
gaps, irregular cluster geometries, and variable-density regions. 
Using a 9k-atom polyethylene system, we identified an operating regime
in which the method closely matches atom-based $\alpha$-shape references
while achieving substantial reductions in runtime and memory usage.

Applications to 180k- and 989k-atom systems showed that these parameters
transfer robustly to more heterogeneous crystallization environments,
including quiescent and flow-driven regimes.
Across snapshots, ClusTEK maintained high agreement with atom-based
clustering and exhibited more stable accuracy than \textsc{DBSCAN} and
\textsc{CLIQUE}, while remaining computationally efficient at the largest
scale tested.
Statistical analysis using the Friedman test further indicated that the
algorithms do not behave equivalently across snapshots for several key
metrics, with ClusTEK consistently achieving the top or near-top
performance ranks.

In general, ClusTEK offers a scalable, physically consistent, and
computationally efficient approach to clustering large MD datasets from spatially embedded scalar structural fields.
Its efficient runtime, modest memory footprint, and robustness to
heterogeneous morphologies make it suitable for long trajectories and for
systems extending to millions of atoms.  
The framework also provides a practical foundation for future extensions,
including parallelization, GPU acceleration, and integration with other
computational analysis tools.

\section*{acknowledgments}
The authors acknowledge the Texas Advanced Computing Center (TACC) at the University of Texas at Austin for providing high-performance computing resources used in the large-scale molecular dynamics simulations. Post-processing and data analysis were performed using computational infrastructure provided by the ISAAC at the University of Tennessee, Knoxville. Financial support was provided by the Materials Research and Innovation Laboratory (MRAIL) at the University of Tennessee, Knoxville.

\section*{Author Declarations}

\subsection*{Conflict of Interest}
The authors have no conflict of interest to disclose.

\subsection*{Author Contributions}
Elyar Tourani: Conceptualization (equal); Methodology (equal); Investigation (lead); Data curation (lead); Visualization (lead); Writing – original draft (equal); Writing – review \& editing (equal).\\
Brian J. Edwards: Conceptualization (equal); Methodology (equal); Formal analysis (equal); Supervision (equal); Writing – review \& editing (equal).\\
Bamin Khomami: Conceptualization (equal); Methodology (equal); Formal analysis (equal); Supervision (equal); Funding acquisition (lead); Writing – review \& editing (equal).

\section*{Code and Data Availability}
The complete implementation of the diffusion-enhanced grid clustering method
(\textsc{ClusTEK}), including all parameter-sweep scripts, configuration files, and
reproducibility utilities used in this study, is publicly available at
\url{https://github.com/etourani/ClusTEK}.

\appendix
\section*{Appendix}
\section{Experimental Environment and Implementation}
\label{app:config}

\paragraph{Hardware and Software.}

All experiments were run on a single-node CPU system with a 13th Gen Intel Core i9\textendash13900K (up to 5.8 \,GHz),
$1\times64$\,GB DDR5 RAM (36 \,MB cache) and a 1 \,TB NVMe SSD.
Our implementation uses Python~3.12.3 with NumPy~1.26.4, Pandas~2.2.2, SciPy~1.11.4, 
scikit\textendash learn~1.5.1, Matplotlib~3.9.2, and Seaborn~0.13.2.
Bayesian optimization (when enabled) relies on \texttt{scikit\textendash optimize} (\texttt{skopt}); 
the pipeline degrades gracefully if \texttt{skopt} is unavailable.

\paragraph{Reproducibility.}
We fix the random seed of the BO optimizer to 11 (\texttt{random\_state=11}). 
All intermediate artifacts are written on disk for auditability: 
Stage~I candidates (\path{stageA_pre_diffusion_candidates.csv}), 
Stage~II candidates (\path{stageB_post_diffusion_candidates.csv}) 
and the final summary (\path{best_params_summary.json}). 
Figures for pre\-/post\-/OC\textendash CCA overlays are also saved under the specified output directory.

\paragraph{Grid suggestion and preprocessing.}
Given 2D points \((x,y)\), we propose a quasi-isotropic cell size via three seeds:
(i) k-NN spacing using \texttt{SciPy}’ \texttt{cKDTree} (\(k{+}1\) query, median of the \(k\)th neighbor);
(ii) occupancy targeting (\(\text{avg.\ occupancy}\approx \texttt{TARGET\_OCC}\) while preserving aspect ratio); and 
(iii) the Freedman-Diaconis rule per-axis (geometric mean across axes). 
We then sweep around the consolidated estimate to generate a small candidate set of \((n_x,n_y)\) grids. 
Binning uses vectorized index arithmetic of \(O(n)\).

\paragraph{Diffusion imputation and boundary conditions.}
On the selected grid, we build a normalized field \(C^{(0)}\in[0,1]\), and form three masks:
Dense (\(C^{(0)}\!>\!C_{\mathrm{thr}}\)), Sparse (\(0\!<\!C^{(0)}\!\le\!C_{\mathrm{thr}}\)), and Empty (\(C^{(0)}\!=\!0\)).
We run explicit weighted diffusion on Sparse cells only,
\[
C^{(n+1)}\big|_{\text{Sparse}} \gets \mathrm{clip}\!\Big(C^{(n)} + \beta\, w \odot (L * C^{(n)}),\,0,\,1\Big),
\]
with Dense clamped to \(1\) and Empty clamped to \(0\) at each step. 
Here \(L\) is the 2D 5-point discrete Laplacian implemented via \texttt{scipy.ndimage.convolve}; 
boundary conditions follow \texttt{mode="wrap"} (periodic) or \texttt{"nearest"} (nonperiodic), exactly matching the 
\texttt{PERIODIC\_CCA} flag. 
We terminate when either \(n\ge n_{\min}\) and \(\max_{\text{Sparse}}\!\lvert C^{(n+1)}{-}C^{(n)}\rvert<\varepsilon\) 
or after the \(N_{\max}\) steps. In all reported runs, we use \(N_{\max}=50{,}000\), \(n_{\min}=60\), 
\(\varepsilon=10^{-6}\), and \texttt{check\_every}\(=10\).

\paragraph{Selection and labeling.}
After diffusion, selected cells are added to the system according to 
\[
\mathcal{S}=\{C^{(0)}\!>\!C_{\mathrm{thr}}\}\ \cup\ \{0\!<\!C^{(0)}\!\le\!C_{\mathrm{thr}} \ \wedge\ C^{(\mathrm{final})}\!>\!C_{\mathrm{sel}}\}~.
\]
We first run standard CCA (union-find) on the Dense mask to obtain seed labels \(L_{\text{seed}}\) under 
4- or 8-connectivity with optional periodic wrapping. 
Then we perform an origin-constrained region growing into \(\mathcal{S}\): each unlabeled cell adopts a label 
if and only if its face-neighborhood contains exactly one distinct seed label, ensuring no post-hoc cluster merging. 
Connectivity uses direct lattice neighbors (no KD-tree).

\paragraph{Scoring and tuning.}
Partitions are scored using a composite \(\mathcal{Q}=w_{\mathrm{sil}}\cdot\mathrm{sil}
+w_{\mathrm{dbi}}\cdot(1/(1+\mathrm{DBI}))+w_{\mathrm{cov}}\cdot\mathrm{coverage}\) (scikit\textendash learn metrics).
Stage~A either (i) scans quantiles to set \(C_{\mathrm{thr}}\) (\texttt{tuning=grid}) or 
(ii) runs 5D BO over \((h,R,w_{\mathrm{sil}},w_{\mathrm{dbi}},w_{\mathrm{cov}})\) (\texttt{tuning=bo}). 
Stage~B keeps the grid and \(C_{\mathrm{thr}}\) fixed and sweeps \((\beta,C_{\mathrm{sel}})\) to maximize \(\mathcal{Q}\).

\section{Density Evolution of MD Systems}
\label{app:density_evolution}
To contextualize the clustering analysis presented in 
Secs.~\ref{sec:results_9k} and \ref{sec:results_large}, 
we report the time evolution of the global number density for the molecular 
dynamics systems studied in this work. 
These density traces are shown solely to demonstrate that the selected 
snapshots correspond to physically meaningful stages of crystallization.

Both systems exhibit a clear increase in density after quenching to 
300~K. The specific time steps selected for the clustering analysis are indicated 
by orange dashed vertical lines. 
The lighter orange dashed lines in panel (b) denote 
additional snapshots that were included in the statistical averaging 
procedures reported in Sect.~\ref{sec:results_large}. 

Although density evolution is not used directly in the clustering 
pipeline, it provides independent validation of the physical regimes 
sampled by the selected snapshots and confirms that the clustering analysis 
is performed on representative states of the crystallization process.

\begin{figure}[h!]
\centering
\includegraphics[width=0.98\textwidth]{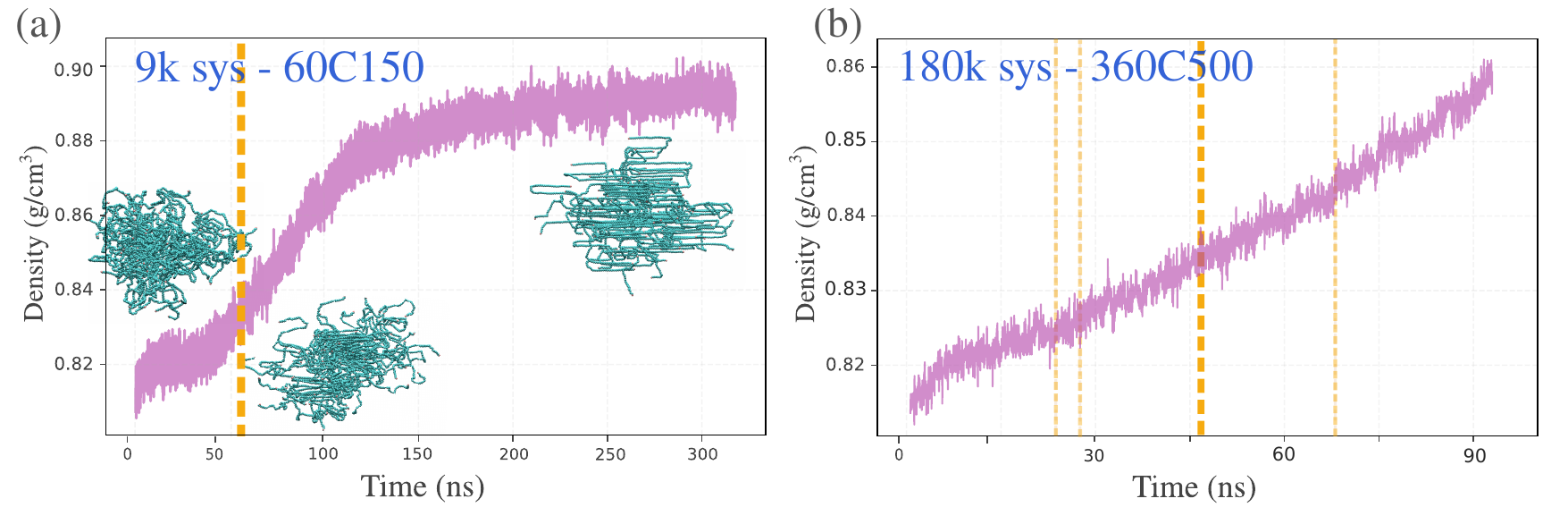}
    \caption{Time evolution of the global number density for the molecular 
    dynamics systems analyzed in this work.
    Orange dashed vertical lines indicate the timesteps selected for clustering 
    analysis, while lighter dashed lines (reduced opacity) denote additional 
    snapshots used for statistical averaging.
    These density traces provide macroscopic context for the clustering results 
    presented in Secs.~\ref{sec:results_9k} and \ref{sec:results_large}, but are 
    not used directly in the clustering pipeline.}
\label{fig:density_time_appendix}
\end{figure}

\section{Selection of the $\alpha$ Parameter for Geometric Cluster Definition}
\label{appendix:alpha_tuning}

The calibration procedure for $\alpha$-parameters follows the same methodology 
introduced in our previous work on directional entropy bands, 
Ref.~\cite{DEB}, and the corresponding density-based diagnostic is reproduced 
here for completeness.

To determine an appropriate $\alpha$ value for geometric surface reconstruction, 
we evaluated the density of $\alpha$-shaped crystalline clusters obtained using 
a range of $\alpha$ values ($\alpha \in [0.01, 1.0]$) throughout the growth 
trajectory. 
For each $\alpha$, the volume of the corresponding $\alpha$ shape was calculated, 
and an effective cluster density was estimated from the number of enclosed atoms, 
normalized by a simulation-specific scaling factor.

\begin{figure}[t]
    \centering
    \includegraphics[width=0.9\linewidth]{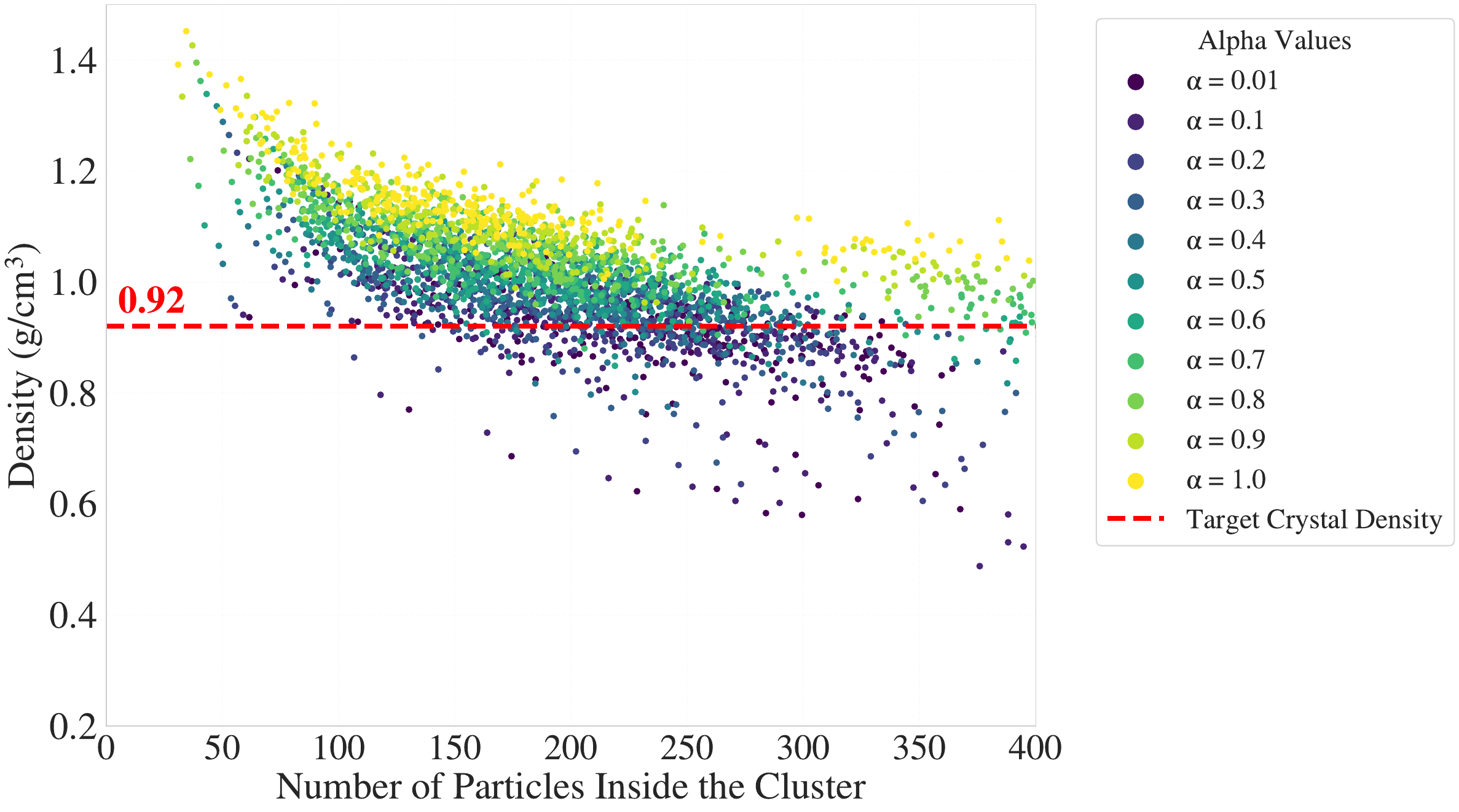}
    \caption{Estimated density of $\alpha$-shaped crystalline clusters as a 
    function of the number of enclosed particles, evaluated across multiple 
    $\alpha$ values throughout the growth trajectory. 
    This diagnostic and calibration procedure was previously introduced in 
    Ref.~\cite{DEB} and is reproduced here for completeness. 
    The red dashed line indicates the equilibrium crystalline density 
    ($0.92~\mathrm{g\,cm^{-3}}$), as obtained from an independently equilibrated 
    bulk simulation using the same force field and molecular model.}
    \label{fig:appendix_alpha_tuning}
\end{figure}

Figure~\ref{fig:appendix_alpha_tuning} reports the resulting density estimates as 
a function of the number of enclosed particles. 
Based on this analysis, $\alpha$ values in the range $0.3 \le \alpha \le 0.7$ 
yield physically consistent density estimates near the independently measured 
crystalline reference value for the simulation setup, 0.92~$\rm{g/cm^3}$.
For geometric comparisons (volume difference and surface area difference), we select $\alpha = 0.5$, which captures a broader set of interfacial atoms while 
avoiding excessive sensitivity to thermal noise. 
We emphasize that $\alpha$ is used exclusively for geometric surface 
reconstruction and does not influence clustering or diffusion-imputation 
procedures.

\section{Additional Heatmap Analysis for the 9k-Atom System}

\begin{figure}
\centering

\begin{minipage}[t]{0.47\linewidth}
    \includegraphics[width=\linewidth]{9k_vol_diff_heatmap_nodiff_highlighted_1.pdf}
    \centering
    {\small (a) Volume difference (\%) for Grid vs.\ Atom.}
\end{minipage}%
\hspace{0.02\linewidth}%
\begin{minipage}[t]{0.47\linewidth}
    \includegraphics[width=\linewidth]{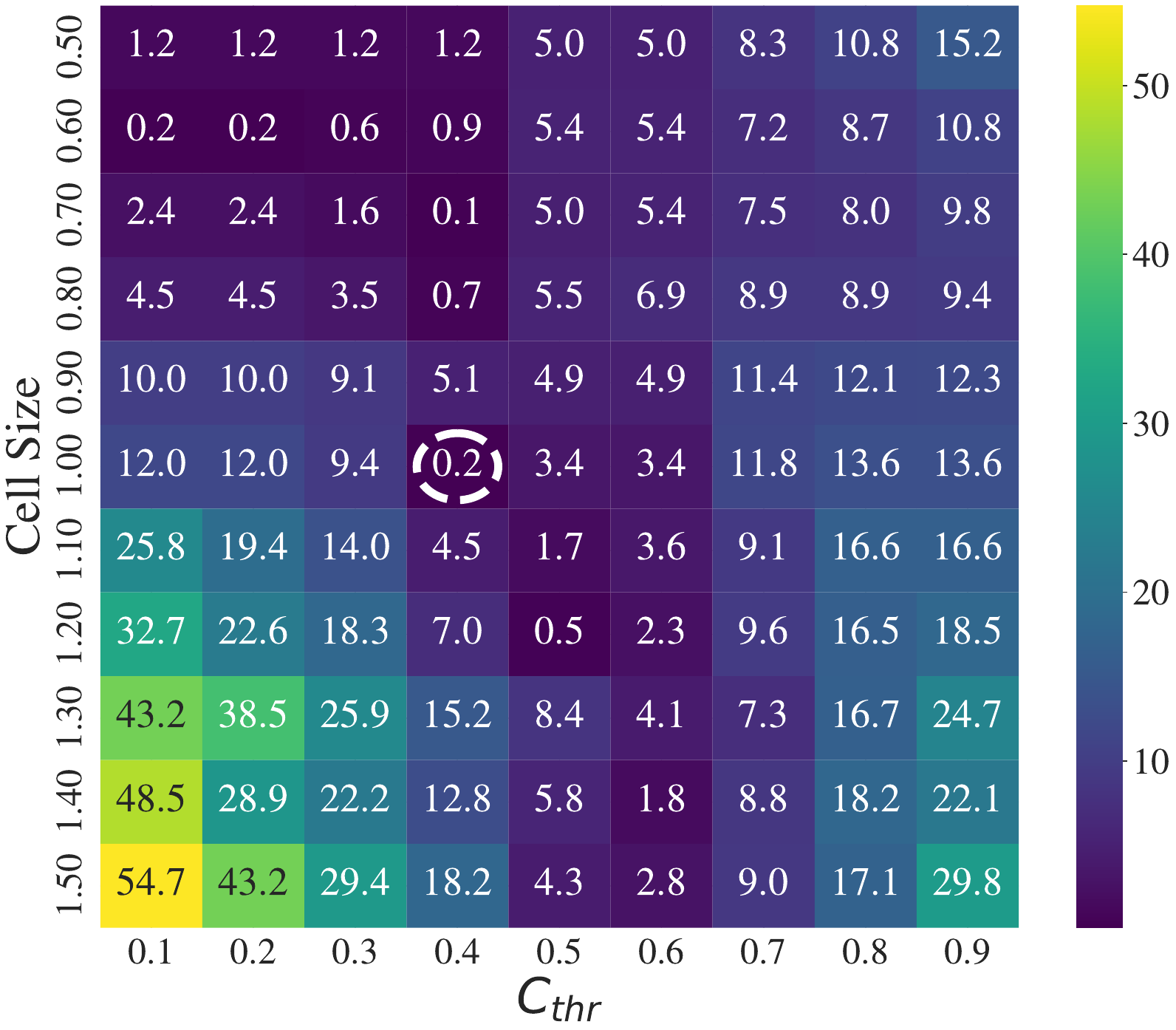}
    \centering
    {\small (b) Volume difference (\%) for Grid+diffusion vs.\ Atom.}
\end{minipage}

\vspace{4pt}

\begin{minipage}[t]{0.47\linewidth}
    \includegraphics[width=\linewidth]{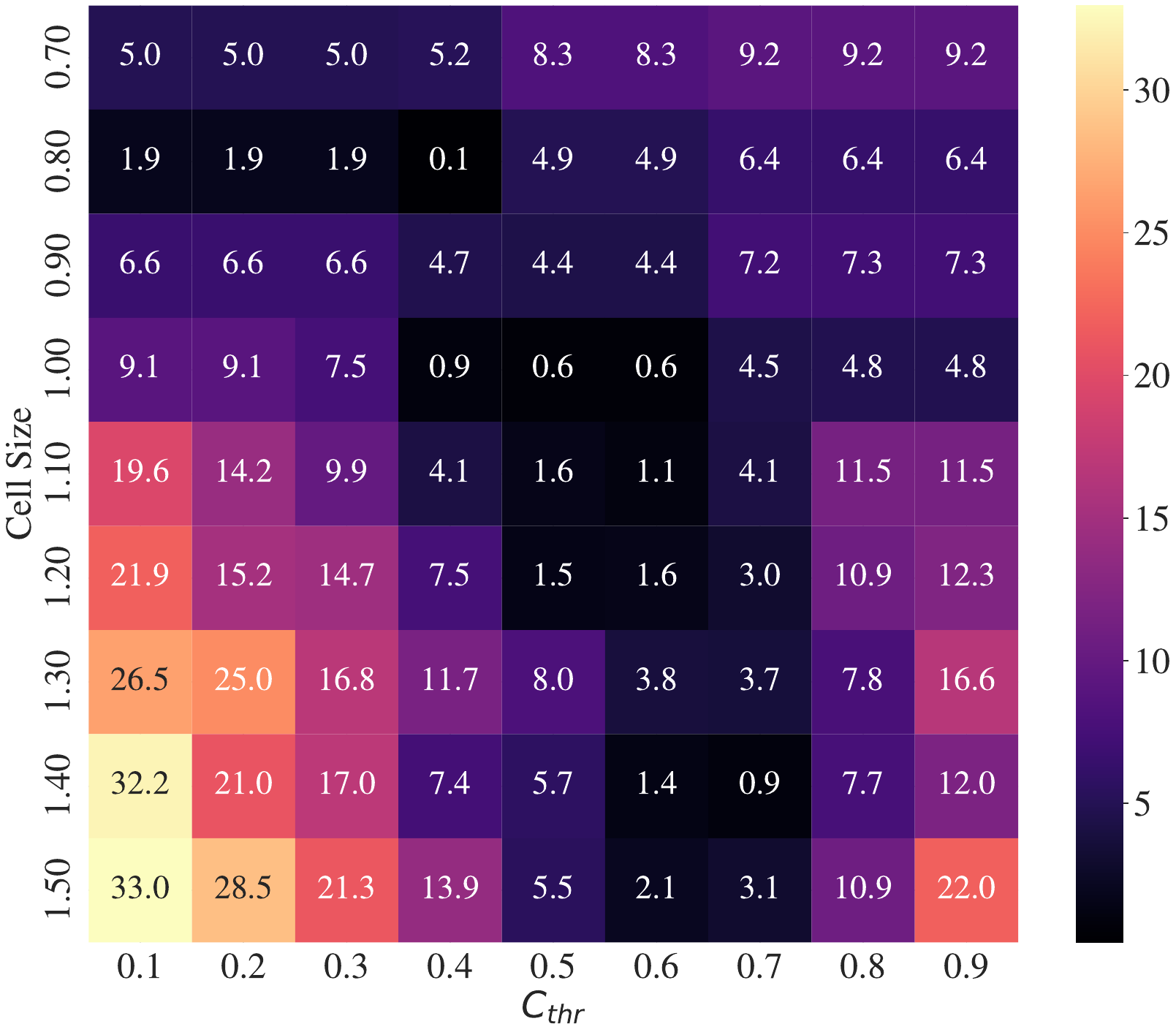}
    \centering
    {\small (c) Surface-area difference (\%) for Grid vs.\ Atom.}
\end{minipage}%
\hspace{0.02\linewidth}%
\begin{minipage}[t]{0.47\linewidth}
    \includegraphics[width=\linewidth]{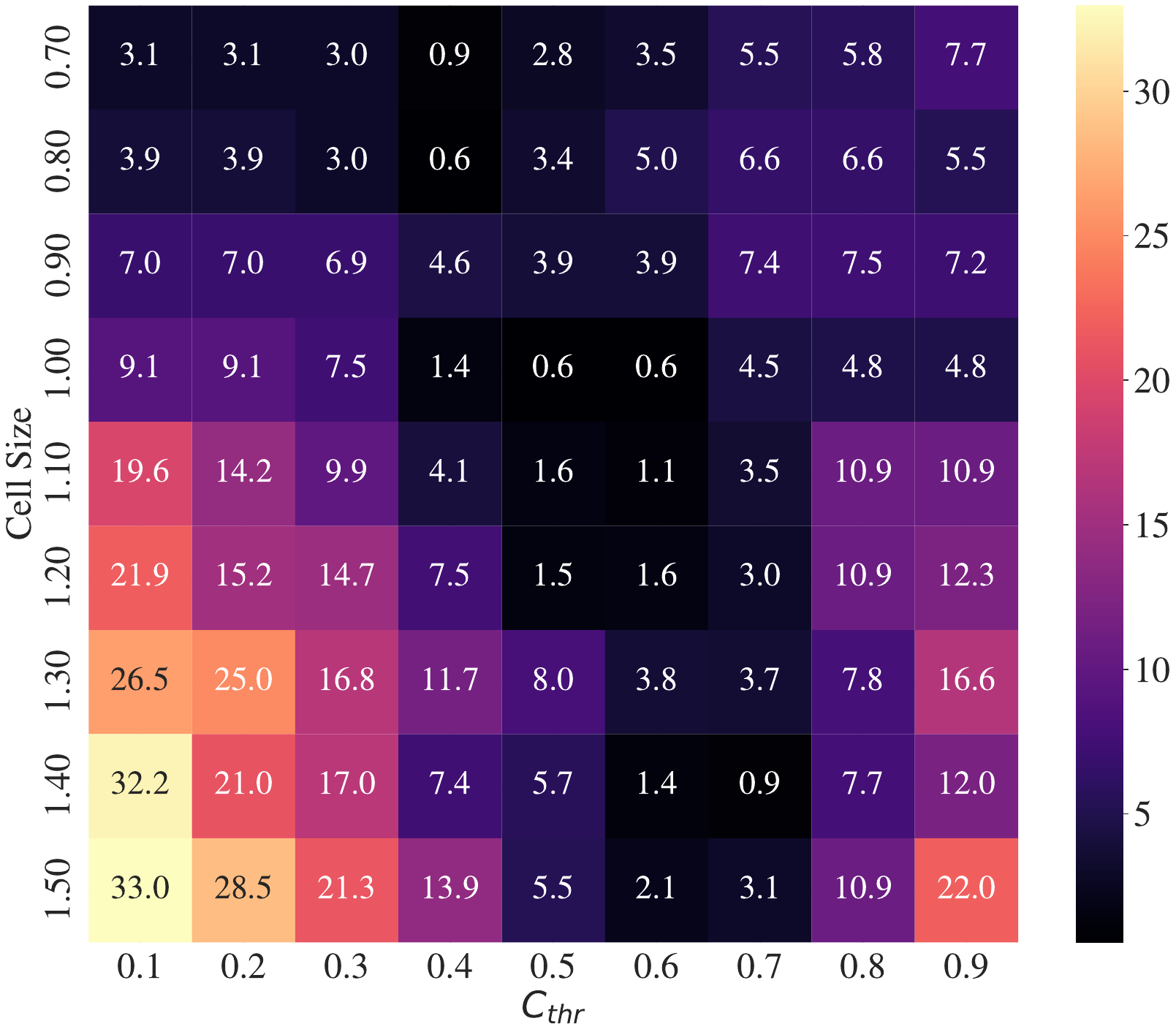}
    \centering
    {\small (d) Surface-area difference (\%) for Grid+diffusion vs.\ Atom.}
\end{minipage}

\caption{Additional error heatmaps for the 9k-atom system. 
    (a) Percent volume difference between grid-based and atom-based cluster 
    $\alpha$-shapes over $(C_\mathrm{thr}, \text{cell size})$ for the non-imputed grid clustering.
    (b) Corresponding percent volume difference for the diffusion-enhanced grid clustering 
    over the same parameter space (the optimal point used in the main text is circled in red).
    (c) Percent surface-area difference for the non-imputed grid clustering.
    (d) Percent surface-area difference for the diffusion-enhanced grid clustering.
    These panels mirror the analysis in Fig.~\ref{fig:gridsize_diffs}(a) and show that conclusions 
    drawn from the volume discrepancy are consistent when surface area and diffusion-imputed clusters 
    are considered.}
\label{fig:appendix_heatmaps_9k}
\end{figure}

To complement the grid-resolution study in Sec.~\ref{sec:results_9k}, 
Fig.~\ref{fig:appendix_heatmaps_9k} presents additional error heat maps for both 
volume and surface-area discrepancies, evaluated over the full parameter space $(C_\mathrm{thr},\text{cell size})$
. These results include both the non-imputed grid clustering and the 
diffusion-enhanced variant. The patterns observed mirror those reported in the main 
text: the optimal region in parameter space is consistent across volume and surface 
metrics, and diffusion imputation improves fidelity while preserving stability across 
a broad range of hyperparameters.

\section{DBSCAN and CLIQUE Three-Dimensional Cluster Visualizations}
\label{app:dbscan_clique_3d}

\begin{figure}
\centering

\begin{minipage}[t]{0.48\linewidth}
    \centering
    \includegraphics[width=\linewidth]{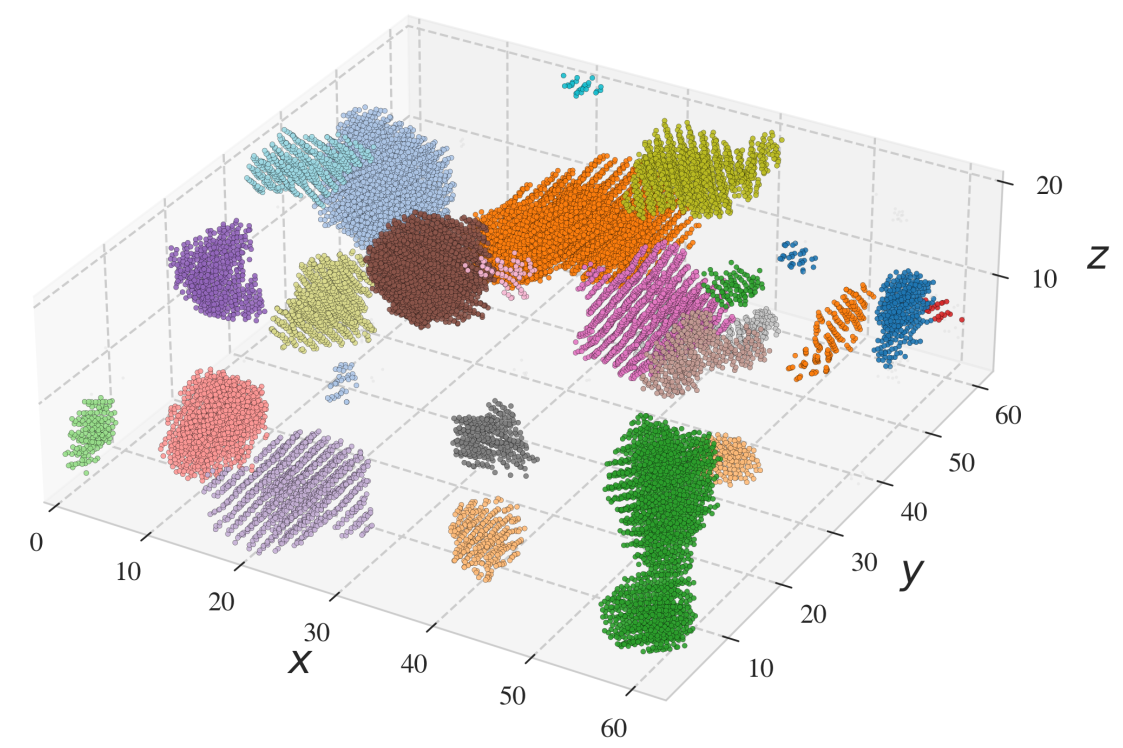}
    \vspace{1mm}
    
    \small (a) \textsc{DBSCAN}, 180k
\end{minipage}
\hfill
\begin{minipage}[t]{0.48\linewidth}
    \centering
    \includegraphics[width=\linewidth]{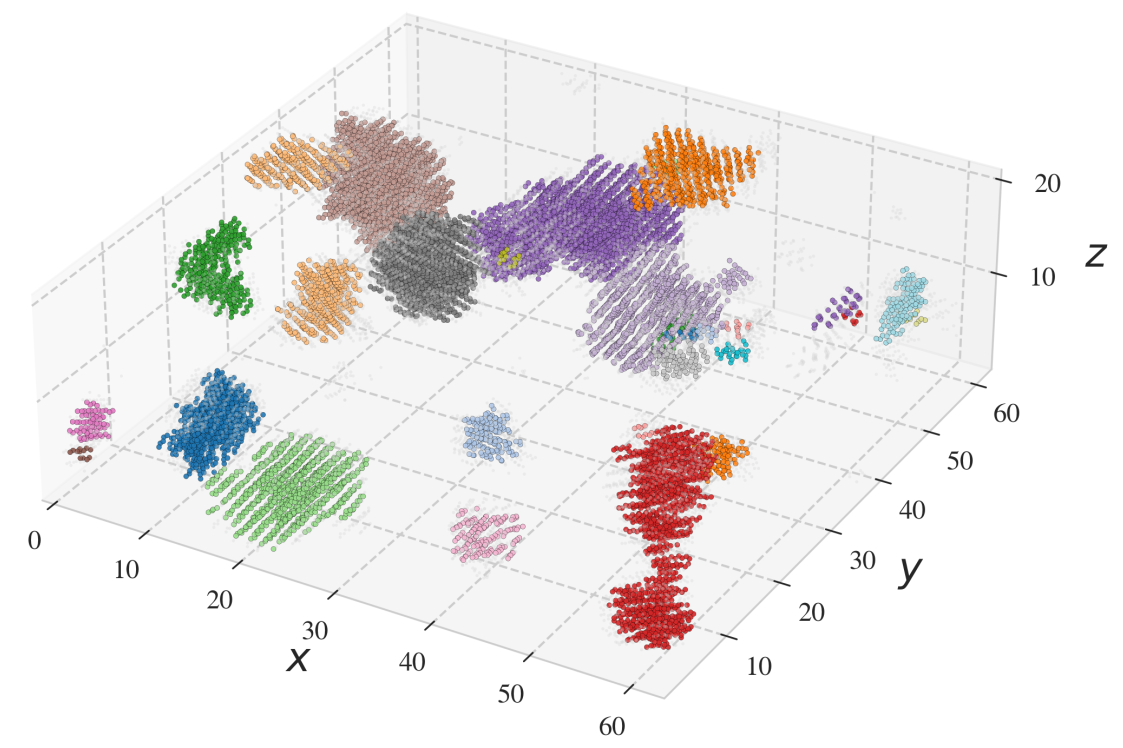}
    \vspace{1mm}
    
    \small (b) \textsc{CLIQUE}, 180k
\end{minipage}

\vspace{4mm}

\begin{minipage}[t]{0.48\linewidth}
    \centering
    \includegraphics[width=\linewidth]{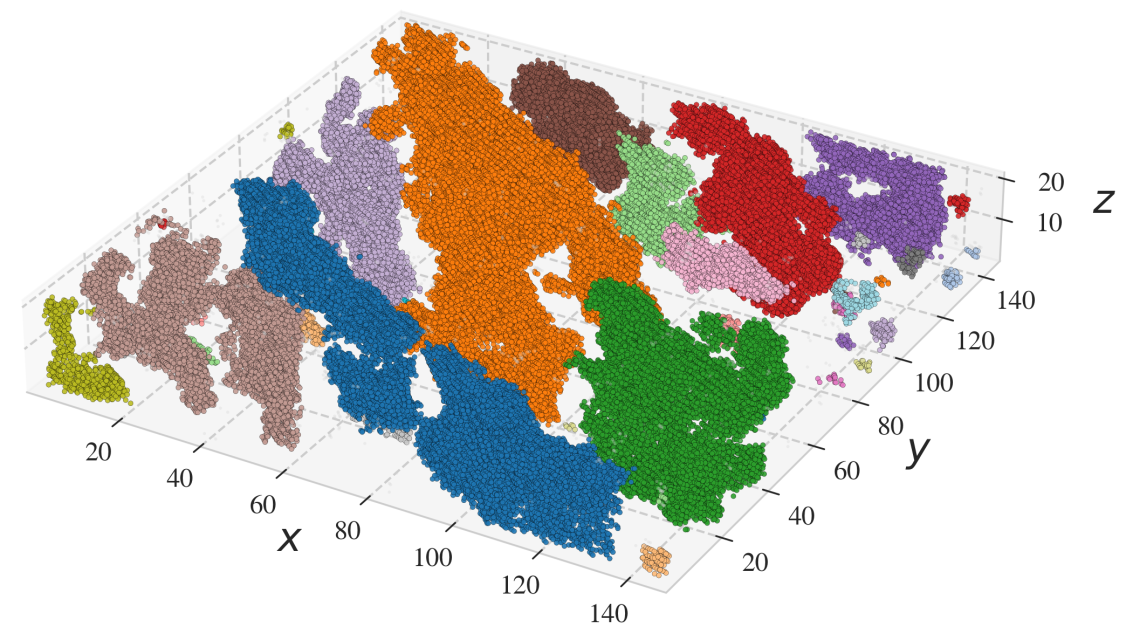}
    \vspace{1mm}
    
    \small (c) \textsc{DBSCAN}, 989k
\end{minipage}
\hfill
\begin{minipage}[t]{0.48\linewidth}
    \centering
    \includegraphics[width=\linewidth]{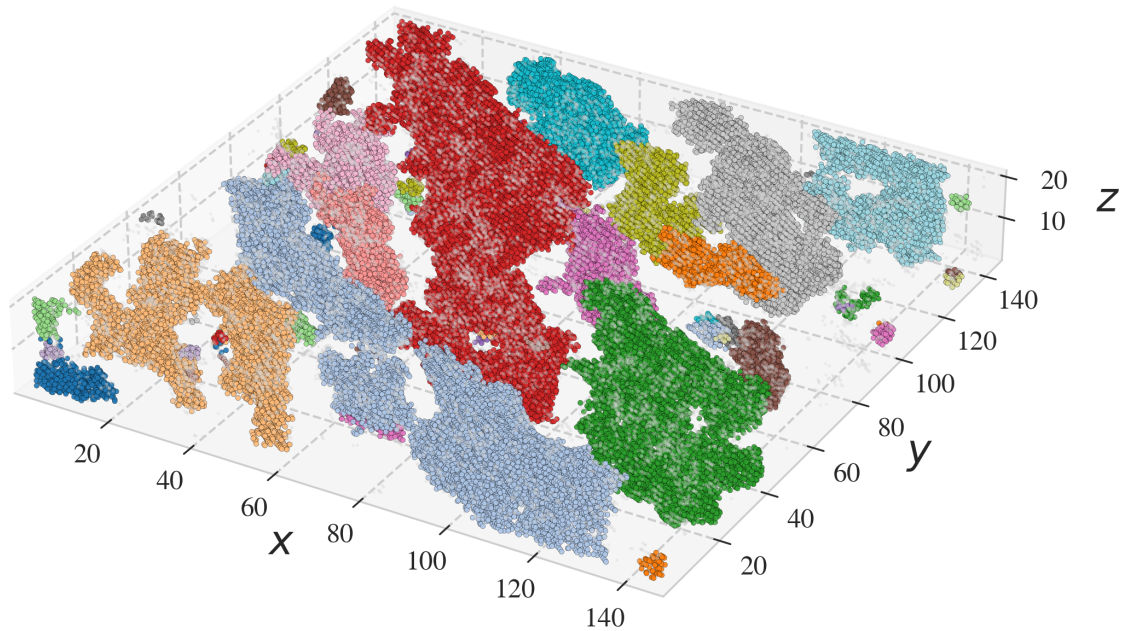}
    \vspace{1mm}
    
    \small (d) \textsc{CLIQUE}, 989k
\end{minipage}

\caption{Three-dimensional visualizations of crystalline clusters identified by 
\textsc{DBSCAN} and \textsc{CLIQUE} for representative snapshots of the 180k 
quiescent system and the 989k polyethylene melt under planar elongational flow. 
\textsc{DBSCAN} exhibits fragmentation of elongated or locally sparse domains, 
while \textsc{CLIQUE} shows grid-induced artifacts and a large number of outlier 
points, particularly in low-density interfacial regions. 
Outliers produced by \textsc{CLIQUE} are rendered in light gray to avoid visual 
occlusion of the dominant crystalline structures; these points are included in 
all quantitative analyses. 
These behaviors contribute to the discrepancies observed in the cluster-size 
distributions of Fig.~\ref{fig:large_systems_overview}.}
\label{fig:dbscan_clique_appendix}
\end{figure}

Figure~\ref{fig:dbscan_clique_appendix} presents fully three-dimensional 
renderings of the crystalline clusters identified by \textsc{DBSCAN} and 
\textsc{CLIQUE} for representative snapshots of the 180k quiescent system 
and the 989k polyethylene melt under planar elongational flow (PEF). 
These visualizations complement the cluster-size distributions shown in 
Fig.~\ref{fig:large_systems_overview} and help to elucidate the sources of the 
observed discrepancies.

In the 180k quiescent system, \textsc{DBSCAN} tends to fragment elongated or 
locally sparse crystalline domains into multiple components, particularly near 
interfaces and low-density bridges. 
\textsc{CLIQUE} occasionally introduces grid-induced artifacts that split 
otherwise continuous structures or suppress thin connections, particularly in 
low-density interfacial regions. 
Similar behaviors are observed in the 989k PEF-driven configuration, where 
flow-induced anisotropy further amplifies the sensitivity to density thresholds and 
grid alignment.

These effects explain the excess of medium-sized clusters and the truncation of 
large components observed in the corresponding cluster-size histograms. 
Although both methods can be tuned to perform well for individual snapshots, their 
limitations become more apparent when applied across heterogeneous morphologies 
and time regimes, motivating the diffusion-enhanced grid strategy adopted in 
ClusTEK3D.

For clarity of visualization, outlier points identified by \textsc{CLIQUE} that 
do not belong to the dominant crystalline components are rendered in light gray. 
The number of such outliers is substantial, particularly in the 989k system, and 
would otherwise obscure the primary cluster structures if plotted with full 
opacity. 
These points are retained in all cluster-size statistics and discrepancy metrics 
reported in the main text; their visual de-emphasis is solely for rendering 
purposes.

\section{Cluster Size Distribution Accuracy Metrics for Large Systems}
\label{app:large_metrics}

Table~\ref{tab:large_emd_ks} reports the full EMD and KS statistics computed for each clustering
method in all sample snapshots of the 180k and 989k systems.  
Lower values indicate better agreement with the atom-based reference.
For the KS values, we also report the per-snapshot $p$-values associated with the KS test. 
Because $p$-values assess statistical significance on a per-snapshot basis and cannot be meaningfully averaged, they are reported individually and are not included in the mean metrics
summarized in the main text.  The mean KS values presented in Table~\ref{tab:mean_emd_ks}
correspond only to the averaged KS statistics, not their $p$-values.

\begin{table}
\caption{EMD and KS statistics comparing each clustering method to the atom-based reference
across all snapshots of the 180k and 989k systems. Lower values indicate better agreement.
KS $p$-values test method vs.\ atom-based reference.}
\label{tab:large_emd_ks}
\centering
\renewcommand{\arraystretch}{1.12}
\setlength{\tabcolsep}{5pt}
\begin{tabular}{ll *{2}{cc} *{2}{cc} *{2}{cc}}
\toprule
& & \multicolumn{2}{c}{\textsc{ClusTEK}} & \multicolumn{2}{c}{\textsc{DBSCAN}} & \multicolumn{2}{c}{\textsc{CLIQUE}} \\
\cmidrule(lr){3-4}\cmidrule(lr){5-6}\cmidrule(lr){7-8}
System & Sample & KS & $p$-value & KS & $p$-value & KS & $p$-value \\
\midrule
\multicolumn{8}{l}{\textbf{KS statistic}}\\
\midrule
180k & 1 & 0.0442 & 1.0000 & 0.0741 & 1.0000 & 0.3370 & 0.0599 \\
180k & 2 & 0.1054 & 0.9945 & 0.0769 & 0.9999 & 0.1883 & 0.7488 \\
180k & 3 & 0.0795 & 1.0000 & 0.1250 & 0.9811 & 0.2381 & 0.4747 \\
989k & 1 & 0.1709 & 0.5497 & 0.1500 & 0.7187 & 0.1688 & 0.6345 \\
989k & 2 & 0.1618 & 0.3016 & 0.1192 & 0.7672 & 0.2059 & 0.0557 \\
989k & 3 & 0.1143 & 0.9758 & 0.1460 & 0.8988 & 0.2886 & 0.1726 \\
\midrule
\multicolumn{8}{l}{\textbf{EMD}}\\
\midrule
180k & 1 & \multicolumn{2}{c}{32.4530}
           & \multicolumn{2}{c}{63.5837}
           & \multicolumn{2}{c}{373.8630} \\
180k & 2 & \multicolumn{2}{c}{67.5084}
           & \multicolumn{2}{c}{21.9385}
           & \multicolumn{2}{c}{160.5547} \\
180k & 3 & \multicolumn{2}{c}{30.8977}
           & \multicolumn{2}{c}{47.9702}
           & \multicolumn{2}{c}{150.0655} \\
989k & 1 & \multicolumn{2}{c}{105.0791}
           & \multicolumn{2}{c}{143.7028}
           & \multicolumn{2}{c}{241.1437} \\
989k & 2 & \multicolumn{2}{c}{515.4534}
           & \multicolumn{2}{c}{1506.4856}
           & \multicolumn{2}{c}{1432.0468} \\
989k & 3 & \multicolumn{2}{c}{322.2595}
           & \multicolumn{2}{c}{1205.1786}
           & \multicolumn{2}{c}{1049.2443} \\
\bottomrule
\end{tabular}
\end{table}


\newpage
\bibliographystyle{unsrt}
\bibliography{ClusTEK_refs}

\end{document}